Technische Universität Braunschweig

Carl-Friedrich-Gauß-Fakultät,
Department Informatik

Institut für Software Systems Engineering

# Modular Description of a Comprehensive Semantics Model for the UML

**Version 2.0**

Informatik-Bericht 2008-06

**Manfred Broy**[1], **María Victoria Cengarle**[1],
**Hans Grönniger**[2] **and Bernhard Rumpe**[2]

[1]Software and Systems Engineering,
Technische Universität München, Germany
[2]Software Systems Engineering,
Technische Universität Braunschweig, Germany

October, 2008

# Contents

# 1 Introduction to the System Model for UML

In this document, we introduce a system model as a semantic domain for the Unified Modeling Language (UML) [OMG07a, OMG07b]. The system model is supposed to form a possible core and foundation of the UML semantics definition. For that purpose, the definitions in this document are targeted towards UML which means that central concepts of UML have been formalized as theories of the system model.

This document is structured as follows: In the rest of Chapter 1, we discuss the general approach and highlight the main decisions. This chapter is important to understand the rest of this document. Chapter 2 contains the definition of the structural part of the system model. Chapters 3 and 4 contain the control and communication related definition definitions which form the basis to describe the state of a system in Chapter 5. Two variants of state transitions systems are introduced to define object behavior in Chapters 6 (event-based) and 7 (timed). Chapter 8 concludes the document.

This document is the second version of the system model which is the result of a major effort to define the structure, behavior and interaction of object-oriented, possibly distributed systems abstract enough to be of general value, but also in sufficient detail for a semantic foundation of the UML. The first version of the system model can be found in [BCR06, BCR07a, BCR07b].

## 1.1 General Approach to Semantics

The semantics of any formal language consists of the following basic parts [Win93]:

- the syntax of the language in question (here: UML) – be it graphical or textual,
- the semantic domain, a domain well-known and understood based on a well-defined mathematic theory, and
- the semantic mapping: a functional or relational definition that connects both, the elements of the syntax and the elements of the semantic domain.

This technique of giving meaning to a language is the basic principle of denotational semantics: every syntactic construct is mapped onto a semantic construct. As discussed in the literature, there are many flavors of these three elements. Syntax can, for example, be specified by grammars or metamodels. To stay formal, our approach intends to use the abstract syntax of UML in a mathematical form that resembles context-free grammars, examples are given in [CGR08b, CGR08a]. In [KRB96] the term system model was used the first time to denominate a semantic domain; it defines a family of systems, describing their structural and behavioral issues. Each concrete syntactic instance (in our case, an individual UML diagram, or even a part of it) is interpreted by the semantic mapping as a predicate over the set of systems defined by the system model. As explained in [HR04] the semantic mapping has the form:

$$Sem : UML \to \mathbb{P}(Systemmodel)$$

and thus functionally relates any item in the syntactic domain to a set of constructs of the semantic domain. The semantics of a model $m \in UML$ is therefore $Sem(m)$.

Given any two models $m, n \in UML$ combined into a complex one $m \oplus n$ (for any composition operator $\oplus$ of the syntactic domain), the semantics of $m \oplus n$ is defined by $Sem(m \oplus n) = Sem(m) \cap Sem(n)$. This definition also works for sets of UML documents which allows an easy treatment of views on a system specified by multiple UML diagrams. The semantics of several views, i.e., several UML documents is given as $Sem(\{doc_1, \ldots, doc_n\}) = Sem(doc_1) \cap \ldots \cap Sem(doc_n)$ A set of UML models *docs* is consistent if systems exist that are are described by the models, so $Sem(docs) \neq \varnothing$. As a consequence, the system model supports both view integration and model consistency verification.

In the same way, $n \in UML$ is a (structural or behavioral) refinement of $m \in UML$, exactly if $Sem(n) \subseteq Sem(m)$. Formally, refinement is the nothing else than "$n$ is providing at least the information about the system that $m$ does". These general mechanisms provide a great advantage, as they simplify any reasoning about composition and refinement operators.

The system model described in this document identifies the set of all possible object-oriented (OO) systems that can be defined using a subset of UML which we call "clean UML" as introduced below. It relies on earlier work on system models [Rum96, KRB96, GKR96, BHH+97, BGH+98, SRS99].

To capture and integrate all the orthogonal aspects of a system modeled in UML, the semantic domain necessarily has to have a certain complexity. Related approaches very often contain a relatively small and specialized semantic domain, such as (pairs of) sets of traces for UML interaction [HHRS05], template semantics based on hierarchical state machines [TA06] or Kripke structures [vdB02] for UML State Machines, or sets of inequations to give semantics to class diagrams focusing on satisfyability of association cardinality [SKU06, CCGM07, MB07, FS07]. However, these approaches fail to give an integrated semantics for different types of UML notations. Approaches with a broader scope are for example [DJPV03] which define a UML subset called krtUML and associates with each model a symbolic transition system. [KGKK02] combine class, object and state machine diagrams using graph transformations. In [ESW07] dynamic metamodeling (also based on graph transformations) is used to define the operational semantics of, e.g., UML activities. Semantics for class and state machine diagrams have been developed for different purposes. [SB06] examines the refinement of associations. [FKdRdB06] provide a compositional semantics that considers activity groups. [Lanar] additionally supports sequences diagrams and considers timing issues. In [ZLQ06] consistency between (simplified) state machines and sequence diagrams is checked using a model checker. Consistency conditions are also proposed [Li06, O'K06].

## 1.2 Structuring the Semantics of UML

Our long term goal is to define the semantics of a comprehensive core of well-defined concepts of UML.

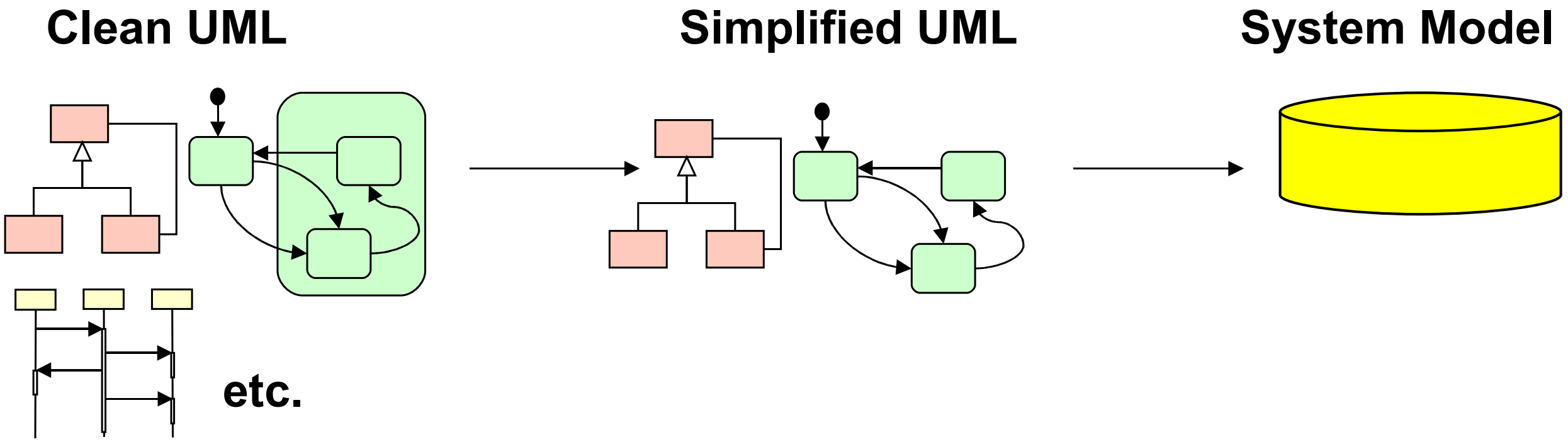

Figure 1.1: General strategy for the definition of the semantics of UML 2.0

The overall strategy of giving semantics to a modeling language is depicted in Figure 1.1. The basic idea expressed by this diagram is as follows:

- Full UML is restricted to a subset (called "clean UML") that can be treated semantically without overly sophisticated constructs.

- Clean UML is mapped by transformations into Simplified UML. In doing so, derived constructs of UML are replaced by their definition in terms of constructs of the core. That way, notational extensions and derived concepts can be eliminated. UML provides a number of derived operators which do not enhance the expressiveness of the language but the comfort of its use. Derived constructs can be defined in terms of constructs of the core as, e.g., state hierarchy of UML's state transition diagrams can be neglected without losing expressiveness.
- Simplified UML, finally, is mapped to the system model using a predicative approach.

The system model describes the "universe (set) of all possible semantic structures (each with its behavior)". The semantic mapping interprets a UML model as a predicate that restricts the universe to a certain set of structures, which represents the meaning of the UML model. To be able to faithfully map concepts from UML to the system model, the system model has to cover a number of basic concepts expressible in UML. Otherwise, the semantic mapping cannot be defined in an adequate manner.

The system model itself is defined in a modular fashion. From a global viewpoint, a system in the system model is a state machine. This semantic universe is introduced in layers of mathematical theories which are shown in Figure 1.2. The links to basic mathematical theories defined in Appendix A (e.g., Function, Logic, etc.) have been left out as they can be used in all theories without an explicit reference.

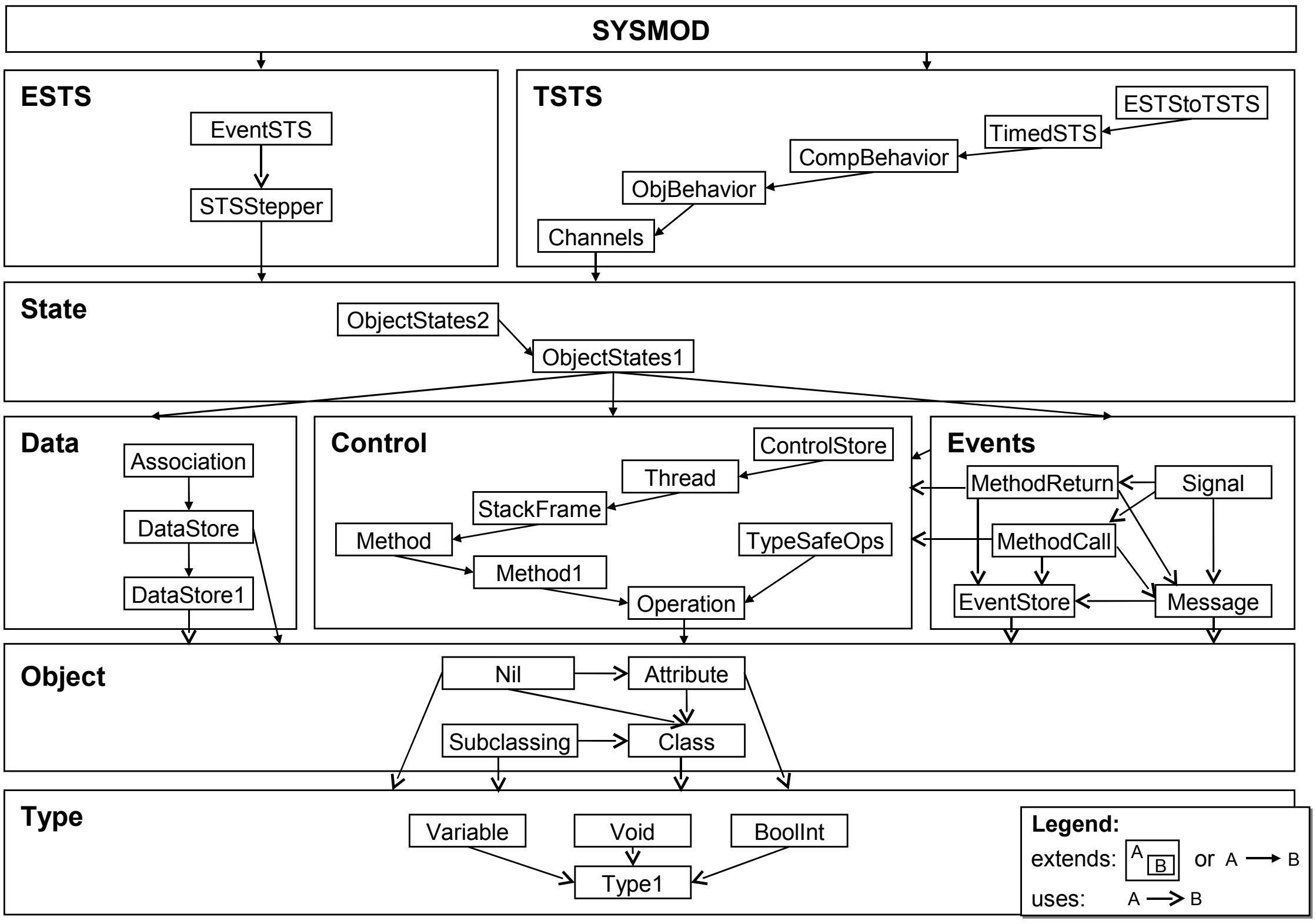

Figure 1.2: Theories that constitute the system model

The rectangles in Figure 1.2 contain names of the theories, whereas arrows show a relationship among concepts that could be paraphrased as "is defined in terms of". For instance, basic theories for types and objects are used to define the data, control, and event state of a system, that in turn are used to define the state space for the transition systems.

When defining the constituents of the system model, we will state the decisions that have to be made, that can be left open or do not even occur when staying informal. We clearly identify those decisions either directly, or mark them as a "variation point" and leave it to the user of the system model to choose or adopt a variation. Those variation points may very nicely correspond to stereotypes on the language side, such that the language designer (and semantics definer) can transfer the freedom of choice to the actual modeler.

## 1.3 The Math behind the System Model

A precise description of the system model calls for a precise instrument. For our purposes, mathematics is exactly appropriate because of its power and flexibility. Admittedly, reading and understanding mathematics is an effort that requires some training, but it allows for precisely and abstractly describing things that cannot be defined using e.g. UML itself. Using UML itself to describe semantics of UML might seem, on the contrary, a pragmatic approach. This approach, however, is somewhat meta-circular and necessarily calls for a kind of bootstrap, typically mathematics again. Moreover, understanding the semantics of UML in terms of UML itself, demands a very good knowledge of the language whose semantics is about to be formally given. Besides, UML does not conveniently provide the appropriate mechanisms we need, e.g., to handle scheduling, distributed systems and to deal with underspecification in a precisely controllable way. Of course, whenever appropriate, we use diagrams to illustrate some mathematically defined concepts, but the diagrams do not replace the mathematical formulas.

Instead of relying on basic mathematics, related work often proposes the use of specialized formalisms. [BF98, ELFR99] translate UML to the formal language Z while [SG06] map to B. Graph transformations are used in [KGKK02]. The process algebra $\pi$-calculus has been proposed to model activities [KKNR06] that also have been formalised using Petri nets [SH05], or Abstract State Machines (ASMs) [SG06]. Trace-based semantics for interactions have been presented in [HHRS05, CK04]. Metamodeling techniques have been employed by [ESW07]. Template semantics [TA06] that are based on state machines allow for describing semantic variation points.

We intentionally avoided the use of more specialized notations such as Z, B, ASMs, etc. for two reasons.

- It is not clear that any of these notations is general and comfortable enough to allow a satisfactory and adequate expression of all concepts in UML.
- Arguably, all these notations have a certain bias (e.g., for state-based formal specification, analysis with a theorem prover, analysis with a model checker); we kept the system model free of this bias to ensure that we obtain a true reference semantics that, if useful, enables the future use of other notations for, e.g., analysis purposes.

Because of these reasons we decided to use only mathematics. The following principles have proven to be useful when defining the system model:

1. Mathematics is used to define the system model. Its sub-theories are built on: numbers, sets, relations, and functions. Additional theories are built in a layered form. That is, only notation and mathematical definitions and neither new syntax nor language are introduced or used in the system model. Diagrams are occasionally used to clarify things, but do not formally contribute to the system model.
2. The system model does not constructively define its elements, but introduces the elements and characterizes their properties. That is, abstract terms are used whenever possible. For instance, instead of using a record to define the structure of an object, we introduce an abstract set of objects and a number of selector functions. Properties of the set are then defined through such selectors. Based on our background and knowledge, we claim that we can transform this system model into a constructive version (and actually do this, cf. [CDGR07]), but that would probably be more awkward to read and less intuitive, as it costs a lot more mathematical machinery. This will satisfy "constructivists" who wish everything being constructive or executable.

3. Everything important is given an appropriate name. For instance, in order to deal with classes, there is a "universe of class names" UCLASS, and similarly there is also a "universe of type names" UTYPE, which however is just a set of names (and not types); see Sects. 2.5 and 2.1 below.

4. To our best knowledge, any underlying assumptions were avoided, according to the slogan: What is not explicitly specified does not need to hold. If we, for instance, do not explicitly state that two sets are disjoint, these two sets might have elements in common. Sometimes these loose (underspecified) ends are helpful to specialize or strengthen the system model and are there on purpose. If you need a property, (a) check whether it is there, (b) if absent, check whether it can be inferred as emerging property, (c) if not, check if it is absolutely necessary, and (d), if yes, you may add it as an additional restriction.

5. Generally, deep embedding (or explicit representation) is used. This means the semantics of the embedded language, i.e., UML, is completely formalized within the supporting language, in our case, mathematics. As one consequence, although there are similar concepts in the language describing the system model (which is mathematics) and the language described (UML), these need not be related. For instance, the system model characterizes the type system of UML, it however does not have and does not need a type system itself.

6. Specific points, where the system model could be further strengthened, have been marked as "variation points". Variation points deal with additional elements that can be defined upon the system model. We may introduce additional machinery that needs not be present in each modeled system. Prominent examples of such variations are the existence of a predefined top-level class called "Object" or an enhanced type system, including, e.g., templates. Furthermore, variations describe changes of definitions, that lead to a slightly different system model. Variation points allow us to describe specialized variants of the system model, that may not be generally valid, but hold for a large part of possible systems. Examples are single inheritance hierarchies or type-safe overriding of operations in subclasses, which may not be assumed in general.

## 1.4 Static and Dynamic Issues

An object-oriented system can basically be described using one of various existing paradigms. We opted for the paradigm of a global state machine in order to accommodate a global (and maybe distributed) state space. The system model, thus, defines a universe of state machines. A state machine is given by its state space, its initial states, and its state transition function. Note that our notion of state machine is more basic and does not directly relate to the state machines/state transition diagrams the UML provides.

The types and classes are static, i.e., they do not change over the lifetime of a system. Similarly, the sets of defined operations, methods, messages, and events do not change. This information is called the static information of a state machine. The set of existing objects, the values of the attributes, the computational state of invoked methods, and dispatched and not yet delivered messages passed from one object to another one are dynamic, i.e., they may change in transition steps. This latter is called the dynamic information of a state machine and is coded in the states of the state machine. In the database realm, the static part is called "schema", and the dynamic part is the "instance". The schema instantiation is changeable while the schema itself is not. Schema changes (usually called "schema evolution" in the literature) are not considered, as they usually do not occur within a running system, but when evolving and/or reconfiguring it.

Summarizing, the state space of the transition system will be defined in terms of the orthogonal constituents data, control, and events. Each of these theories contributes static and dynamic information to the system model definitions.

## 1.5 What is the System Model?

A system model provides a means to define the semantics of any UML model. A system $sm \in SYSMOD$ is defined in terms of a larger number of mathematically defined elements that are subsequently introduced. In general, we introduce our elements in a bottom-up fashion, but we may refer to elements defined later.

Formally, when speaking about a system of the system model, we speak of an instance $sm \in SYSMOD$ as defined in Definition 7.4.1. Hence, a universe *UTYPE* of type names (as will be introduced in Chapter 2 below) defined for $sm \in SYSMOD$, is not necessarily the same in all systems. Therefore, *UTYPE* is a shorthand for $sm.UTYPE$ meaning that *UTYPE* is the universe of type names of the system *sm*. We simply abbreviate to *UTYPE* whenever *sm* is clear from the context which in all the following definitions will be the case, as we will talk about properties of an element *sm* of the system model *SYSMOD*. The same is true for the state machine definition of the system. Each system is equipped with a state machine, i.e., a set of initial states and a transition relation that build upon the defined universes of types, events, etc. So, formally, the property definition starts with $\forall\, sm \in SYSMOD : \ldots$ and ends with the definition of the universe of systems *SYSMOD* in Definition 7.4.1.

The global state machine, if detailed enough, is perfectly appropriate to model parallel, independent and distributed computations. In principle, a system of communicating, elementary state machines could be considered more convenient than a single, global machine for describing the semantics of UML models. It is also possible to construct a global state machine by integrating elementary ones; however, this is a non-trivial operation. Therefore, it is more appropriate to employ the concept/metaphor of one state machine at a higher, non-elementary level. In fact, we introduce a composition operator on state machines representing fragments of larger systems, such that these state machines can be composed, leading to larger state machines.

## 1.6 Notational Conventions

For a good structuring of the mathematical theories we define in the following, we use the following conventions.

Definitions will be given as shown in 1.6.1. They usually contribute new elements to the system model and/or add constraints between these elements. The definitions can be referred to by using *DefinitionName*

**Definition 1.6.1 (This is a definition)**

| *DefinitionName* |
|---|
| extend and use statements referring to other definitions (optional) |
| introduction of new elements (sets, functions, ...) |
| Notation:<br>additional notational abbreviations (optional) |
| definition of properties that hold |
| informal, textual explanation (optional) |

in "use" and "extend" statements. The first (optional) compartment describes on which definitions the theory relies. The symbols from the definitions both in the extend and the use statements, can be used and constrained in the new theory. However, imported symbols are re-exported only if imported through the extend statement. The extend and use statements define a hierarchy of mathematical theories that constitute the system model. In rare occasions, we repeat the imported symbols and their signatures, especially if they are important in the forthcoming definition; we also may specify a context in which the definition is valid.

Noteworthy derived properties following from a definition will be stated as a Lemma with a structure similar to that of a definition. For an example see Lemma 2.5.2.

In addition to the construction of the system model theory, we use examples like depicted in 1.6.2 to demonstrate how this system model can be used. These examples are not formally part of the system model and are referenced only by other examples. Note that an example has essentially the same compartments as a definition.

**Example 1.6.2 (This is how an example looks like)**

| [*ExampleName*] |
| --- |
| import and use statements referring to other definitions, examples (optional) |
| introduction of new elements (sets, functions, ...) |
| definition of properties that hold |
| informal, textual explanation (optional) |

We employ some mathematical machinery to simplify definitions. For example, if a value is not necessary but needs to exist, we use a wildcard $*$. For instance $\forall a : P(a, *, *)$ is equal to $\forall a : \exists y, z : P(a, y, z)$ for otherwise unused variables $y, z$ that are existentially quantified at the innermost level. We also assume a number of container structures, such as $\mathbb{P}(.)$ for powerset, $\mathbb{P}_f(.)$ for finite powerset, $List(.)$, $Stack(.)$, and $Buffer(.)$ defined in mathematical terms with appropriate manipulation and selection functions. This basic mathematics is defined in Appendix A.

## 1.7 System Model Evolution

This document contains the structural part, control (processes, communication, etc.) and a state-based/interaction definition for the system model. Parallel to the process of designing the system model, we are using it to define semantics for some of the most important notational concepts of the UML. Along with this process of defining UML semantics, we hope to be able to enhance the system model defined in here, to lay a solid and also generally acceptable semantic basis for the UML.

# 2 Static Part of the System Model

In this chapter, we introduce the fundamental static part of systems in the system model that will serve to define the semantics of UML models.

The static part is composed of, among other things, some universes of elements, which we assume given and not fully describe here. We define properties and relationships between those universes. For instance:

- the universe of type names *UTYPE*,
- the universe of values *UVAL*,
- a relation *CAR* that associates type names and their possible values,
- the universe of class names *UCLASS*, and
- the universe of object identifiers *UOID*.

Note that we do not further prescribe what "names" are, we take them as primitives.

## 2.1 Type Names and Their Carrier Sets

A type name identifies a carrier set which contains simple or complex data elements called members or values of (or associated with) the type name. Members of all type names are gathered in the universe *UVAL* of values as given in Definition 2.1.1.

**Definition 2.1.1 (Types and values)**

*Type*1

*UTYPE*
*UVAL*
$CAR : UTYPE \to \mathbb{P}(UVAL)$

$\forall\, u \in UTYPE : CAR(u) \neq \varnothing$

*UTYPE* is the universe of type names
*UVAL* is the universe of values
*CAR* maps type names to associated non-empty carrier sets, carrier sets need not be disjoint.

The word "type" has two meanings. On the one hand, a type is a name intuitively understood as a type of any (object-oriented) programming language, whose members do not necessarily own an identity, and which is characterized by the operations it has associated. On the other hand, within the system model we also have a notion of type used to conveniently describe sets of various kinds like, e.g., records and cartesian products (beyond the UML notion of type).

Although we do not deal with peculiarities of various type systems, strong or weak typing, etc., we outline basic assumptions on the underlying type system, as we need to map the type information of UML to this type system.

Type names $T$ of this universe *UTYPE* are normally not detailed further. Although $T \in UTYPE$ models a type, $T$ actually stands for a name, and in short we say type $T$ for it. In that respect, we use a deep embedding of the type system of UML, by representing it through type names and a universe of values only. By deep embedding, we mean that we do not map types of the UML to a type system of the underlying mathematical structure, but explicitly model types as first-class elements.

### 2.1.1 Variation Point: typeOf

In a very general fashion, we do not enforce carrier sets to be disjoint or values to know to which carrier set(s) they belong. For certain type names we may even assume that their carrier sets are identical or in a subset relation. This notion of type allows the subsumption of object types and value types as well as reference types. We may, however, enforce values (or just members of certain types) to have a single (or most specific) type, for instance, by means of a function *typeOf* as defined in 2.1.2 as a partial assignment of a type for each value. In ordinary object-oriented programming languages, objects usually have an assigned type (even though there is subtyping, the assigned type is the class the object is instance of), but special values like *Nil* usually do not.

**Variation Point 2.1.2 (Values having unique types assigned)**

[*typeOf*]

use *Type*1

$typeOf : UVAL \rightharpoonup UTYPE$

$\forall\, v \in UVAL : v \in \mathrm{dom}(typeOf) \Rightarrow v \in CAR(typeof(v))$

*typeOf* partially assigns a type to values, either because this is the "minimal" type or the type is indeed encoded in the value.

A variant of a *typeOf* function, especially suited when no default type is to be assigned to values and the carrier sets are not disjoint, is the introduction of values paired with their type information $(e, T)$ such that $e \in CAR(T)$. So for instance $(3, Int)$ and $(3, Float)$ can be distinguished.

As an aside, note that, in a proper typing system, families of types, together with their functions, form algebras with specific signatures. For details see the concept of abstract data types [LEW97].

### 2.1.2 Typing Examples

The above definitions leave open quite a number of possibilities to characterize types. We will show a few examples, which are not formal part of the system model. In Example 2.1.3 below, *UTYPE* denotes a single type only.

**Example 2.1.3 (Types and values: simple *UTYPE*)**

use *Type*1

$UTYPE = \{Int\}$
$UVAL = \mathbb{Z}$
$CAR(Int) = \mathbb{Z}$

*UVAL* contains integers only. Note that this example contradicts definitions where e.g. Bool is defined as a member of UTYPE and is thus just an example.

Example 2.1.4 shows that the type of a value needs not be unique. This allows, e.g., polymorphic use of functions on values.

**Example 2.1.4 (Types and values: polymorphic values)**

use $Type1$

$Int, Float \in UTYPE$

$\mathbb{R} \subseteq UVAL$
$CAR(Float) = \mathbb{R}$
$CAR(Int) = \mathbb{Z} \subseteq \mathbb{R}$

In this example, *UTYPE* defines types Int and Float, Int values are also Float values.

## 2.2 Basic Type Names and Type Name Constructors

We assume that a number of basic type names for basic values such as Boolean and integer values are given, see Definition 2.2.1. We moreover assume the typical operations on values associated with basic type

**Definition 2.2.1 (Basic types)**

*BoolInt*

use $Type1$

$Bool, Int \in UTYPE$
$true, false \in UVAL$

$CAR(Bool) = \{true, false\}$
$true \neq false$
$CAR(Int) = \mathbb{Z} \subseteq UVAL$

*UTYPE* (*UVAL*) at least contains Boolean and integer (values).

names such as, e.g., logical connectives or arithmetic operators, but do not detail those within the system model.

A special type name is *Void* (see Definition 2.2.2), whose carrier set is a singleton. The value *void* is usually needed for giving semantics to procedures or methods with no return value. This is customary in the semantics of programming languages.

### 2.2.1 Variation Point: Basic Types

Further basic type names –e.g., Real, Character or String and their subtyping relations, if any– are neither assumed nor detailed in this system model, but are natural variation points. Actually, the concrete choice of these types may depend on many factors like, e.g., the hardware platform or the processor. For example the processor may be restricted to integer arithmetics and overflow errors as well as exceptions might be an issue to be modeled here.

It is also possible to model untyped systems or systems with no static type system. In this case, we just introduce the universe of all elements.

**Definition 2.2.2 (Basic type Void)**

*Void*

use $Type1$

$Void \in UTYPE$
$void \in UVAL$

$CAR(Void) = \{void\}$

*void* can for example be used to describe that control is transferred without an actual return value being sent.

It might also be interesting to introduce a notion of equivalence on type names in a form like $T1 \approx T2$ to express that $T1$ and $T2$ represent the same carrier sets, i.e., $CAR(T1) = CAR(T2)$.

## 2.3 Variables

In order to give semantics to attributes of objects, parameters and local variables of method calls and of executions, we introduce a notion of variable names. (For an account on records, cf. *VarAssign* in Definition 2.3.1, see Appendix A.3.)

**Definition 2.3.1 (Variables, attributes, parameters)**

*Variable*

use $Type1$

$UVAR$
$vtype : UVAR \to UTYPE$
$vsort : UVAR \to \mathbb{P}(UVAL)$
$VarAssign = RECORD(UVAR, vsort)$

Notation:
$a : T$ denotes a typed variable and the name and type of the variable are stated explicitly. Note that $vtype(a : T) = T$.

$\forall\, v \in UVAR : vsort(v) = CAR(vtype(v))$

*UVAR* is the set of all variable names in the system model. For simplicity, we assume that each variable name has a unique type assigned.
*VarAssign* is the set of all total and partial variable assignments for variables from *UVAR*.

### 2.3.1 Visibility and Unique Variable Names

As specified in Definition 2.3.1, we assume that each variable name has a unique type assigned. In practice, it would be relatively unhandy if every variable name could only be used once in a program. We then would see a global namespace and thus not have any hiding concepts in the language. In the system model, however, we may accept such a restriction, and handle it as follows. Like in ordinary programming languages, variables shadow each other when a new variable with the same name is introduced in an inner scope. We assume static binding, thus each variable name can be statically resolved (as opposed to dynamic binding of variables by which the resolution of a variable name depends not on the environment of its definition but on the environment of its use, and thus variable resolution can only occur at run time). Generally, we assume that in the modeling languages we deal with, a consistent and model-wide redefinition of variable names is possible in such a way that each variable is used only once. Then variable shadowing does not occur and any variable is unique. We may handle that systematically through encoding the place of definition or the namespace within each variable. Quite the same is done by many compilers anyway. Example 2.3.2 demonstrates how this can be achieved by use of dot notation, i.e., prefixing the namespace, e.g., the class containing an attribute $age : Int$ or the class and method name for a parameter $age : Days$ of method *buy*.

**Example 2.3.2 (Unique variables distinguished by namespace)**

use $Type1, Variable, BoolInt$

$Person.age, Fruit.buy.age \in UVAR$

$vtype(Person.age) = Int$
$vtype(Fruit.buy.age) = Days$

Names from *UVAR* include syntactically resolvable namespaces.

If necessary, class names can be further qualified with e.g. package name.

## 2.4 Summary of Types, Variables and Values

The theory built so far constitutes the basic constraints for values, types and variables. The theory *Type* (Definition C.1.1 in Appendix C) summarizes these definitions. We build all further theories on them as a basis.

Figure 2.1 graphically illustrates the theory dependencies. It is an excerpt of Figure 1.2 but additionally shows the variation points for the current theory.

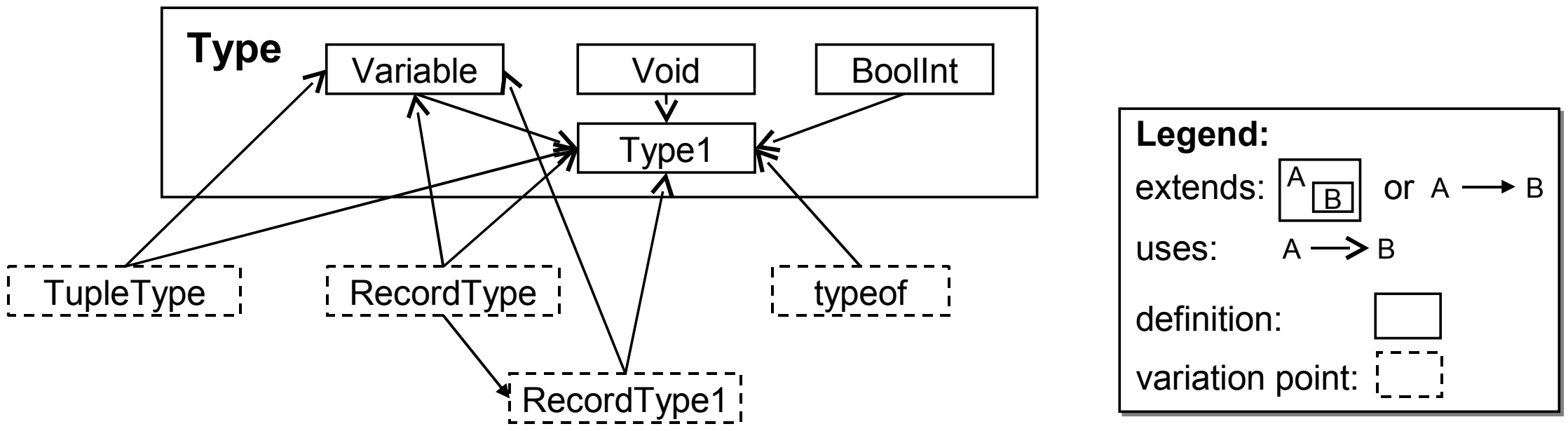

Figure 2.1: Theory *Type* and its dependencies.

### 2.4.1 Variation Point: Records

Records can be defined using the following Variation Point 2.4.2, which relies on the definition of mathematical records in the Appendix A.3. There we already defined the notion of records, but not a type constructor. Note that records are structurally rather similar to classes, but serve different purposes. We therefore do not mix those two concepts. Furthermore, the explicit notion of the element types is not necessary, as each variable is unique and has a unique type assigned via function *vtype*. The definition of *SRec* can be found in Appendix A.3.

**Variation Point 2.4.1 (Basic structure of record types)**

*RecordType1*

use *Type1*, *Variable*

$TRec : \mathbb{P}_f(UVAR) \to UTYPE$
$RECORD(UVAR, vsort) \subseteq UVAL$

Notation:
$Rec\{a_1 : T_1, \ldots, a_n : T_n\}$ is shorthand for the type
$TRec(\{(a_1, T_1), \ldots, (a_n, T_n)\})$

$\forall\, i : vtype(a_i) = T_i \Rightarrow$
$CAR(Rec\{a_1 : T_1, \ldots, a_n : T_n\}) =$
$SRec(\{(a_1, \ldots, a_n\}, vsort)$

The notations used for record values $[a_1 = v_1, \ldots, a_n = v_n]$ and for record types $Rec\{\ldots\}$ provide a common notation.

**Variation Point 2.4.2 (Records including attribute selection)**

*RecordType*

extend *RecordType1*
use *Type1*, *Variable*

$TRECORD \subseteq UTYPE$
$attr : RECORD(UVAR, vsort) \to \mathbb{P}_f(UVAR)$
$attr : TRECORD \to \mathbb{P}(UVAR)$

$TRECORD = \{Rec\{a_1 : T_1, \ldots, a_n : T_n\} \mid n \in \mathbb{N}_0, a_i \in UVAR, T_i \in UTYPE\}$
$attr(Rec\{a_1 : T_1, \ldots, a_n : T_n\}) = \{a_1, \ldots, a_n\}$
$attr([a_1 = x_1, \ldots, a_n = x_n]) = \{a_1, \ldots, a_n\}$

*attr* is the list of attribute names.
*TRECORD* contains all record types.

For access of record variables we define auxiliary functions in Variation Point 2.4.2.

Any list of type names can be composed into record type names. The variables $a_i$ are called the attributes of the record type name. Notice that, as *Rec* is defined on (finite) sets of pairs, the definition of *Rec* does not rely on the ordering of its attributes, thus $Rec\{a : T, b : S\}$ and $Rec\{b : S, a : T\}$ describes the same type name.

### 2.4.2 Variation Point: Cartesian Products

Some languages also provide cartesian products (also called “cross products” or “tuples”) as types. In the following Variation Point 2.4.3, we introduce tuple types of arbitrary size. *STuple* is introduced in Appendix A.4.

## 2.5 Class Names and Objects

Given a number of mathematical prerequisites, we now build the notion of objects and classes on top.

**Variation Point 2.4.3 (Cartesian products)**

*TupleType*

use $Type1, Variable$

$Tuple : List(UTYPE) \to UTYPE$
$TUPLE(UVAL) \subseteq UVAL$

$CAR(Tuple[T_1, \ldots, T_n]) = STuple[CAR(T_1), \ldots, CAR(T_n)]$

$Tuple[\ldots]$ acts as type constructor.

A class name defines attributes and methods, and may be related (by associations) to other class names. At first we concentrate on the structure defined for class names. As Definition 2.5.1 shows, each class has a set of object identifiers and a set of attributes associated. This is sufficient to define the structure of objects belonging to a class in form of a tuple, consisting of object identifier (*this*) and the record of all attributes.

**Definition 2.5.1 (Classes and instances)**

*Class*

use $Type$

$UCLASS, \quad UOID, \quad INSTANCE$
$attr : UCLASS \to \mathbb{P}_f(UVAR)$
$oids : UCLASS \to \mathbb{P}(UOID)$
$objects : UCLASS \to \mathbb{P}(INSTANCE)$
$objects : UOID \to \mathbb{P}(INSTANCE)$
$classOf : INSTANCE \to UCLASS$
$classOf : UOID \to UCLASS$

$\forall\, C \in UCLASS, oid \in UOID :$
$objects(oid) = \{(oid, r) \mid r \in VarAssign \wedge attr(r) = attr(C)\}$
$objects(C) = \bigcup_{oid \in oids(C)} objects(oid)$
$\forall\, oid \in oids(C) : classOf(oid) = C$
$\forall\, o \in objects(C) : classOf(o) = C$

*UOID* contains the universe of object identifiers, *UCLASS* class names and *INSTANCE* objects.
*attr* assigns attributes to each class.
*oids* assigns a set of object identifiers to a class.
*classOf* ensures that each object and each identifier knows its class.

Object identifiers uniquely point to objects and we do not have dangling references, so there is a bijection between object identifiers and objects. This allows to uniquely define the class of an object identifier (except the below introduced *Nil*). This means, an object knows its identifier and its class. As a consequence of this definition each object belongs exactly to one class. For handling of polymorphism, see Section 2.6. This also ensures that structurally equivalent classes can be distinguished. Furthermore, this “belonging” also does not vary over time, whereas the object value can vary and a dereferencing from object identifier to the object value is state dependent.

Note that *UOID* contains references to all possible objects and, in a similar way, *INSTANCE* contains all possible objects. These sets are usually infinite because they resemble the possible existence of objects. Furthermore, *INSTANCE* contains all object values thus describing many different object values with the same identifier. At each point of time only a finite subset of objects will actually exist in the data store (see Section 2.8 below) and there will be at most one instance for any identifier.

In an earlier version, we considered a class itself to be a type [BCR06, BCR07a, BCR07b, CDGR07] and objects to be values. However, this led to a number of tricky encodings of the special variable *this*, which is taken out of the variable assignment and now stored as an extra part of the object. Objects still know about their identity as well as their class, but *this* is treated in a special form.

From Definition 2.5.1 we can derive Lemma 2.5.2.

**Lemma 2.5.2 (*oids* is disjoint)**

use *Class*

$$\forall\, oid \in UOID : oid \in oids(classOf(oid))$$
$$\forall\, C_1 \neq C_2 \in UCLASS : oids(C_1) \cap oids(C_2) = \varnothing$$
$$\forall (oid, r) \in objects(C) : classOf(oid, r) = classOf(oid)$$

follows from the existence of *classOf*.

In Definition 2.5.1, we have not yet decided whether classes will be types and if yes what their carrier sets are, and whether objects or object identifiers are values. This will be done in subsequent definitions. Objects will not be forced to be values: the identifiers are passed around as argument and handled as values, not the objects themselves.

Definition 2.5.3 handles access to attributes within an object. While *this* is not an attribute and thus does not appear in $attr(C)$, it can however be treated as it were an attribute. This does not enforce to associate a type with *this* and we therefore get out of a bunch of problems starting with, e.g., recursive type definitions.

The following Definition 2.5.4 introduces the special identifier *Nil* and constrains *UOID* to exactly consist of object identifiers and *INSTANCE* objects only.

## 2.6 Subclassing

*Subclassing* (also called *inheritance*) is a basic feature in object-oriented programming. To indicate that a class $C_1$ inherits from a class $C$, we introduce binary subclass relation *sub* on the universe of types in Definition 2.6.1.

Given the subclass relationship we are also able to precisely define what we understand under the type of a class: We type the object identifiers (instead of the object themselves). Object identifiers are stored in variables, can be passed as parameters, etc. So we use these identifiers as values, and leave open whether objects are also to be treated as values (see below for an appropriate variation point).

The above definition is sufficient to capture subclassing from the structural viewpoint. As an important consequence, it shows that the type $C^{\&}$ has as carrier set all object identifiers belonging to that class or any subclass. Therefore the carrier sets of subclasses are included in those from superclasses (see derived lemma, following Definition 2.6.1). This allows to polymorphically store subclass identifiers in places where superclass identifiers are expected.

However, the definition also leaves quite a few things open for refinement. For instance the binary relation *sub* is not enforced to be antisymmetric (although no implementation language supports this today). Furthermore, subclassing is not based on a structural definition: two classes $C_1$ and $C_2$ may have the same attributes, but still be in no relationship at all.

With this technique on defining a subset relation on object identifiers instead of objects, we get a great simplification on the type system within the system model. Furthermore, it allows us to redefine attribute structures in subclasses without an otherwise necessary loss of the substitution principle.

**Definition 2.5.3 (Attribute access)**

*Attribute*

use *Class*, *Type*

$this : INSTANCE \rightarrow UOID$
$getAttr : INSTANCE \times UVAR \rightharpoonup UVAL$
$attr : INSTANCE \rightarrow \mathbb{P}_f(UVAR)$
$attr : UOID \rightarrow \mathbb{P}_f(UVAR)$

Notation:
$o.this$ is shorthand for $this(o)$
$o.a$ is shorthand for $getAttr(o,a)$

$this((oid,r)) = oid$
$getAttr((oid,r),a) = r.a$
$attr(oid) = attr(classOf(oid))$
$attr(o) = attr(classOf(o))$

*o.this* is written in the spirit of attribute selection, but treated differently. *this* is not an actual attribute of the class.

Derived lemma:
$\forall\, o \in INSTANCE : classOf(o.this) = classOf(o) \wedge o.this \in oids(classOf(o))$
$\forall\, o \in INSTANCE : o.this = oid \Leftrightarrow o \in objects(oid)$

**Definition 2.5.4 (Introduction of *Nil*)**

*Nil*

use *Class*, *Type*, *Attribute*

$Nil \in UOID$

$\forall\, C \in UCLASS : Nil \notin oids(C)$
$\forall\, o \in INSTANCE : o.this \neq Nil$
$UOID = \{Nil\} \cup \bigcup_{C \in UCLASS} oids(C)$
$INSTANCE = \bigcup_{C \in UCLASS} objects(C)$

*Nil* is a special oid and the only one not associated to a class or an object. *UOID* and *INSTANCE* only consist of identifiers resp. objects.

Derived lemma:
$\forall\, o \in INSTANCE : o \in objects(classOf(o))$
$INSTANCE = \bigcup_{oid \in UOID, oid \neq Nil} objects(oid)$

Remark: Multiple inheritance allows a class to inherit features from more than one class. While a constructive class definition inherits from several classes, from a relational point of view multiple inheritance is covered by several binary inheritance relationships.

To avoid name conflicts that arise when attributes of different superclasses (or of a superclass and of the extension) are homonyms, we simply assume that all attribute definitions introduce different names.

**Definition 2.6.1 (Subclassing)**

*Subclassing*

use *Class*, *Nil*, *Type*

$sub \subseteq UCLASS \times UCLASS$
$.^{\&} : UCLASS \rightarrow UTYPE$

$UOID \subseteq UVAL$

$transitive(sub) \wedge reflexive(sub)$
$\forall\, C \in UCLASS : CAR(C^{\&}) = \{Nil\} \cup \bigcup_{C_1\ sub\ C} oids(C_1)$

*sub* is the transitive and reflexive subclass relation.
Type $C^{\&}$ contains all object identifiers that belong to class $C$ or any of its subclasses.

Derived lemma:
$\forall\, C \in UCLASS : Nil \in CAR(C^{\&}) \wedge oids(C) \subseteq CAR(C^{\&})$
$UOID = \cup_{C \in UCLASS} CAR(C^{\&})$
$\forall\, C_1\ sub\ C : CAR(C_1^{\&}) \subseteq CAR(C^{\&})$

Semantically, this convention means no restriction because attribute access is always resolved statically and there is no dynamic lookup for attributes.

## 2.7 Summary of Classes and Objects

We summarize the theory built so far, defining objects, identifiers and classes in theory *Object* (Definition C.1.2).

Figure 2.2 illustrates the theory dependencies and variation points.

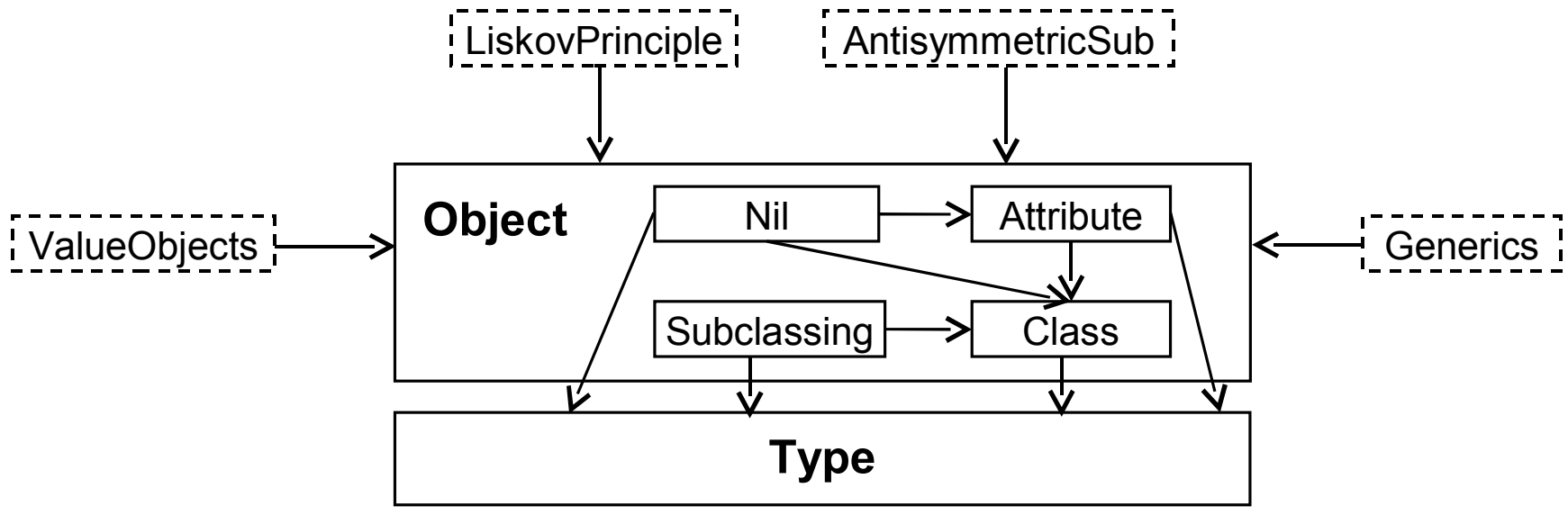

Figure 2.2: Theory *Object* and its dependencies.

As mentioned earlier, quite a number of variation points arise to extend and specialize the theory *Object*.

### 2.7.1 Variation Point: Subclassing Respects Structure

From the definition, we can see that a subclass can have a different attribute structure. This is methodically somewhat questionable and often cannot be reflected in an implementation anyway. Therefore, the constraints may be refined as a variation point. The substitution principle [LW94] enforces object identifiers of subclass $C_1$ to be special cases of class $C$. That can easily be enforced by introducing a set inclusion on attributes for classes in subclass relation as given in Variation Point 2.7.1.

**Variation Point 2.7.1 (Subclassing respects structure)**

[*LiskovPrinciple*]
use *Object*

$$\forall C_1, C_2 \in UCLASS : C_1 \, sub \, C_2 \Rightarrow attr(C_2) \subseteq attr(C_1)$$

It is enough to enforce attributes being included, as they exhibit the same types by Definition 2.3.1.

### 2.7.2 Variation Point: Antisymmetric Subclassing

Variation Point 2.7.2 enforces the subclass relation to be antisymmetric which implies that there are no inheritance cycles.

**Variation Point 2.7.2 (Antisymmetric subclass relation)**

[*AntisymmetricSub*]
use *Object*

$$\forall C_1, C_2 \in UCLASS : C_1 \, sub \, C_2 \wedge C_2 \, sub \, C_1 \Rightarrow C_1 = C_2$$

### 2.7.3 Dynamic Reclassification of Objects

Notice that an object may be regarded as instance of more than one class along the subtyping hierarchy, even though its object identifier is uniquely tied to a fixed, unchangeable class. Still an object may be dynamically reclassified by its context according to the given subclass hierarchy. However, this only changes the external viewpoint on an object, but neither its internal structure (attributes) nor its behavior.

In UML 2.0 dynamic reclassification for classes is introduced in a very general way. The system model does not reflect this capability of reclassification, because we assume that this concept should be mapped to the system model through introduction of additional infrastructure. For instance, possible implementations of dynamic reclassification could introduce an additional superclass that contains all attributes and a flag to indicate which behavior is currently active. Even more flexibility becomes possible when change of dynamic behavior is realized through delegation of behavior to other objects.

### 2.7.4 Variation Point: Objects are Values

We have decided not to enforce objects to be values because there was no necessity. However, if “value objects” are desired that are, e.g., passed around directly instead of their identifiers, then an corresponding variation may be defined.

During our considerations we found it clearer to explicitly distinguish between a class name and the associated type it induces: A class *Person*, e.g., induces the type $Person^{\&}$ of object identifiers of that class. Now we also add the type $Person^{*}$ of objects.

We use $.^{\&}$ and $.^{*}$ as type constructors to exhibit similarities to C++, but use them as suffix to avoid confusion. Furthermore, with this approach, we do not get a recursive type definition for types $C^{*}$ and $C^{\&}$ since they formally are not related to each other. This reflects implementation very straightforwardly, as in there we also distinguish object identifiers quite clearly from the objects themselves.

Note that we can derive that there is no inclusion of object sets in this variation point and thus no substitution of objects. However, another variation point could make that differently. Also note that

**Variation Point 2.7.3 (Objects are values)**

[*ValueObjects*]

use *Object*

$INSTANCES \subseteq UVAL$

$.^* : UCLASS \to UTYPE$

$\forall\, C \in UCLASS : CAR(C^*) = objects(C)$

$C^*$ is a type denoting the objects of class $C$.

Derived lemma:

$\forall\, C_1 \neq C_2 : CAR(C_1^*) \cap CAR(C_2^*) = \varnothing$

when *typeOf* was defined like in Variation Point 2.1.2, we have: $classOf(oid)^{\&} = typeOf(oid)$ and $classOf(obj)^* = typeOf(o)$.

### 2.7.5 Variation Point: Generic Type System

Further constructs for building type names are possible. For instance, an array type or a subtyping structure beyond the subclassing concept inherent to object orientation may be available. We also did not deal either with parametric polymorphism or generic classes within the system model, which was introduced in Java 1.5 in form of instantiable templates. A type system is an enhanced syntactic concept and can therefore be handled together with the concrete syntax of the models.

Variation Point 2.7.4 describes a simplified version of generics, where constraints on the classes that can be instantiated are not described. We model "generic classes" not directly as classes, but as functions yielding classes when instantiated.

**Variation Point 2.7.4 (A sketch for Generic Classes)**

[*Generics*]

use *Object*

$GENERICS$

$parcount : GENERICS \to \mathbb{N}$

$attr : GENERICS \to \mathbb{P}_f(UVAR)$

$build : GENERICS \times List(UTYPE) \rightharpoonup UCLASS$

Notation:

$parcount(C) = n \Rightarrow C(T_1, \ldots, T_n)$ is the class $build(C, [T_1, \ldots, T_n])$

$attr(C(T_1, \ldots, T_n)) = attr(C)$

This is a sketch how to introduce generics. (More definitions are necessary.)

Note that in case of generics, attribute names may have different types in different contexts. This can be handled by attaching type names to variables and actually leads to a slightly more involved definition of $attr(C)$ than stated in Variation Point 2.7.4.

## 2.8 Data Store Structure

In the system model, we abstract away a number of details, such as storage layout and physical distribution. We use an abstract global store to denote the state of an object system. Even if there is no such concept in the real, possibly distributed system, we can conceptually model the system that way by organizing all instances in this single global store. We also allow interleaving, as well as concurrent activities, as can be seen in the control part of the system model in Chapter 3.

Intuitively, the data store models the data state of a system at a certain point in time. Normally, at each point of time the store contains real objects for a finite subset of the universe *UOID* of all object identifiers. In Chapter 5 we will, however, see that the data store is not enough to describe the system, but a control store and an event store need to be added. In these stores time progress is modeled by state transitions of the overall state machine.

A data store is a snapshot of the data state of a running system. Definition 2.8.1 introduces a stores as a set of objects assigned to their identifiers. As an important restriction on *DataStore*, we enforce that the mapping assigns an object *o* to identifier *oid* only if this is the identifier of that object *o.this*.

**Definition 2.8.1 (The data store)**

*DataStore*1

use *Object*

$DataStore \subseteq (UOID \rightarrow INSTANCE)$
$oids : DataStore \rightarrow \mathbb{P}(UOID)$

$\forall\, ds \in DataStore : oids(ds) = \mathrm{dom}(ds)$
$\forall\, o \in UOID, ds \in DataStore : ds(o).this = o$

*DataStore* is the set of snapshot values.
*oids* is the set of existing objects, given a data store.

Derived lemma:
$\forall\, ds \in DataStore : Nil \notin oids(ds)$

It is convenient to have a number of retrieval and update functions for the data store at hand, as given in Definition 2.8.2. They basically deal with lookup and change of attribute values as well as "creating" a new object in the store.

Again various restrictions on the use of retrieval and update functions apply. This involves the use of values of appropriate type, attributes that actually exist in a class, etc. However, we refrain from defining these specifications here.

At each point in time, i.e., in each state of the state machine, when an instance exists, we assume that its attributes are present and their values are defined (including *Nil*), but it is not necessarily the case that we do know about these values. They may be left underspecified. In particular it may be that, after creation of an instance, its attributes still need to be initialized, i.e., come into a known (and thus well-defined) state. Note that this is a usual modeling technique used, e.g., in verification systems to avoid an explicit handling of a pseudo-value "undefined" [NPW02]. It also resembles reality, e.g., when an uninitialized variable of type *int* is accessed, we do know that it contains an integer, but we do not have a clue which one it is.

**Definition 2.8.2 (DataStore Infrastructure)**

*DataStore*

extend *Object*;
extend *DataStore*1

---

$val : DataStore \times UOID \times UVAR \rightharpoonup UVAL$
$setval : DataStore \times UOID \times UVAR \times UVAL \rightharpoonup DataStore$
$addobj : DataStore \times INSTANCE \rightarrow DataStore$

---

Notation:
$ds(oid.at)$ is shorthand for $val(ds, oid, at)$
$ds[oid.at = v]$ is shorthand for $setval(oid, at, v)$

---

$\forall\, ds \in DataStore, oid \in oids(ds), at \in attr(oid), v \in CAR(vtype(at)))$ :
$val(ds, oid, at) = ds(oid).at$
$setval(ds, oid, at, v) = ds \oplus [oid = (oid, \pi_2(ds(oid)) \oplus [at = v])]$
$o.this \notin oids(ds) \Rightarrow addobj(ds, o) = ds \oplus [o.this = o]$

---

*val* retrieves the value for a given object and attribute.
*setval* updates a value for a given object and attribute.
*addobj* adds a new object.

### 2.8.1 Variation Point: Finite Object System

The elements of *DataStore* are usually partial mappings. However, the system model does not enforce the mapping to be finite in any snapshot of the computation. As, however, in practical implementations this is usually the case, the following Variation Point 2.8.3 is introduced.

**Variation Point 2.8.3 (Set of objects is finite at every time)**

[*FiniteObjectSet*]

use *DataStore*1

---

$\forall\, ds \in DataStore : \#(oids(ds)) \in \mathbb{N}$

### 2.8.2 Variation Point: Locations

Although it is error prone to use locations, they are common in some programming languages in order to pass around pointers to parts of the data store as first-class values. Locations allow reading and modifying the space they refer to. We demonstrate locations as an extension to the basic data structures in Variation Point 2.8.4.

Explicit introduction of locations also allows modeling other effects such as shared variables between objects and “futures” in order to return a value to a sending object in a distributed message passing environment while both are operating concurrently. However, the use of locations as shared variables needs to be properly defined or at best completely avoided as it spoils compositionality of objects as defined in Chapter 5.

We denote by *Loc T* the type name whose associated values are locations for values associated with type name $T$. Note that we allow arbitrary combinations of types such as *Loc Loc T*. By $ULOC \subseteq UVAL$ locations can be passed around and stored like ordinary values. Both dereferencing of the location to the contained

**Variation Point 2.8.4 (Locations as pointers to mutable store)**

[*Locations*]

use *DataStore*

$loc : UOID \times UVAR \rightharpoonup ULOC$
$ULOC \subseteq UVAL$
$Loc : UTYPE \rightarrow UTYPE$
$val : DataStore \times ULOC \rightharpoonup UVAL$
$setval : DataStore \times ULOC \times UVAL \rightharpoonup DataStore$

$\forall\, T \in UTYPE : CAR(Loc\ T) \subseteq ULOC$
$\forall\, ds \in DataStore, oid \in oids(ds), at \in attr(oid), v \in CAR(vtype(at)))$ :
  $(oid, at) \in dom(loc) \Rightarrow$
    $val(ds, loc(oid, at)) = val(ds, oid, at)$
    $setval(ds, loc(oid, at), v) = setval(ds, oid, at, v)$
    $typeOf(loc(oid, at)) = Loc(vtype(at))$

Notation:
$ds(l)$ is shorthand for $val(ds, l)$
$ds[l = val]$ is shorthand for $setval(l, val)$

*ULOC* is the universe of locations.
*Loc T* denotes the type of locations that store data of type *T*.
*val* and *setval* again deal with retrieving and setting of a value.

value and updating to an new value ("get" and "set") is done in the context of a data store. However, further locations for local variables, parameters or places in arrays, etc., are possible.

By construction and definition *loc* is injective. Basically, it can be understood as a wrapper encoding object identifier and attribute name into a unique location (address). This close relation can be seen in structurally rather equivalent definitions of the *val* and *setval* functions.

The partiality of the wrapping function *loc* allows to control which attributes are locations and therefore possibly accessed from outside and which attributes are safe.

As a further variation point, we could introduce a comparison of locations or even operations to allow nasty things such as pointer arithmetic.

### 2.8.3 Variation Point: Reference Types

In the core of the system model, we also do not need references. In an earlier version [BCR06, BCR07a, BCR07b], we had defined object identifiers based on references, because they are essentially similar concepts. In this version, object identifiers are defined directly. If desired, references to other types can be included as well.

A reference is either *Nil* or an identifier for one value in the carrier set of a given type name. Let *T* be an arbitrary type name. Then *Ref T* is a type name whose carrier set consists of references to values of type *T* and the distinguished reference *Nil*.

Given any type name *T*, the carrier set of type name *Ref T* has a rather limited set of operations. References basically allow for comparison (i.e., test for equality) and provide the special reference *Nil*. By dereferencing we obtain the actual value from a reference. If values change over time, dereferencing is state dependent and thus is basically equivalent to the location concept of above. If we assume that the referenced value is immutable, then we can define references as given in Variation Point 2.8.5.

**Variation Point 2.8.5 (Reference types)**

$[References]$
use *DataStore*

---

$Ref : UTYPE \rightarrow UTYPE$
$deref : CAR(Ref\ T) \rightharpoonup CAR(T) \quad \text{for all } T \in UTYPE$

---

$\forall\, T \in UTYPE : Nil \in CAR(Ref\ T)$
$\forall\, T \in UTYPE : \text{dom}(deref) = CAR(Ref\ T) \setminus \{Nil\}$

---

*Ref T* contains references with no additional internal structure.

---

Given a reference $r \in CAR(Ref\ T)$, its dereference $deref(r) \in CAR(T)$ is defined only if $r \neq Nil$. Note that not every value needs to have a reference on it and furthermore, there may be many references to the same value that nevertheless can be distinguished (as they are values on their own).

Note that locations and references are really similar concepts, but serve different purposes. A reference points to a value and this relation is static, i.e., independent of the state of the system. A location contains a value (or its content) and is dependent on the state.

## 2.9 Class Variables and Constants

While attributes are by far the most commonly used elements to store values, there are two further types of elements present in an object-oriented setting.

Constants on one hand are values with a name, such that the name can be used instead of the value. We do not need to represent constants explicitly in the system model: Their associated values are present in the universe of values and the mapping of names to values as well as their visibility is not part of the system model but part of the mapping from UML to the system model.

A second concept that we have not explicitly represented so far is the concept of static attributes. These are attributes that can be regarded as shared between all objects of the class. Indeed they exist independently of any object, but can only be accessed from within a limited scope. As the system model does not deal with visibility of a static attribute, we just need to identify a place where to store such an attribute. In the following Variation Point 2.9.1, we do this by introducing a special object that plays the role of a static, singleton instance hosting all static attributes. Note that this is, again, just a conceptual model and by no means intended to be implemented that way. A more modular implementation could introduce such a static object for each class separately.

Still, we believe that static attributes should be avoided anyway since they may lead to uncontrolled side effects that degrade compositionality.

## 2.10 Associations

One of the core concepts of UML are associations. Associations are relations between classes; and links, which can be regarded as instances of associations, are the corresponding relations between object identifiers at runtime. While associations are mostly binary, they may be of any arity, in addition they may be qualified in various ways and may have additional attributes on their own. Furthermore, an association can be “owned” by one or more of the participating objects/classes or can stand on its own, not owned by any of the related objects. In an implementation a basic mechanism for managing those relations is to use direct links or Collection classes but there are other possibilities as well. To semantically capture different variants of realizations of associations, we use a generalized, extensible approach: Retrieval functions extract links

**Variation Point 2.9.1 (Modeling static attributes)**

[*StaticAttributes*]

use *DataStore*

$StaticC \in UCLASS$
$staticOid \in UOID$
$StaticAttr \subseteq UVAR$
$val : DataStore \times StaticAttr \rightharpoonup UVAL$

$oids(StaticC) = \{staticOid\}$
$\forall\, ds \in DataStore : staticOid \in oids(ds)$
$attr(StaticC) = StaticAttr$
$val(ds, at) = val(ds, staticOid, at)$

The only instance of Class *StaticC* has identifier *staticOid* and stores a value for each attribute of *StaticAttr*.

from the store. We allow for a variety of realizations of these functions. This approach is very flexible as it, on the one hand, abstracts away from the owner of associations as well as from how associations are stored and, on the other hand, does not restrict possible forms of an association. As a big disadvantage of this approach, we cannot capture all forms of associations in one uniform characterization, but need to provide a number of standard patterns that cover the most important cases. If no standard case applies, e.g., for a new stereotype for associations, then the stereotype developer has to describe his/her interpretation of the stereotype directly in the terms of the system model. We demonstrate this approach by defining variants of binary associations below.

According to Definition 2.10.1, any association has a name $R$, a signature given by a list of classes $[C_1, \ldots, C_n]$, possibly additional values of that association $extraVals(R)$ and a relation retrieval function $relOf(R)$.

Note that the use of $CAR(C_i^{\&})$ includes relations between object identifiers of subclasses of $C_i$ which is usually intended by associations, but would not be covered if we used $oids(C_i)$ directly.

Also note that with this approach it is possible to model qualified associations by interpreting one (or more) of the additional attributes as the qualifier as well as to model non-unique associations by introducing a value as distinguishing flag. Some examples for association mappings are given below.

In UML class diagrams, associations usually define certain restrictions on their changeability. This can only be stated if sequences of *DataStore*s are used to compare behavior over time. Thus, the semantics of a class diagram cannot fully be defined on one snapshot of the *DataStore*, but needs to compare two snapshots from different times.

The retrieval function *relOf* depends on the concrete realization of the association. Even after quite a number of years of studying formalizations of object orientation, there is so far not a really satisfactory approach describing all variants of association implementations. Therefore, we provide this abstract function and impose some properties on the function without discussing the internal storage structure. The only decision we made so far is that associations are somehow contained within the store, i.e., they are somehow part of objects and association relations do not extend the store. This is pretty much in the spirit of the system model where higher-level concepts are explained using lower-level concepts. In order to retrieve the links of an association, the state of multiple objects may have to be examined. From the viewpoint of a single object, this is not possible since it only has access to its own state. Hence, we assume that links may be retrieved using an "API", i.e., special methods that can be called by an object and that return the links.

With *classes* we assign a list of classes to an association because the order of the classes is relevant. For example, think of a reflexive *1-to-**-association like "parent-child".

**Definition 2.10.1 (Basic definitions for associations)**

*Association*

extend *DataStore*

---

$UASSOC$
$classes : UASSOC \to List(UCLASS)$
$extraVals : UASSOC \to \mathbb{P}(UVAL)$
$relOf : UASSOC \times DataStore \to \mathbb{P}(UVAL \times UVAL)$

---

$$\forall R \in UASSOC, i, C_i \in UCLASS, ds \in DataStore :$$
$$classes(R) = \{C_1, \ldots, C_n\} \Rightarrow$$
$$relOf(R, ds) \subseteq (CAR(C_1^{\&}) \times \ldots \times CAR(C_i^{\&})) \times extraVals(R)$$

---

*UASSOC* is the universe of association names,
*classes* describes which classes it relates,
*extraVals* describes the possible set of distinguishing values, and
*relOf* is the retrieval function to derive the actual links for an n-ary association based on the current store.

Note that we do not constrain the relationship between *UASSOC* and *UTYPE*. In particular one might regard each association manifesting itself as type ($UASSOC \subseteq UTYPE$) or only a few associations being realized as types. This freedom allows modeling simple relations as attributes only, without having to attach a type to them within the system model.

We also do not constrain the set of attached values *extraVals*, but would assume that in many associations, this set either contains a single value only (which then can be neglected as unimportant) or values of a certain type. In particular, this type may be a variable assignment allowing to model any set of attributes to be added to the relation.

Subsequently, we define various functions that help to define associations quite easily as well as a number of variation points constraining associations. These functions cover standard situations for associations, but there will probably be much more "standard" implementations.

## 2.10.1 Variation Point: Simple Associations Only

Variation Point 2.10.2 constrains the possible set of implementations in a number of ways. Each of the constraints may be imposed individually.

**Variation Point 2.10.2 (Simplified associations)**

[*SimplAssociation*]

use *Association*

---

$$\forall R \in UASSOC, ds \in DataStore : \#relOf(R, ds) \in \mathbb{N}$$

---

$$\forall R \in UASSOC : \#extraVals(R) = 1$$

---

$$\forall R \in UASSOC : \#classes(R) = 2$$

---

$$\forall R \in UASSOC, ds \in DataStore, oid \in UOID :$$
$$\#\{(oid_1, oid_2, x) \in relOf(R, ds) \mid oid = oid_1\} = 1$$

These constraints restrict to:

- finite sets of links for an association only;

- no extra attributes for associations;
- all associations are binary;
- all associations are *1-to-** or even *1-to-1*, but not **-to-**.

### 2.10.2 Variation Point: Plain Binary Associations

Many associations are binary without any additional attributes. For those we can use functions given in Variation Point 2.10.3. Besides the definition functions such as *binaryAssoc* we also provide retrieval functions for the actual relation such as *binaryRelOf* with appropriate signature and additional machinery to conveniently handle ordinary binary associations.

**Variation Point 2.10.3 (Binary associations)**

[*BinaryAssociation*]

use *Association*

$$binaryAssoc : UASSOC \times UCLASS \times UCLASS \to Bool$$
$$binaryRelOf : UASSOC \times DataStore \to \mathbb{P}(UOID \times UOID)$$
$$sources, destinations : UASSOC \times DataStore \times UOID \to \mathbb{P}(UOID)$$

$$\begin{array}{l} \forall R \in UASSOC, C_1, C_2 \in UCLASS, ds \in DataStore : \\ \quad binaryAssoc(R, C_1, C_2) \Rightarrow \\ \quad\quad classes(R) = [C_1, C_2] \wedge \\ \quad\quad \#extraVals(R) = 1 \wedge \\ \quad\quad binaryRelOf(R, ds) = \{(oid_1, oid_2) \mid (oid_1, oid_2, *) \in relOf(R, ds)\} \wedge \\ \quad\quad sources(R, ds, oid_1) = \{oid_2 \mid (oid_1, oid_2, *) \in binaryRelOf(R, ds)\} \wedge \\ \quad\quad destinations(R, ds, oid_2) = \{oid_1 \mid (oid_1, oid_2, *) \in binaryRelOf(R, ds)\} \end{array}$$

$$binaryAssocMultis : UASSOC \times UCLASS \times UCLASS \times \mathbb{P}(\mathbb{N}) \times \mathbb{P}(\mathbb{N}) \to Bool$$
$$binaryAssoc^{1to*} : UASSOC \times UCLASS \times UCLASS \to Bool$$
$$binaryAssoc^{1to(0-1)} : UASSOC \times UCLASS \times UCLASS \to Bool$$
$$binaryAssoc^{1to1} : UASSOC \times UCLASS \times UCLASS \to Bool$$

$$\begin{array}{l} \forall R \in UASSOC, C_i \in UCLASS, N_i \subseteq \mathbb{N}, ds \in DataStore : \\ binaryAssocMultis(R, C_1, C_2, N_1, N_2) \Rightarrow binaryAssoc(R, C_1, C_2) \wedge \\ \quad \forall o_1 \in CAR(C_1^{\&}), o_2 \in CAR(C_2^{\&}) : \\ \quad\quad \#destinations(R, ds, o_1) \in N_1 \wedge \#sources(R, ds, o_2) \in N_2 \\ binaryAssoc^{1to*}(R, C_1, C_2) \Rightarrow binaryAssocMultis(R, C_1, C_2, \{1\}, \mathbb{N}) \\ binaryAssoc^{1to(0-1)}(R, C_1, C_2) \Rightarrow binaryAssocMultis(R, C_1, C_2, \{1\}, \{0, 1\}) \\ binaryAssoc^{1to1}(R, C_1, C_2) \Rightarrow binaryAssocMultis(R, C_1, C_2, \{1\}, \{1\}) \end{array}$$

*binaryAssoc* handles binary functions in general.

With *binaryRelOf* we can retrieve the overall relation (without the then unnecessary extra value), and with *sources* and *destinations* we can retrieve the objects a specific object is linked with. With *binaryAssocMultis* specific restrictions for multiplicities apply.

### 2.10.3 Variation Point: Realization Techniques for Binary Associations

Retrieval functions are not further specified yet because they may have quite a number of different realizations. For clarification, we define some standard functions below, covering standard cases and provide them as variation points. However, more variations are possible.

Variation Point 2.10.4 realizes a binary association through an attribute.

**Variation Point 2.10.4 (A realization for associations as attributes)**

[*AttributeAssociation*]

use *Association*, *BinaryAssociation*

$attributeAssoc : UASSOC \times UCLASS \times UCLASS \times UVAR \rightarrow Bool$

$$\forall R \in UASSOC, C_i \in UCLASS, ds \in DataStore, at \in UVAR :$$
$$attributeAssoc(R, C_1, C_2, at) \Rightarrow$$
$$binaryAssoc(R, C_1, C_2) \wedge at \in attr(C_1) \wedge vtype(at) = C_2 \wedge$$
$$binaryRelOf(R, ds) = \{(o_1, ds(o_1.at)) \mid o_1 \in CAR(C_1^{\&}) \cap oids(ds)\}$$

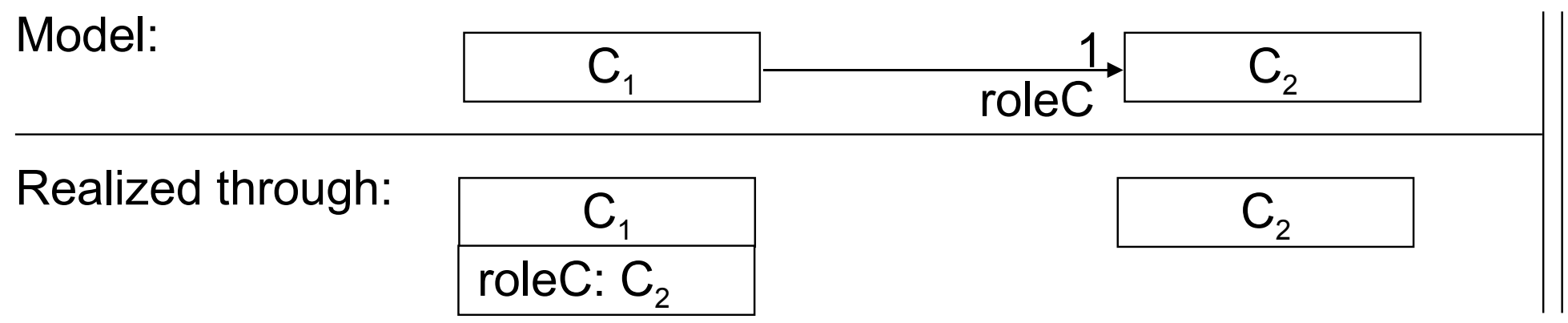

Binary association realized through an attribute.

Derived:

$destinations(R, ds, o_1) = ds(o_1.at)$

Variation Point 2.10.5 for a binary **-to-**-association works quite similarly for *n*-ary associations of arbitrary size, and uses an additional class whose objects describe each one link. Note that although this is a correct realization, inspired by relational database techniques, it may not be the most efficient one with respect to navigation.

The above two definitions demonstrate that the issue of owning a link can quite generally be covered through the use of retrieval functions as given in 2.10.1. In the first definition, the objects own the links, in the second, the links are separated from the objects. Of course also combinations are possible, as shown in the following third Variation Point 2.10.6.

If collections are used in an implementation, the corresponding retrieval functions should be an abstraction of what collections actually store. Note that these functions are mathematical constructs that make the intentions of collection classes explicit but need not be actually implemented.

It is important to note that the effect of, e.g., an action on the links of an association can be described by using the retrieval function, without having to look at the actual representation in the system model. It is not even necessary to provide such a representation; it suffices to know there is one. This principle comes from abstract data types in algebra where the changes on data structures are also purely defined on the effect on access functions.

## 2.10.4 Variation Point: Qualified and Ordered Binary Associations

The above given retrieval functions neither regard ordering nor qualified associations. An appropriate extension for qualification of associations and ordering is given by the more detailed retrieval function in 2.10.7.

Please note that ordering is to some extent a property that is defined over time. Many properties of

**Variation Point 2.10.5 (A realization for associations using association classes)**

[*MtoMAssociation*]

use *Association*, *BinaryAssociation*

$$
\begin{aligned}
&\mathit{mtomAssoc} : \mathit{UASSOC} \times \mathit{UCLASS} \times \mathit{UCLASS}\\
&\qquad \times (\mathit{UCLASS} \times \mathit{UVAR} \times \mathit{UVAR}) \to \mathit{Bool}
\end{aligned}
$$

$$
\begin{aligned}
&\forall R \in \mathit{UASSOC}, C_i, M \in \mathit{UCLASS}, ds \in \mathit{DataStore}, a, b \in \mathit{UVAR} :\\
&\quad \mathit{mtomAssoc}(R, C_1, C_2, (M, a, b)) \Rightarrow\\
&\qquad \mathit{binaryAssoc}(R, C_1, C_2) \wedge\\
&\qquad a, b \in \mathit{attr}(M) \wedge \mathit{vtype}(a) = C_1 \wedge \mathit{vtype}(b) = C_2 \wedge\\
&\qquad \mathit{binaryRelOf}(R, ds) = \{(ds(m.a), ds(m.b)) \mid m \in \mathit{CAR}(M^{\&}) \cap \mathit{oids}(ds)\}
\end{aligned}
$$

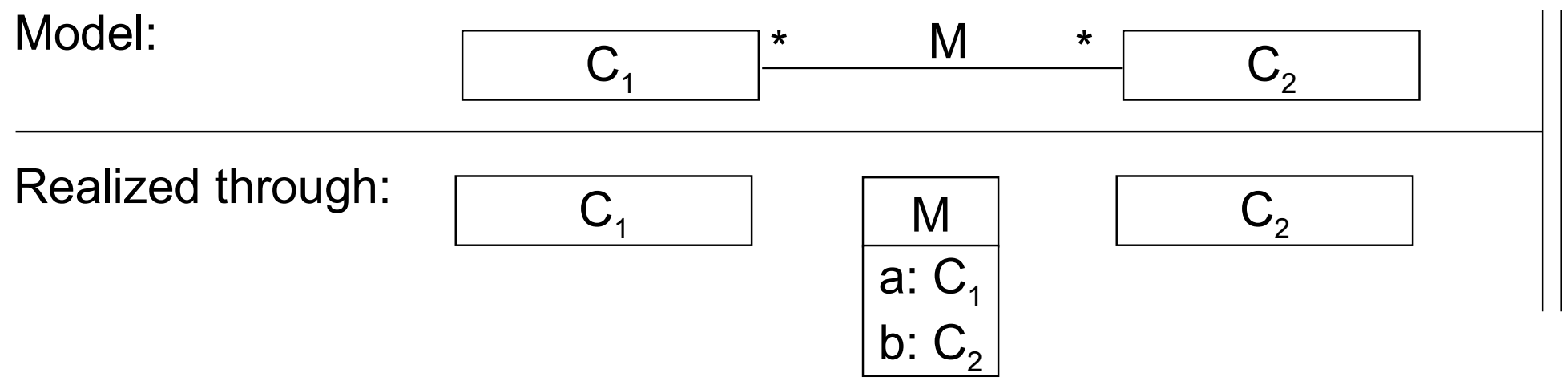

Binary **-to-**-association realized through extra class *M*.

Derived:

$\mathit{destinations}(R, ds, ds(m, a)) = \{ds(m.b)\}$

$\mathit{sources}(R, ds, ds(m, b)) = \{ds(m.a)\}$

ordered specifications cannot be described on one snapshot only but are described through the operators that are allowed on ordered associations, e.g., retrieving elements at a position, deleting, inserting and other list-based operations. Thus, for ordering we cannot define much more than the signature at the moment.

For general qualified associations this situation is similar. However, one case of qualification results form the use of an attribute of the linked-to class as qualifier. We provide a specific definition for this case.

In the qualified case, we retrieve triples, where two object identifiers define the link and the value in the middle defines the qualifier. If the qualifier is an attribute *at* of the linked-to class, then *attrqualifiedAssoc* from Variation Point 2.10.8 can be used. The qualifier identifies at most one object, a generalization to arbitrary multiplicities would easily be possible.

## 2.11 Summary of the Data State of the System Model

All elements of this chapter together provide a description of the data state of the system. Definition C.1.3 contains the full theory *Data* of the data store and associations that model the data state of the system.

Figure 2.3 illustrates the theory dependencies and variation points.

**Variation Point 2.10.6 (A realization for associations using collections)**

[*MtoOneAssociation*]

use *Association*, *BinaryAssociation*, *AttributeAssociation*

$mtooneCollectionAssoc : UASSOC \times UCLASS \times UCLASS$
$\quad \times (UVAR \times UVAR) \rightarrow Bool$

$\forall R \in UASSOC, C_i \in UCLASS, ds \in DataStore, a, b \in UVAR :$
$\quad mtooneCollectionAssoc(R, C_1, C_2, (a, b)) \Rightarrow$
$\quad\quad attributeAssoc(R, C_1, C_2, a) \wedge$
$\quad\quad b \in attr(C_2) \wedge vtype(b)\, sub\, Collection(C_2) \wedge$
$\quad\quad \forall o_2 \in oids(ds) \cap CAR(C_2^{\&}) : collectionValues(ds(o_2.b)) = sources(R, ds, o_2);$

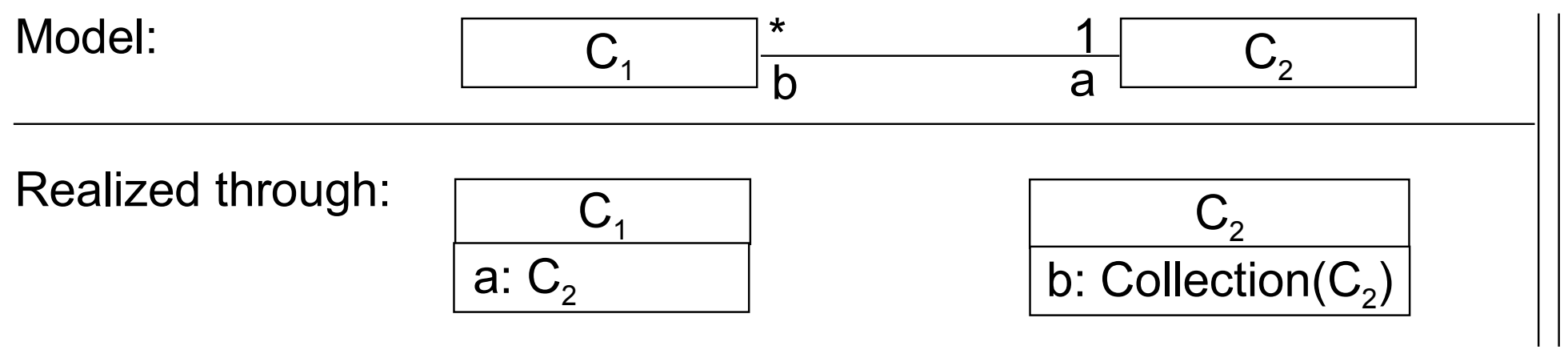

This binary **-to-1*-association is realized redundantly.

**Variation Point 2.10.7 (Qualified and ordered association)**

[*QualifiedAssociation*]

use *Association*, *BinaryAssociation*

$orderedAssoc : UASSOC \times UCLASS \times UCLASS \rightarrow Bool$
$orderedRelOf : UASSOC \times DataStore \rightarrow (UOID \rightarrow List(UOID))$

$\forall R \in UASSOC, C_i \in UCLASS, ds \in DataStore :$
$\quad orderedAssoc(R, C_1, C_2) \Rightarrow binaryAssoc(R, C_1, C_2) \wedge$
$\quad\quad binaryRelOf(R, ds) = \{(oid_1, oid_2) \mid oid_2 \in orderedRelOf(R, ds)(oid_1)\}$

$qualifiedAssoc : UASSOC \times UCLASS \times UCLASS \rightarrow Bool$
$qualifiedRelOf : UASSOC \times DataStore \rightarrow (UOID \times UVAL \times UOID)$

$\forall R \in UASSOC, C_i \in UCLASS, ds \in DataStore :$
$\quad qualifiedAssoc(R, C_1, C_2) \Rightarrow binaryAssoc(R, C_1, C_2) \wedge$
$\quad\quad binaryRelOf(R, ds) = \{(oid_1, oid_2) \mid (oid_1, q, oid_2) \in qualifiedRelOf(R, ds)\} \wedge$
$\quad\quad \forall oid_1 \in UOID, q \in UVAL : \exists^{\leq 1} oid_2 : (oid_1, q, oid_2) \in qualifiedRelOf(R, ds)$

## 2.12 An Example for the Structural System Model

Having defined quite a number of modular parts of the system model, it is now time to populate this structural part of the system model with a nontrivial example.

Note that the example basically maps some given UML models into maths. It necessarily is less designed for compactness and therefore not very intuitive to read. But on the one hand it is precise and on the other not UML users but language and semantics developers are meant to look at it.

**Variation Point 2.10.8 (Qualified Associations using Attributes)**

[*AttrQualifiedAssociation*]

use *Association*, *BinaryAssociation*, *QualifiedAssociation*

$attrqualifiedAssoc : UASSOC \times UCLASS \times UCLASS \times UVAR \to Bool$

$\forall R \in UASSOC, C_i \in UCLASS, ds \in DataStore :$
$attrqualifiedAssoc(R, C_1, C_2, at) \Rightarrow qualifiedAssoc(R, C_1, C_2) \wedge$
$\quad \forall (oid_1, q, oid_2) \in qualifiedRelOf(R, ds) : q = ds.oid_2.at$

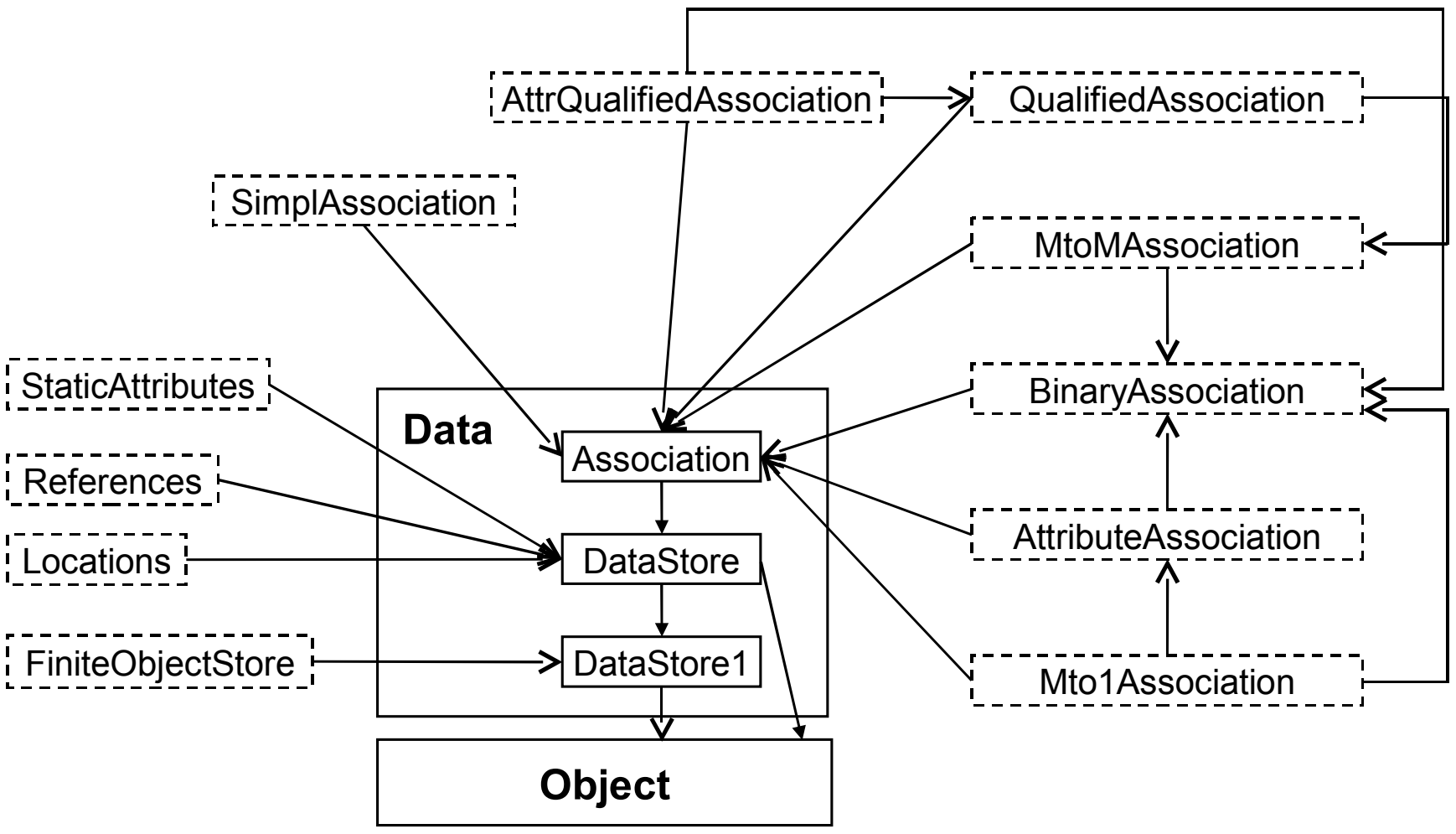

Figure 2.3: Theory *Data* and its dependencies.

Figure 2.4 illustrates the system model example as a UML class diagram. In part 1 of the formalization (Example 2.12.1), we can observe the following:

- A primitive type *String* is introduced.
- The attributes and their corresponding types are defined.
- The classes from Figure 2.4 are elements of *UCLASS*. The function *attr* assigns the right attributes to each class. Object identifiers are assumed to consist of the name of the class and a unique id. The abstract class *Entry* does not have object identifiers on its own since it cannot be instantiated.
- The subclassing relation is defined and it becomes clear that instances of *Meeting* and *Appointment* can be used whenever an *Entry* is expected since the carrier set of the type $Entry^{\&}$ contains at least all identifiers for *Meeting* and *Appointment*.

Example 2.12.2 continues the example by formalizing the given associations:

- Most of the associations are assumed to be realized as attributes. This is detailed only for association *owns* since the rest is defined analogously.
- The anonymous association between *User* and *Meeting* is resolved by introducing a name *Aparticipants* for it. It demonstrates the variation point for many-to-many associations. An association class *Part*2*Meet* and attributes corresponding to the association ends are defined. The predicate *mtomAssoc* ensures that the realization has the desired properties. Both names *Aparticipants* and *Part*2*Meet* are introduced as fresh names. Whether they remain anonymous or can be used by the developer, to describe further properties is another, yet open variation point. In case, e.g., class *Part2Meet* is to be exposed to developers, there need to be clear rules how to build that name from the association or the associated class names (which is not the case in this example).

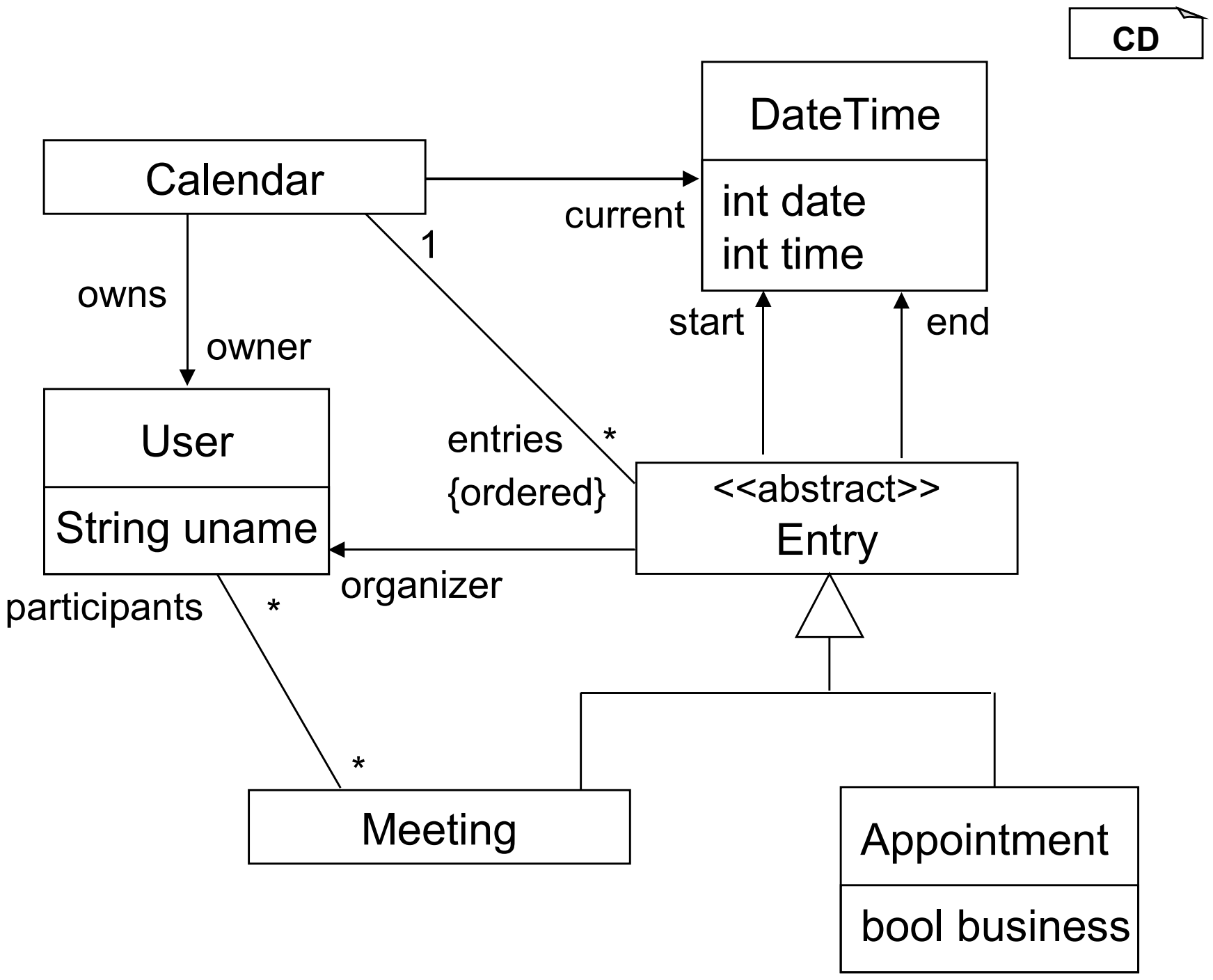

Figure 2.4: Example structure of a system, modeled as a UML class diagram

- The ordered association *Aentries* combines several variation points. First, it is a many-to-many association realized as another (potentially anonymous) association class *Cal2Entries*. Then, it is also a binary association where multiplicities are restricted to 1-to-many. And finally, it is an ordered association where the relation *orderedRelOf* is given in such a way that the list is sorted by start date.

We furthermore demonstrate how the data store might look like in a system run, providing a snapshot of the data state of the system. Figure 2.5 shows the snapshot as a UML object diagram.

In Example 2.12.3, an example data store for the object diagram of Figure 2.5 is given. Most links have become attribute values of the corresponding classes, e.g., the owner attribute of the object of class *Calendar* points to the user instance $oid_1^{User}$.

The link *entry*1 from the Calendar object to the Meeting object is represented in the store as an instance of class *Cal2Entries*, connecting the objects through appropriate attribute values. Similarly, the links *participant*1 and *participant*2 are realized by two objects of class *Part2Meet*.

**Example 2.12.1 (Structure example, part 1)**

extend *Data*

// Additional primitive types:
$String \in UTYPE$

// Variables:
$date, time, uname, business \in UVAR$
$vtype(date) = vtype(time) = int$
$vtype(uname) = String$
$vtype(business) = bool$

// Classes and object identifiers:
$Calendar, DateTime, User, Entry, Meeting, Appointment \in UCLASS$
$date, time \in attr(DateTime)$
$uname \in attr(User)$
$business \in attr(Appointment)$
// A definition for oids:
$oids(Entry) = \varnothing$
$\forall c \in \{Calendar, DateTime, User, Meeting, Appointment\}$ :
$oids(c) = \{oid_i^c \mid i \in \mathbb{N}\}$
$\forall oid_i^c \in oids(c) : classOf(oid_i^c) = c$

// Subclassing:
$(Meeting, Entry) \in sub$
$(Appointment, Entry) \in sub$
$CAR(Entry\&) \subseteq oids(Meeting) \cup oids(Appointment)$

This is a system model based definition of the class diagram (part 1).

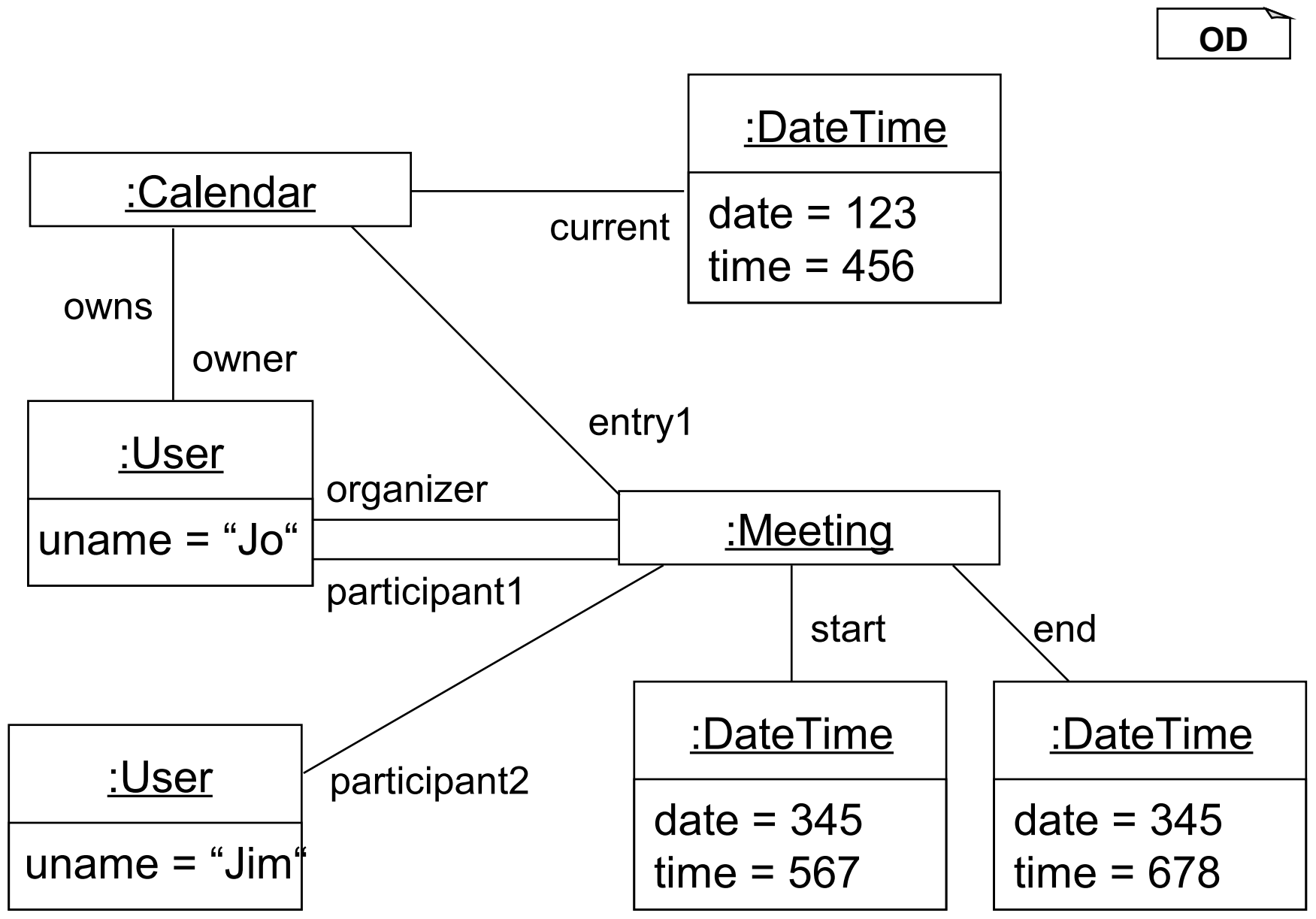

Figure 2.5: Example data snapshots of a system model as a UML object diagram.

**Example 2.12.2 (Structure example, part 2)**

extend *Data*

// Association owns:
$owns \in UASSOC$
$owner \in UVAR$
$attributeAssoc(Aowner, Calendar, User, owner)$

// Associations for current, start, end, organizer defined analogously

// Association participants:
$Aparticipants \in UASSOC$
$Part2Meet \in UCLASS$
$participants, meetings \in UVAR$
$mtomAssoc(Aparticipants, User, Meeting, (Part2Meet, meetings, participants))$

// Association entries:
$Aentries \in UASSOC$
$Cal2Entries \in UCLASS$
$cal, entries \in UVAR$
$mtomAssoc(Aentries, Calendar, Entry, (Cal2Entries, cal, entries))$
$binaryAssoc^{1to*}(Aentries, Calendar, Entry)$
$orderedAssoc(Aentries, Calendar, Entry)$ where
$\forall\, ds \in DataStore, oid1 \in CAR(Calendar\&) \cap oids(ds)$ :
  $orderedRelOf(Aentries, ds)(oid1) = L \wedge$
  $\forall\, oid2 \in L$ :
    $(oid1, oid2) \in binaryRelOf(Aentries, ds) \wedge$
  $\forall\, i$ :
    $d_i = ds(ds(L[i].start).date) \wedge$
    $t_i = ds(ds(L[i].start).time) \wedge$
    $(d_i < d_{i+1}) \vee (d_i = d_{i+1} \wedge t_i \leq t_{i+1})$

This is a system model based definition of the class diagram (part 2).

**Example 2.12.3 (Example data store)**

extend *Data*

$$
\begin{aligned}
& ds = [ \\
& \quad oid_1^{Calendar} = \\
& \quad\quad (oid_1^{Calendar}, [owner = oid_1^{User}, current = oid_1^{DateTime}]), \\
& \quad oid_1^{DateTime} = \\
& \quad\quad (oid_1^{DateTime}, [date = 123, time = 456]), \\
& \quad oid_1^{User} = \\
& \quad\quad (oid_1^{User}, [name = \text{“}Jo\text{“}]), \\
& \quad oid_2^{User} = \\
& \quad\quad (oid_2^{User}, [name = \text{“}Jim\text{“}]), \\
& \quad oid_1^{Meeting} = \\
& \quad\quad (oid_1^{Meeting}, [organizer = oid_1^{User}, \\
& \quad\quad\quad\quad start = oid_2^{DateTime}, end = oid_3^{DateTime}]), \\
& \quad oid_2^{DateTime} = \\
& \quad\quad (oid_2^{DateTime}, [date = 345, time = 567]), \\
& \quad oid_3^{DateTime} = \\
& \quad\quad (oid_3^{DateTime}, [date = 345, time = 678]), \\
& \quad oid_1^{Cal2Entries} = \\
& \quad\quad (oid_1^{Cal2Entries}, [cal = oid_1^{Calendar}, entries = oid_1^{Meeting}]), \\
& \quad oid_1^{Part2Meet} = \\
& \quad\quad (oid_1^{Part2Meet}, [participants = oid_1^{User}, \\
& \quad\quad\quad\quad meetings = oid_1^{Meeting}]), \\
& \quad oid_2^{Part2Meet} = \\
& \quad\quad (oid_2^{Part2Meet}, [participants = oid_2^{User}, \\
& \quad\quad\quad\quad meetings = oid_1^{Meeting}]) \\
& ]
\end{aligned}
$$

This is a system model based definition of the object diagram.

# 3 Control Part of the System Model

Having defined the data part, this and the following chapters focus on the control part of the system model. The control part defines the constituents of the structure used to store control information. The control store contains additional information needed to determine the state of the system during computation. In particular, we provide means to express:

- how control flows (as part of method calls) through active and passive objects,
- what it means for an object to be active or passive,
- how messages are passed, delayed and handled,
- how events are handled,
- how threads work in a distributed setting, and
- how synchronization of all these concepts takes place.

Besides the rather general rules that specify how control and data structure fit together, the theories of this chapter also define variation points, e.g., for single-threaded or completely distributed and asynchronously communicating systems.

One result of this chapter is a flexible mechanism to describe control structures of various kinds resembling quite a number of implementation languages. This variability is enforced by the UML and leads to a rather complex formalization of control. In fact, UML does not allow us to abstract away from control primitives. In the systems we describe with UML, we do not only have various types of control and interaction, but also very often their combinations within a single system.

In addition to its data store as introduced in Definition 2.8.2, a state machine of the system model has a control store. The control store contains information about the behavior of the intended system and is used by the state machine in order to decide which transition to perform next. A control store consists of:

- a stack of method/operation calls, each with its arguments and local variables,
- the progress of the running program (e.g., a program counter), and
- possibly information about one or more threads.

In any setting, be it distributed or not, any state machine of the system model also has to deal with receiving and sending of events that trigger activities in objects. General events, such as “message arrived” or “timeout”, must be handled by any object. As a first step, these events are put into an event store, which consists of an event buffer for each object where handling of events is managed. The event store, which is the last constituent of an objects state, is defined in Chapter 4.

Unfortunately, the rather detailed definition of stacks, events, and threads is not very elegant and does not give us much abstraction. However, this lack of elegance accurately covers the lack of elegance in distributed object-oriented systems where method calls, asynchronous signals and threads of activity are orthogonal concepts that can be mixed in various ways. On the one hand, these concepts provide the system developer with great flexibility. On the other hand, they make it difficult to understand the behavior of the resulting systems. In addition, many orthogonal concepts make it very awkward to describe a system model that uses all of them, because any combination (useful or not) needs to be covered. The resulting complexity becomes apparent in modeling the control part of the system model.

## 3.1 Operations

Objects are accessed through their methods/operations. Here, the term operation refers to the signature whereas the term method refers (also) to the implementation (or body). Operations can be called and they may provide a return value as given by the corresponding implementation. Each operation has a name and a signature (which includes arguments and a return value that may be of type *Void*).

Definition 3.1.1 specifies signatures, which consist of a (possibly empty) list of types for parameters and a type for the return value. Note that parameter names are not present in the signature; parameter names are only part of the implementation. For each operation, its signature and its implementation, as well as the class it belongs to, are uniquely specified. This approach does not explicitly specify overloading, signature and implementation inheritance, overriding and dynamic binding but allows specializations in a flexible way to various actually used mechanisms of method binding. This even includes binding mechanisms such as that in Modula-3. These concepts, thus, are to be decided and defined by the time the mapping from UML to the system model is devised.

**Definition 3.1.1 (Definition of operations)**

*Operation*

extend *Object*

---

$UOPN, UOMNAME$
$nameOf : UOPN \to UOMNAME$
$classOf : UOPN \to UCLASS$
$parTypes : UOPN \to List(UTYPE)$
$params : UOPN \to TUPLE(UVAL)$
$resType : UOPN \to UTYPE$

---

$params(op) = STuple(parTypes(op))$

---

*UOPN* denotes the universe of operations,
*UOMNAME* denotes the universe of operation (or method) names,
*nameOf* returns the name of the operation,
*classOf* returns the class an operation belongs to,
*parTypes* is the list of parameters (their types),
*params* returns the operation arguments (as tuple),
*resType* returns the result type of the operation,

In order to complete the definition of signatures, we clarify how these signatures fit together with the subclassing mechanism. Subclassing ($c\ sub\ d$) defines a constraint on signatures and, in many languages, also on externally promised behavior of its related classes $c$ and $d$.

In UML, interestingly, subclassing does not impose clear constraints on the implementation, as the implementation may be redefined according to some "compatibility" notion. This notion, however, is a semantic variation point that we therefore also leave open to a semantic specialization, e.g., by adding additional constraints for redefined method behavior.

Subclassing in general allows for renaming of parameters in the implementation, as those are not part of the signature. The signatures (in the form of lists of types), however, are either equal or in a generalization/specialization relation. The types of parameters can be generalized, and the type of the return value can be specialized. This is the well-known co/contra-variant way [Mey97] that ensures type safety in a language and modeled in Definition 3.1.2.

It is unclear whether in UML operation parameters can actually be redefined, as, e.g., Java allows, so we

**Definition 3.1.2 (Definition of type safety on operations)**

*TypeSafeOps*
extend *Operation*

$$\begin{aligned}&\forall op_1 \in UOPN, c \in UCLASS : c\ sub\ classOf(op_1) \Rightarrow\\&\quad \exists op_2 \in UOPN : classOf(op_2) = c \wedge\\&\qquad nameOf(op_1) = nameOf(op_2) \wedge\\&\qquad CAR(resType(op_1)) \supseteq CAR(resType(op_2)) \wedge\\&\qquad params(op_1) \subseteq params(op_2)\end{aligned}$$

For any subclass $c$ we type safely inherit $op_1$ from the superclass.

just have it in the system model, although UML may not need it. Variation Point 3.1.3 contains a version of type safety with a non-changeable signature.

**Variation Point 3.1.3 (Stronger version of type safety on operations)**

[*TypeSafeOps*2]
extend *TypeSafeOps*

$$\begin{aligned}&\forall op_1 \in UOPN, c \in UCLASS : c\ sub\ classOf(op_1) \Rightarrow\\&\quad \exists op_2 \in UOPN : classOf(op_2) = c \wedge\\&\qquad nameOf(op_1) = nameOf(op_2) \wedge\\&\qquad params(op_1) = params(op_2) \wedge\\&\qquad resType(op_1) = resType(op_2)\end{aligned}$$

Type safety with no specialization of signatures.

Although, Definition 3.1.2 is rather general it must not necessarily hold in all object-oriented languages. In particular, languages such as Smalltalk exhibiting “Message not understood” errors to which a program can react, do not enforce this type safety requirement.

UML furthermore provides “out” and “in/out” parameters. Many authors however advise against the use of (in/)out parameters. The recommendation in the present context is to use a variation point where, if several “out”-values are to be assigned, each of these is assigned through method call or message passing. In this way, object encapsulation is kept. However, if needed, the system model allows to encode these parameters by passing locations of the variables where the “out”-values are to be stored. Locations are discussed in Variation Point 2.8.4 above.

In the system model operations have exactly one return value. Multiple return values can be encoded in such a way that they are packed in a separate class or record.

In UML there is also the notion of “object behavior”, which, strictly speaking, is not a method. However, for simplification we assume that “object behavior” can be encoded as a special kind of operation associated with the object whose parameters define the signature of the operation.

## 3.2 Methods

Recall that the term operation only refers to the signature whereas the term method refers (also) to the implementation. Methods, thus, have a signature and an internal implementation. The signature of a method consists of a list of parameter names with their types. Projected on the list of types, this list coincides with the parameter type list of the associated operation(s).

To provide all information necessary for a detailed understanding of method interactions, we need an abstract notion of a program counter, a binding between argument values and the corresponding formal parameter variables, and a store for local variables as given in Definition 3.2.1. Furthermore, a method is equipped with the class name, to which it belongs and where it is implemented. Note that *localsOf* and *parOf* result in variable assignments that contain mappings of variables to appropriate values.

**Definition 3.2.1 (Definition of methods)**

*Method*1

extend *Operation*

$UMETH, UPC$
$nameOf : UMETH \to UOMNAME$
$definedIn : UMETH \to UCLASS$
$parNames, localNames : UMETH \to List(UVAR)$
$parOf, localsOf : UMETH \to VarAssign$
$resType : UMETH \to UTYPE$
$pcOf : UMETH \to \mathbb{P}_f(UPC)$

$parOf(m) = SRec(parNames(m), vsort)$
$localsOf(m) = SRec(localNames(m), vsort)$
$parNames(m) \cap localsNames(m) = \varnothing$
$parNames(m) \cap attr(definedIn(m)) = \varnothing$
$localsNames(m) \cap attr(definedIn(m)) = \varnothing$

*UMETH*: universe of methods;
*UPC*: universe of program counter values.
*definedIn*: the class the method (implementation) belongs to (and was defined);
*parNames*: formal parameter variables;
*localNames*: local variables used;
*parOf*: assignment of parameters;
*localsOf*: assignment of local variables;
*resType*: result type;
*pcOf* set of possible program counters of a method;
For convenience we assume the parameters, attributes and local variables are disjoint (syntactic resolution allows that).

Here we have fully decoupled the concept of method (implementation) and operation (signature) to describe them independently. However, there usually is a strong link between methods and operations: A method can only implement operations with compatible signatures. However, as implementations can be inherited, multiple operations can refer to the same method as its implementation. In this way, on the one hand the operation signature can be adapted (e.g., made more specific) without changing the implementations, and on the other the implementation can be redefined using a new method in a subclass. Definition 3.2.2 describes this relation through function *impl* that associates a method to a signature; if the class can be instantiated, all operations of that class need to have implementations.

Note that the external signature of an operation is defined using a tuple ($params(op)$) while the parameter list of the corresponding method implementation also includes variable names to refer to these parameters ($parNames(m)$). Knowing which names the parameters have (from *parNames*), we can use the bidirectional mappings *rec* and *tuple* between these two structures.

The UML specification documents sometimes say “behavior executions” are kinds of objects on their own. We do not require every “behavior execution” to be an object on its own, but if necessary, it is not a

**Definition 3.2.2 (Relationship between method and operation)**

*Method*

extend *Method*1

$$impl : UOPN \rightharpoonup UMETH$$

$$\begin{aligned}
&\forall\, op \in UOPN : m = impl(op) \Rightarrow \\
&\quad nameOf(m) = nameOf(op) \wedge \\
&\quad classOf(op)\; sub\; definedIn(m) \wedge \\
&\quad CAR(resType(m)) \subseteq CAR(resType(op)) \wedge \\
&\quad tuple[parNames(m)]^{*}(parOf(m)) \supseteq params(op) \\
&\forall\, c \in UCLASS : \\
&\quad oids(c) \neq \varnothing \Rightarrow \\
&\quad\; \forall\, op \in \{op \in UOPN \mid classOf(op) = c\} : op \in dom(impl)
\end{aligned}$$

*impl* assigns a method implementation to each operation.

problem to encode behavior executions as objects.

## 3.3 Stacked Method Calls

A stack is a well-known mechanism to store the structure necessary to handle chained and (mutually) recursive method calls. In order to describe nested operation calls and, in particular, object recursion[1], we cannot abstract from the control stack. Object recursion is a common principle in object orientation and provides much flexibility and expressiveness. Almost all design patterns [GHJV95] as well as callback-mechanisms of frameworks [FPR01] rely on this principle.

Actually, to resume a computation after a method call, the information where computation is to be continued must be available. Therefore, the system model provides an abstract notion of stack frames, including (abstract) program counters.

To introduce the general, multi-threaded case in an understandable fashion, we start by introducing the simplified case with one thread only. According to Definition 3.3.1, a stack frame consists of the identifier of the object that the method being (or to be) executed belongs to, the method name, the program counter of the method, the identifier of the invoking object and a record value for parameters and local values. *StackFrame* is a rather general definition and many additional conditions can be added to furhter constrain the actual set of states in the system model.

The method may fork several control flows. Nevertheless, frames have only one program counter. When a fork takes place, a new thread is started. Each thread has its own stack of frames, each of which contains again only one program counter. Therefore, the frames do not differ in the multi-threaded case, but we have more than one stack of such frames.

In the case of a single threaded system, the only existing thread could then be defined as element of type $Stack(FRAME)$.

In the following section, we extend this definition to model distributed threads running in parallel or somehow interleaved. For this purpose, we develop two different and isomorphic views.

[1] That is, a method calls another method of the same object. This is in contrast to calling the same method (of an other object), which is method, but not object recursion.

**Definition 3.3.1 (Stack frames)**

*StackFrame*

extend *Method*

$FRAME = UOID \times UOMNAME \times VarAssign \times UPC \times UOID$
$framesOf : UMETH \to \mathbb{P}(FRAME)$

$framesOf(m) = \{(callee, nameOf(m), vars, pc, caller) \mid$
$\quad \exists\, op \in UOPN : m = impl(op) \wedge$
$\quad callee \in oids(classOf(op)) \wedge pc \in pcOf(m) \wedge$
$\quad vars \in parOf(m) \oplus localsOf(m)\}$

*FRAME* is the universe of frames;
*framesOf* is the set of possible frames for a given method.

Derived:
$framesOf(m) =$
$\bigcup_{op \in UOPN, m = impl(op)} oids(classOf(op)) \times \{nameOf(m)\} \times$
$\quad (parOf(m) \oplus localsOf(m)) \times pcOf(m)$

## 3.4 Multiple-thread Computation, Centralized View

There are quite a number of approaches to combine object orientation and concurrency. Some approaches argue that each object is a unit of concurrency on its own. Others group passive objects into regions around single active objects, allowing operation calls only within a region and message passing only between regions. The programming languages that are commonly used today, however, have concurrency concepts that are completely orthogonal to objects. This means, various concurrent threads may independently and even simultaneously "enter" the very same object. In the following, we add a model of threads to our system model that handles this general case and allows specialization to all these approaches.

We do, however, have the basic assumption that there is a notion of atomic action. These atomic actions are the basic units for concurrency; their exact definition is deferred to the UML actions definitions. On top of atomic actions we assume forms of concurrency control that are provided through appropriate concepts in UML (like "synchronized" in Java). However, UML currently does not provide sufficient mechanisms to actually define scheduling and atomicity of actions conveniently. Possible units of concurrency, for example, would be a variable assignment or an operation invocation.

To model multiple threads, we introduce the (abstract) universe of possibly infinitely many threads in Definition 3.4.1. The control store maps a stack of frames to each thread. The condition ensures that any calling object was the called object in the previous frame. We leave open which *oid* belongs to the starting Frame1 and regard the callee of the last frame to be the currently operating object. An illustration of the central control store with concurrently executing threads is given in Example 3.4.2.

## 3.5 Multiple-thread Computation, Object-Centric View

The model of threads defined above is rather general, but so far does not cover how concurrent threads are executed within an object. To enable a general mechanism for scheduling and definition of priorities, we rearrange the representation of thread-based stacks providing a different view on threads (i.e., without changing the described model). The key idea is to use an object-centric view of stacks instead of the current thread-centric view as shown in Definition 3.5.1. As an important side effect, objects are then described in a

**Definition 3.4.1 (The control store in centralized version)**

*Thread*

extend *StackFrame*

---

*UTHREAD*

$CentralControlStore \subseteq (UTHREAD \rightarrow Stack(FRAME))$

---

$\forall\, ccs \in CentralControlStore, t \in UTHREAD$ :
$\quad \forall\, n < \#ccs(t) : \exists\, oid \in UOID$ :
$\quad\quad ccs(t)[n] = (oid, *, *, *, *) \wedge ccs(t)[n+1] = (*, *, *, *, oid)$

---

*UTHREAD* is the universe of threads.
*CentralControlStore* assigns a stack to each thread.

**Example 3.4.2 (Centralized view on concurrently executing threads)** The figure below illustrates the situation where two threads are active, and both object recursion as well as concurrency occurs. Here "Frame$x.y$" denotes that the frame is in thread $x$ at position $y$, where the highest $y$-numbers denote the active frames:

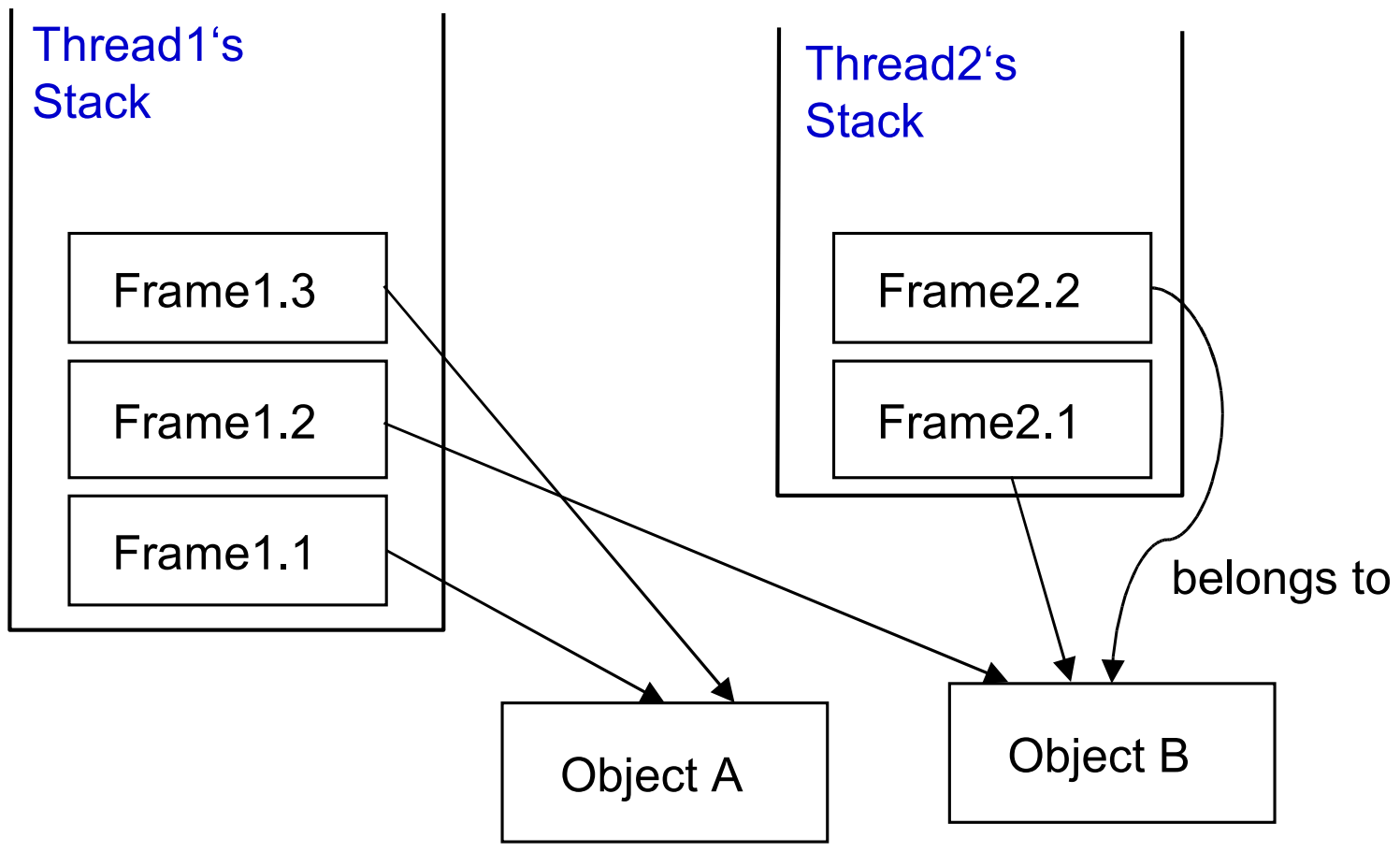

self-contained way. This means that the control information in the system and the object state in full provide a compositional view on object-oriented systems.

For a *ControlStore* $cs$ the stack $cs(oid)(t)$ contains exactly those frames where a method from object *oid* was called in thread $t$.

Note that the relation $. \sim .$ defines an isomorphism as formulated in Lemma 3.5.2. The decentralization into a control store is by definition a function. However, the original stacks can also be uniquely reconstructed because the caller object identifier is part of the frame of the called object. So both representations of the control store provide exactly the same information differently arranged.

Example 3.5.3 shows the above Example 3.4.2 as represented by the object-centered control store.

According to the definition of method calls below (c.f., 4.2.1), we will see that an object is able to recognize that it is being called a second time within the same thread. This is important when, for instance, scheduling or blocking incoming messages from other threads. The Java synchronization model distinguishes recursive calls from other threads and calls from the same threads, and blocks the former but not the latter.

**Definition 3.5.1 (The control store in object-centric version)**

*ControlStore*
extend *Thread*

$ControlStore \subseteq (UOID \rightarrow UTHREAD \rightarrow Stack(FRAME))$
$. \sim . \subseteq CentralControlStore \times ControlStore$

$ccs \sim cs \Leftrightarrow$
$\forall\, oid \in UOID, t \in UTHREAD :$
$\quad cs(oid)(t) = filter(\{(oid, *, *, *, *)\}, ccs(t))$

*ControlStore* splits each stack in parts that belong to objects.
$. \sim .$ relates two representations of the control store by essentially filtering the centralized stack with regards to individual objects.

**Lemma 3.5.2 (Control store representations are equivalent)**

use *ControlStore*

$\forall\, ccs \in CentralControlStore : \exists^1\, cs \in ControlStore : ccs \sim cs$
$\forall\, cs \in ControlStore : \exists^1\, ccs \in CentralControlStore : ccs \sim cs$

**Example 3.5.3 (Object-centric view on concurrently executing threads)**

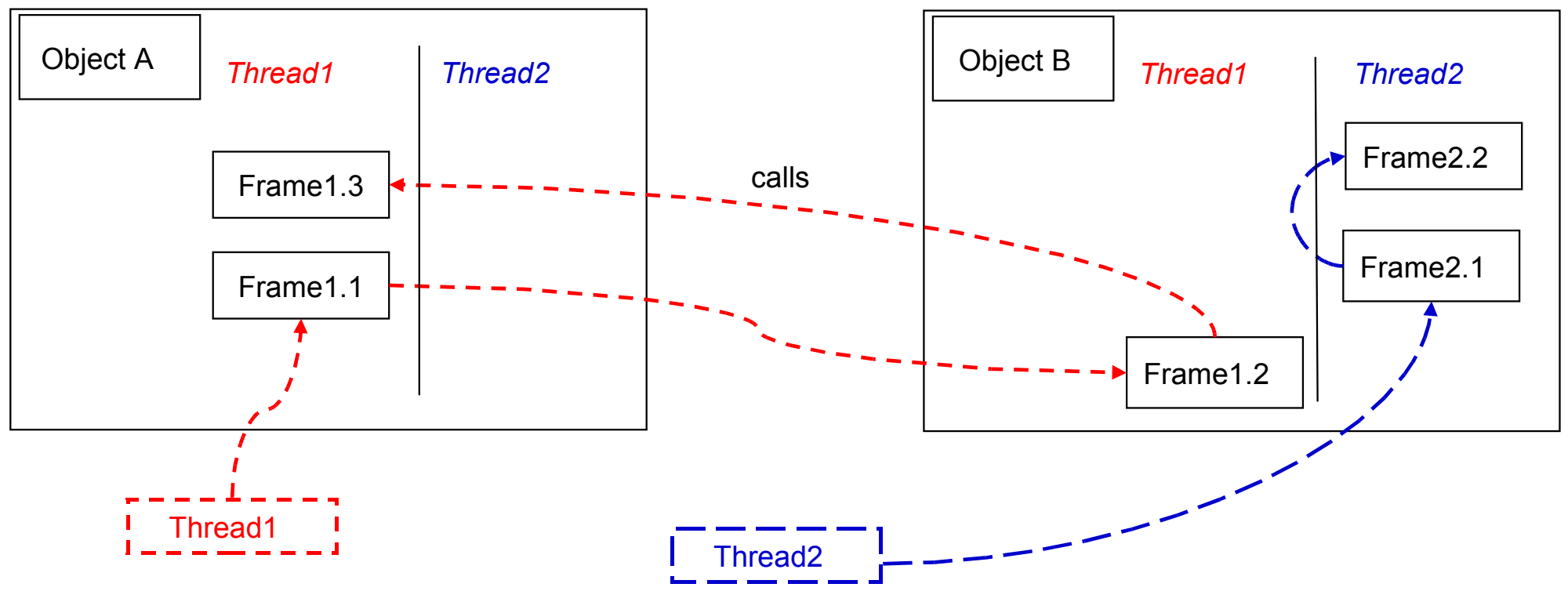

## 3.6 Summary of Threads and Stacks

We summarize the theory built so far, defining threads, stacks and method frames as well as the operations and methods in theory *Control* (Definition C.1.4). This definition relies on Theories *Type* (and *Object*) but is independent from *Data*.

Figure 3.1 illustrates the theory dependencies and variation points.

## 3.7 Example for Operations, Methods, and ControlStore

To get a better understanding of how this works, we define an example with operations and methods in the system model that correspond to the methods that have been added to the class diagram in Figure 3.2.

There are three operations with name *check* to ensure that each subclass of *Entry* provides the required operations (for type-safety). The implementation (i.e., method) *Mcheck* is reused by all three operations and therefore inherited but not overridden. The definitions are straightforward and so only an excerpt of the full definition is shown in Example 3.7.1.

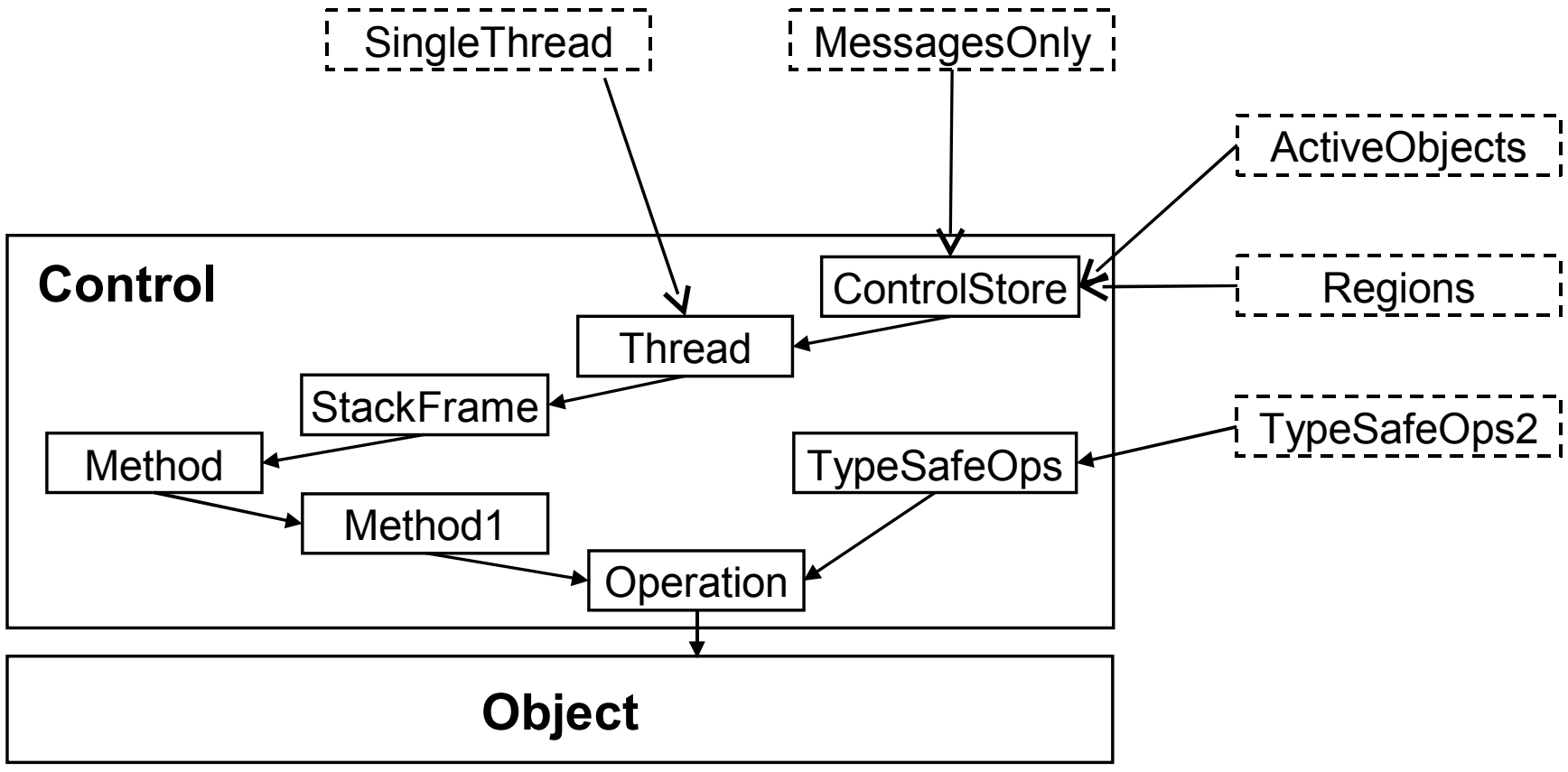

Figure 3.1: Theory *Control* and its dependencies.

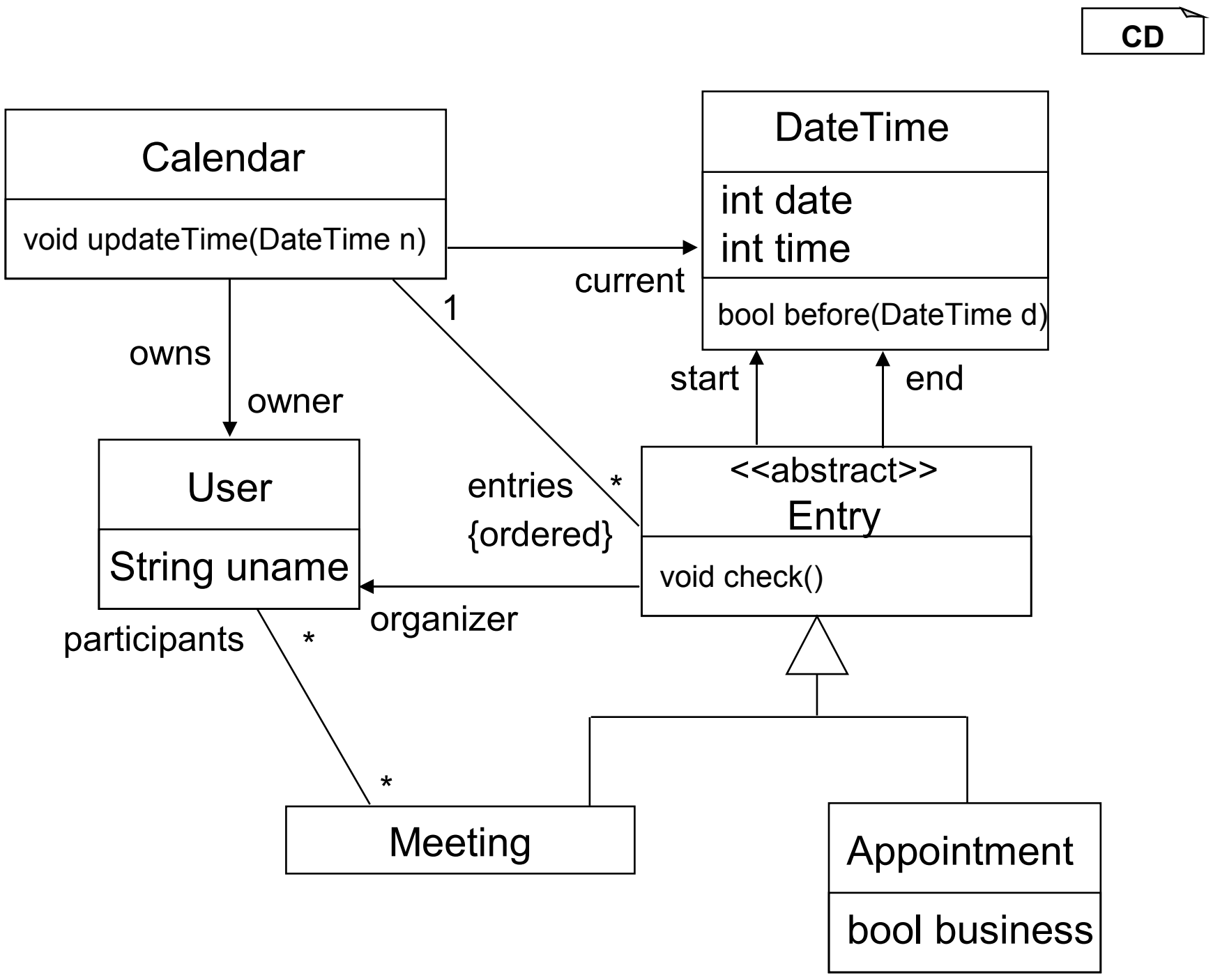

Figure 3.2: Extended class diagram with methods.

Next, we continue the example and show how a control store might look like at a specific point of time during a system run. We assume a data state as given in Example 2.12 and the existence of a thread *th*1. We model the control state in which operation *OcheckM* has been called by a user that in turn is assumed to call operation *Obefore*. The current program counter in method *Mcheck* is hence assumed to be a "wait point". The control state of the user that called the method is omitted in the example.

Figure3.3 provides the isomorphic thread-centric view. The only thread *th*1 has two frames. The frame for the currently active method *before* is on top. After finishing the method, that frame will be removed from the stack . Then frame 1 becomes the currently active frame and hence method *before* the currently executing method for thread *th*1.

**Example 3.7.1 (Control example, part 1)**

extend *Control*

// Operations:
$OupdatetTime, Obefore, Ocheck, OcheckA, OcheckM \in UOPN$
$nameOf(OupdateTime) = updateTime$
$classOf(OupdateTime) = Calendar$
$parTypes(OupdateTime) = [DateTime]$
$resType(OupdateTime) = Void$
$nameOf(Ocheck) = check = nameOf(OcheckA) = nameOf(OcheckM)$
$classOf(Ocheck) = Entry$
$classOf(OcheckA) = Appointment$
$classOf(OcheckM) = Meeting$
$parTypes(Ocheck) = [] = parTypes(OcheckA) = parTypes(OcheckM)$
$resType(Ocheck) = Void = resType(OcheckA) = resType(OcheckM)$
(similar for operation *Obefore*)

// Methods:
$MupdateTime, Mbefore, Mcheck \in UMETH$
$nameOf(MupdateTime) = updateTime$
$definedIn(MupdateTime) = Calendar$
$parNames(MupdateTime) = [n]$
$localNames(MupdateTime) = []$
$resType(MupdateTime) = Void$
$pcOf(MupdateTime) = pcs$
(definitions similar for *Mbefore*, *Mcheck*)

// Assignment of operations to methods:
$impl(OupdateTime) = MupdateTime$
$impl(Ocheck) = Mcheck = impl(OcheckA) = impl(OcheckM)$
$impl(Obefore) = Mbefore$

This is a system model based definition of the operations and methods from the class diagram.

**Example 3.7.2 (Control example, part 2)**

extend *Control*

$cs = [$
$oid_1^{Meeting} =$
$[th1 = Stack[oid_1^{Meeting}, check, [], waitPC, oid_1^{User}]],$
$oid_2^{DateTime} =$
$[th1 = Stack[oid_2^{DateTime}, before, [d = oid_3^{DateTime}], *, oid_1^{Meeting}]]$
$]$

This is an example control store of a system given by the class diagram.

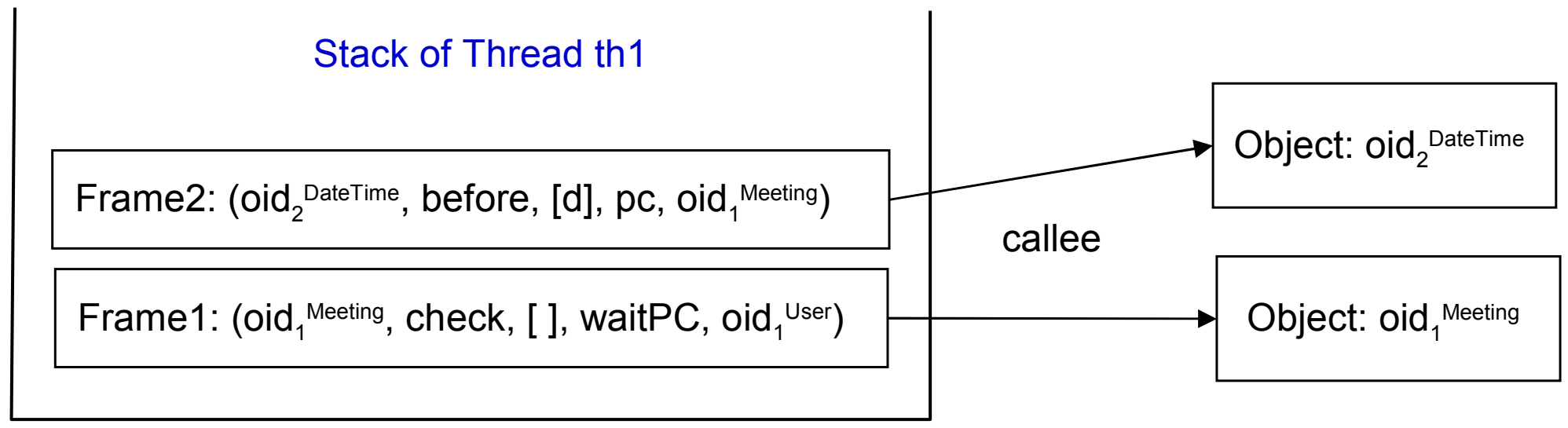

Figure 3.3: Thread-centric view for the example control store.

### 3.7.1 Variation Point: Single Thread Only

There are a number of variations and specializations possible. For example, a restriction to a single threaded (classical sequential) execution is enforced by Variation Point 3.7.3.

**Variation Point 3.7.3 (A single thread only)**

[*SingleThread*]
use *Thread*

$\#UTHREAD = 1$

### 3.7.2 Variation Point: Message Passing Objects Only

Quite the contrary, we can assume that every object is active and does not call any other object via method calls, but only via asynchronous message passing which we will define later. Variation Point 3.7.4 makes all existing objects active and method calls occur only within an object.

**Variation Point 3.7.4 (Objects communicate via messages only)**

[*MessagesOnly*]
use *ControlStore*

$UTHREAD = UOID$
$\forall\, cs \in ControlStore, oid \in UOID, n : cs(oid)(oid)[n] = (oid, *, *, *, *, *)$

Derived:
$\forall\, cs \in ControlStore, t, oid \in UOID : t \neq oid \Rightarrow cs(oid)(t) = []$

### 3.7.3 Variation Point: Active vs. Passive Objects

UML provides the notion of “active” vs. “passive” objects. An active object owns a scheduler that is capable of handling asynchronously received messages. A passive object, on the contrary, only reacts to incoming method calls. However, method calls between active and passive objects are possible. Thus, the thread concept is orthogonal to the concept of objects being active.

In Variation Point 3.7.5 we formalize an “active” object as an object that has a thread associated which can be retrieved through a partial function *thread*. Furthermore, the first frame on the stack of the control store $ControlStore(oid, thread(oid))$ is a permanently running method called *run* that should observe and control the input message queue of object *oid*.

**Variation Point 3.7.5 (Some objects are regarded active)**

$[ActiveObjects]$
use *ControlStore*

$thread : UOID \rightharpoonup UTHREAD$
$run \in UOMNAME$

$\forall\, cs \in ControlStore, oid \in dom(thread) \cap dom(cs) :$
$\quad cs(oid, thread(oid))[1] = (oid, run, *, *, *, *)$

### 3.7.4 Variation Point: Objects located in Regions

The normal case lays somewhere in between: In a distributed system, there are regions of objects belonging together and sharing threads, whereas objects in different regions (processes, processors) only communicate via message passing. Variation Point 3.7.6 models this.

**Variation Point 3.7.6 (Threads can be regionally localized)**

$[Regions]$
use *ControlStore*

$UREGION$
$owner : UOID \to UREGION$

$\forall\, cs \in ControlStore, oid \in UOID, t \in UTHREAD, n :$
$\quad cs(oid)(t)[n] = (callee, *, *, *, *, caller) \Rightarrow$
$\qquad region(caller) = region(callee)$

As a side effect, threads are then also owned by a region, because they never leave these objects. While this definition does not principally simplify reasoning on object behavior, it de facto simplifies reasoning quite well since we can control the number of threads in a region and thereby control and limit the interactions between threads.

# 4 Messages and Events in the System Model

In this Chapter, we specify messages and events and how they are stored and handled within objects.

## 4.1 Messages, Events and the Event Store

UML provides a rather general notion of events. Examples for events are

- sending and reception of messages, which resembles method calls with parameters or return values,
- a timeout,
- a simple signal, or
- a spontaneous state change.

In general, we assume that events may be handled by an operation being executed, or continued (in case of a return event), or ignored. Events are stored in the event buffer. They need not be consumed in the order they appear, a more sophisticated management (scheduling) is allowed individually for each object.

Messages are specific kinds of events. This leads to a uniform handling of events and messages and the more general concept that we chose to call "Events". In general, handling of event occurrences may be delayed, ignored or stored until it is possible to handle it. To capture this rather general notion of event, we introduce a universe of events that can occur in systems, without structuring it further at this point. Later we introduce several kinds of messages, like method call and return, as special forms of events, but leave open which other kinds of events exist.

Based on the universe of events in Definition 4.1.1, we introduce the *EventStore* where events that have occurred are buffered for processing. A buffer is a rather general structure to store and handle messages, deal with priorities etc. Note that the UML specification distinguishes between event (types) and event occurrences (c.f, [OMG07b, Sect. 6.4.2]). Event occurrences are instances of events that may store information like the time the instance occurred and possibly other state information. In our model, event occurrences then correspond to system states in which the event has just been added (sent) or just has been removed (received) from the event store.

In the following Definition 4.1.2 we introduce the universe of messages and with it the special kind of event *MsgEvent*. Messages are a general mechanism to encode any kind of synchronous method call, as well as asynchronous message passing. Here we describe that any message has a unique sender and receiver. However, we do not enforce any further distinction between the various possible forms of messages. This approach does not allow us to model broadcasting or multicasting directly. Multicasting, for example, can be simulated by sending many times the same message to different addressees.

We assume that a Buffer is a given, rather general structure for storing and retrieving occurring events. It also allows the scheduler to rearrange the order of their processing with respect to priorities. In a simplified version, a Buffer behaves like a FIFO-queue without any priority mechanisms. In the case of single threaded programs without specific event structures, the Buffer will only contain incoming (message) events. These events trigger method calls and returns. Because of single threaded computing, the Buffer never contains more than one element and can safely be ignored.

**Definition 4.1.1 (EventStore and object event signature)**

*EventStore*

use *Object*

$UEVENT$
$eventsIn : UOID \rightarrow \mathbb{P}(UEVENT)$
$eventsOut : UOID \rightarrow \mathbb{P}(UEVENT)$
$EventStore \subseteq (UOID \rightarrow Buffer(UEVENT))$

$events(oid) = eventsIn(oid) \cup eventsOut(oid)$
$\forall\, es \in EventStore : es(oid) \in Buffer(events(oid))$

*UEVENT* is the universe of events;
*eventIn* are the events that an object may receive;
*eventOut* are the events that an object may generate;
*EventStore* maps an object identifier to a buffer of processable events.

**Definition 4.1.2 (Object message signature)**

*Message*

use *EventStore*, *Object*

$UMESSAGE$
$MsgEvent : UMESSAGE \rightarrow UEVENT$
$sender, receiver : UMESSAGE \rightarrow UOID$
$msgIn, msgOut : UOID \rightarrow \mathbb{P}(UMESSAGE)$

$\forall\, m \in UMESSAGE, oid \in UOID :$
$\quad sender(m) = oid \Leftrightarrow MsgEvent(m) \in eventsOut(oid)$
$\quad receiver(m) = oid \Leftrightarrow MsgEvent(m) \in eventsIn(oid)$
$\quad msgIn(oid) = \{m \mid receiver(m) = oid\}$
$\quad msgOut(oid) = \{m \mid sender(m) = oid\}$

*UMESSAGE* is the universe of messages;
*MsgEvent* wraps messages into events that can then be stored in the event store; and *sender* and *receiver* enforce that any message has a unique sender and receiver.

## 4.2 Method Call and Return Messages

We introduce some very common kinds of messages: namely messages that describe method call and return. Method call and return are encoded into messages which is a well-known technique in distributed systems that support “remote procedure calls”.

Call messages carry the usual information, like called object, method name, parameter values, as well as the caller and the thread. All the parameters of a call that may arrive to invoke a message in an object are packed by the function *callsOf* into an appropriate message. Definition 4.2.1 introduces *callsOf*.

Return messages carry the return value, the thread, the sender and receiver of the result value. So the Definition 4.2.2 differs only slightly from the previous definition of method calls.

Note that according to Definition 4.2.2 receiver $r$ was the sender of the original method call and now receives the return answer.

Given the definition of message structures above, it is actually possible to unify the concepts of method

**Definition 4.2.1 (Method call messages)**

*MethodCall*

use $EventStore, Message, Control$

$callsOf : UOID \times UOPN \times UOID \times UTHREAD \to \mathbb{P}(UMESSAGE)$
$callsOf : UOID \to \mathbb{P}(UMESSAGE)$

$\forall r, s \in UOID, op \in UOPN, th \in UTHREAD :$
$callsOf(r, op, s, th) \subseteq UOID \times UOMNAME \times TUPLE \times UOID \times UTHREAD$
$callsOf(r, op, s, th) = \{(r, nameOf(op), pars, s, th) \mid$
$\quad r \in oids(classOf(op)) \wedge$
$\quad pars \in params(op)$
$callsOf(r, op, s, th) \subseteq msgIn(r)$
$callsOf(r, op, s, th) \subseteq msgOut(s)$

$callsOf(r) = \bigcup_{s \in UOID, th \in UTHREAD, op \in UOP} callsOf(r, op, s, t)$

*callsOf* defines the set of method calls from object $s$ to $r$ with operation signature $op$ and run in thread $th$.

Derived:

$m \in callsOf(r, *, s, *) \Rightarrow receiver(m) = r \wedge sender(m) = s$

**Definition 4.2.2 (Return messages)**

*MethodReturn*

use $EventStore, Message, Control$

$returnsOf : UOID \times UOPN \times UOID \times UTHREAD \to \mathbb{P}(UMESSAGE)$
$returnsOf : UOID \to \mathbb{P}(UMESSAGE)$

$\forall r, s \in UOID, op \in UOPN, th \in UTHREAD :$
$returnsOf(r, op, s, th) \subseteq UOID \times UVAL \times UOID \times UTHREAD$
$returnsOf(r, op, s, th) = \{(r, v, s, th) \mid$
$\quad s \in oids(classOf(op)) \wedge v \in CAR(resType(op))\}$
$returnsOf(r, op, s, th) \subseteq msgIn(r)$
$returnsOf(r, op, s, th) \subseteq msgOut(s)$

$returnsOf(r) = \bigcup_{s \in UOID, t \in UTHREAD, op \in UOP} returnsOf(r, op, s, t)$

*returnsOf* defines the set of returns from object $s$ to $r$ that may occur as response to a method call in thread $th$.

Derived:

$m \in returnsOf(r, *, s, *) \Rightarrow receiver(m) = r \wedge sender(m) = s$

calls and returns, on the one hand, and of messages, on the other, into one single concept of message passing. This allows the handling of composition of objects and provides a clear interface definition for objects and object groups. Method calls and returns are then just special kinds of messages and can be treated together with other kinds of incoming messages.

## 4.3 Asynchronous and Broadcast Messages

Not every message needs to carry a thread marker. There may be signals that an object may accept. Formally, signals are just asynchronous messages that do not transfer any control. In this case, an object needs to be "active" in the sense that it already has an internal thread to process the message. Variation Point 3.7.5 defines a way to identify active objects. However, it may be that an object is not in itself active, but it belongs to a group of objects that has a common scheduling concept for the processing of messages that come from outside. This concept resembles the situation in classical language realizations where one process contains many objects. We can use the concept of regions from Variation Point 3.7.6 to describe such a common scheduling strategy.

To be able to explicitly talk about signals, we introduce a subset of messages to be asynchronous in Definition 4.3.1.

**Definition 4.3.1 (Signals as asynchronous messages)**

*Signal*

use *Message*, *MethodCall*, *MethodReturn*

$USIGNAL \subseteq UMESSAGE$

$callsOf(*,*,*,*) \cap USIGNAL = \varnothing$
$returnsOf(*,*,*,*) \cap USIGNAL = \varnothing$

### 4.3.1 Example: Handling Signals

Signals can be treated in a variety of ways. In the following Example 4.3.2 we provide a specific instantiation that assumes all messages are encoded as values of a specific type *Signal* and are handled by a message handling method *signalHandler*, which is called when a signal is to be processed.

As we cannot model the behavioral aspects of such a handling right now, we just concentrate on the structural elements. We leave open whether type *Signal* is a primitive type, a record, or a class.

**Example 4.3.2 (A handler method for signals)**

use *Object*, *ActiveObjects*, *Signal*

$signalHandler \in UOMNAME$
$Signal \in UTYPE$

$CAR(Signal) \subseteq USIGNAL$
$\forall\, oid \in dom(thread) : \exists\, op \in UOPN :$
$\quad classOf(op) = classOf(oid) \wedge nameOf(op) = signalHandler \wedge$
$\quad parTypes(op) = [Signal]$

Depending on the kind of system, asynchronously sent messages can be treated as normal values like in Definition 4.3.1, or they can be handled as method calls. When treated as ordinary call messages a special thread marker for the "asynchronous case" could be used, indicating that no answer is desired.

### 4.3.2 Variation Point: Multicast

Messages can be multicasted and even broadcasted. Multicasted messages are put in the set of input events *eventsIn*$(o)$ of many objects, broadcasted messages in the set of input messages of all objects currently present in the system. Multicasted messages, before being replicated, do not carry the identifier of a receiver; they are, however, associated to a number of intended addressees. Broadcasted messages need not be associated to any receiver, since they are to be sent to every object in the system or in the region.

Multicasted and broadcasted messages are normally not anticipated by the multiple recipients and therefore by nature they are asynchronous and, as explained above, do not carry a thread identifier.

A special kind of events models timeouts. Timeout events can be defined as ordinary messages sent by a special timer object to tell the receiving object that a certain amount of time has passed. Usually timeout events have a high priority in event buffers. As an alternative, the formalization of state machines in the system model are timed. Thus an object can react to passing time, for instance by counting time slices and reacting via creation of a timeout event that is buffered at first and handled (possibly immediately afterwards) like an ordinary event.

### 4.3.3 Computation and Scheduling

The organization of event arrival and storage in the event buffer is very much underspecified at the moment. This is part of the scheduling strategy that a system, a subsystem, a region, a component or even a single object may have.

A scheduling strategy determines the next step of an object. In fact, a centralized scheduling strategy (e.g., for all objects of a processor) may be modeled in the system model as easily as a decentralized version where in the extreme case each object has its own scheduler. The scheduling may also be defined for groups of objects (belonging to the same processor, virtual processor, scheduling domain, etc.). We cannot directly define specific scheduling strategies here, as these scheduling strategies rely on the notion of actions as well as the state machine model which will only be introduced in a later chapter.

## 4.4 Example for Events and Messages

In Example 4.4.1, we briefly show how an event store might look like that could be part of a system described by the class digram in Figure 3.2. The example models the situation in which an object of class *Calendar* has a message in its buffer that corresponds to a method call to method *MupdateTime*.

## 4.5 Summary of Messages and Events

Summarizing, we have defined events and messages as well as two specific kinds of messages to model method calls in theory *Events* in Definition C.1.5. Figure 4.1 illustrates the theory dependencies and variation points.

The complexity of the theories *Data*, *Control*, and *Events* of the system model has shown that the integration of objects, threads, concurrency and communication is either not solved very well in object-oriented programming languages, or it is by nature complex. All three concepts are somehow orthogonal and can be used in quite a variety of ways. This flexibility offers great opportunities but also great methodological challenges to ensure the quality of the modeled system. It is particularly complex to model the possible interferences between these concepts and one or a few standardized solutions would greatly help. Instead, UML today tries to allow any combination and thus leads to a larger set of variation points, and no appropriate overview of those is yet available.

**Example 4.4.1 (Event store example)**

extend *Events*, *Control*

Messages:

$m = (oid_1^{Calendar}, updateTime, oid_1^{DateTime}, *, th2)$
$\in callsOf(oid_1^{Calendar}, OupdateTime, *, th)$
$\subseteq UMESSAGE$

Events:

$MsgEvent\ m \in eventsIn(oid_1^{Calendar}) \subseteq UEVENTS$

Example control store:

$cs = [$
$oid_1^{Calendar} =$
$Buffer[(oid_1^{Calendar}, updateTime, oid_1^{DateTime}, *, th2)]$
$]$

An example event store.

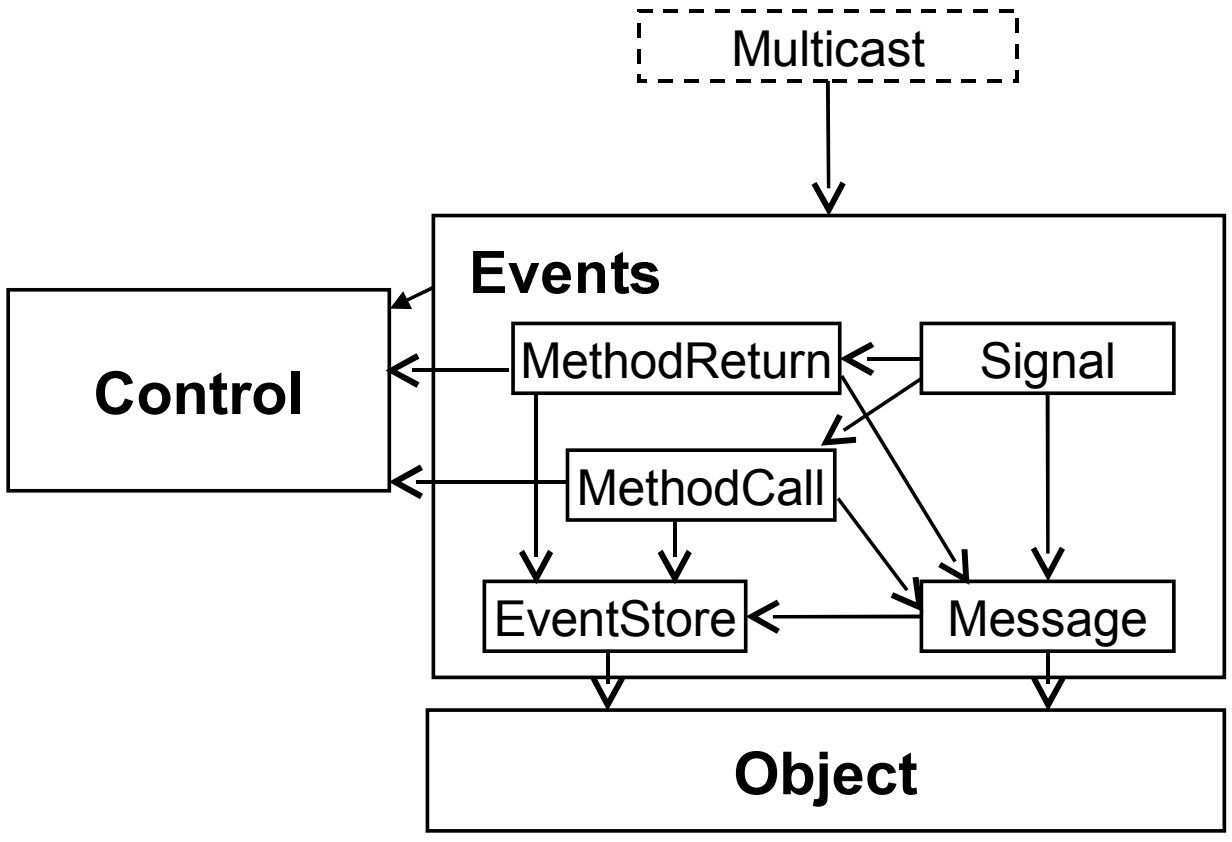

Figure 4.1: Theory *Events* and its dependencies.

# 5 Object State

## 5.1 Individual Object States

The signature and the state space of an object can now be defined completely. It comprises data, control and event stores. Recalling the definitions, we see that all three stores are defined as mappings from *UOID* to the respective state elements. Thus, the state of an object is fully described by a value of *OSTATE* as given in Definition 5.1.1.

**Definition 5.1.1 (State space of an individual object)**

*ObjectStates*1

extend *Data*, *Control*, *Events*

$STATE \subseteq DataStore \times ControlStore \times EventStore$
$oids : STATE \to \mathbb{P}(UOID)$
$OSTATE = INSTANCE \times (UTHREAD \to Stack(FRAME))$
$\quad \times Buffer(UEVENT)$
$state : STATE \times UOID \to OSTATE$
$states : UOID \to \mathbb{P}(OSTATE)$

$STATE = \{(ds, cs, es) \mid dom(ds) = dom(cs) = dom(es)\}$
$oids(ds, cs, es) = oids(ds) = dom(ds)$
$\forall\, oid \in oids(us) : state((ds, cs, es), oid) = (ds(oid), cs(oid), es(oid))$
$states(oid) = \{state(us, oid) \mid us \in STATE \wedge oid \in oids(us)\}$

The state of an object consists of its actual attribute values, the events and the threads belonging to an object. *states* defines the potential states of an object.

Derived:

$oids(ds, cs, es) = dom(ds) = dom(cs) = dom(es)$

## 5.2 Grouped Object States

In Definition 5.2.1 both functions *state* and *states* can be generalized to define the actual and potential set of states for groups of objects. These generalizations, however, use a mapping from object identifiers to their respective contents and are thus structurally equivalent to *STATE*. The structural equivalence of *STATE* and $states(os)$ raises the possibility to use a composition on object states in Lemma 5.2.2 that we can furthermore use to compose state machines in the following chapter. In particular is $f \oplus g$ well-defined, as $state(us, os_i)$ is equal on the common objects $os_1 \cup os_2$. This also allows to regard the possible set of object states in *states* as a cross product, where the common object identifiers need to coincide in their state.

Definition 5.2.1 also identifies $states(o)$ and $states(\{o\})$ as equivalent, as the latter is a function with a single value domain only.

**Definition 5.2.1 (State space of sets of objects)**

*ObjectStates*2

extend *ObjectStates*1

$state : STATE \times \mathbb{P}(UOID) \rightarrow (UOID \rightarrow OSTATE)$
$states : \mathbb{P}(UOID) \rightarrow \mathbb{P}(UOID \rightarrow OSTATE)$

$\forall\, os \subseteq UOID, us \in STATE, oid \in UOID :$
$\quad state(us, os)(oid) = state(us, oid)$
$\forall\, os \subseteq UOID :$
$\quad states(os) = \{state(us, os) \mid us \in STATE \wedge os \subseteq oids(us)\}$

Function *state* and *states* can be generalized to define the actual and potential set of states for groups of objects.

Derived:

$\forall\, os \subseteq UOID, us \in STATE : dom(state(us, os)) = os \cap dom(us)$
$\forall\, os \subseteq UOID, f \in states(os) : dom(f) = os \cap dom(us)$

**Lemma 5.2.2 (State space composition)**

*ObjectStates*

extend *ObjectStates*2

$\forall\, os_1, os_2 \subseteq UOID, us \in STATE :$
$\quad state(us, os_1 \cup os_2) = state(us, os_1) \oplus state(us, os_2)$
$\forall\, os_1, os_2 \subseteq UOID, os_1 \cap os_2 = \varnothing \Rightarrow$
$\quad states(os_1 \cup os_2) = \{f_1 \oplus f_2 \mid f_i \in states(os_i), i = 1, 2\}$

Function *state* and *states* are compositional wrt. the state of objects.

Derived:

$\forall\, os, os_1, os_2 \subseteq UOID : os = os_1 \cap os_2 \Rightarrow$
$states(os_1 \cup os_2)$
$= \{f_1 \oplus (f_2 |_{os_2 \setminus os_1}) \mid f_i \in states(os_i)\}$
$= \{(f_1 |_{os_1 \setminus os_2}) \oplus f_2 \mid f_i \in states(os_i)\}$
$= \{(f_1 \oplus f_2) \mid f_i \in states(os_i) \wedge f_1 |_{os} = f_2 |_{os}\}$

### 5.2.1 Example: Layout of an Object Structure

Example 5.2.3 demonstrates the structure of an object state.

## 5.3 Summary of Object State

Definition 5.2.2 already summarizes all functions related to object states. Therefore, we just briefly recapitulate this in Definition C.1.6 in theory *State*. Figure 5.1 illustrates the theory dependencies.

**Example 5.2.3 (Object state, comprising of data, control, and events)**

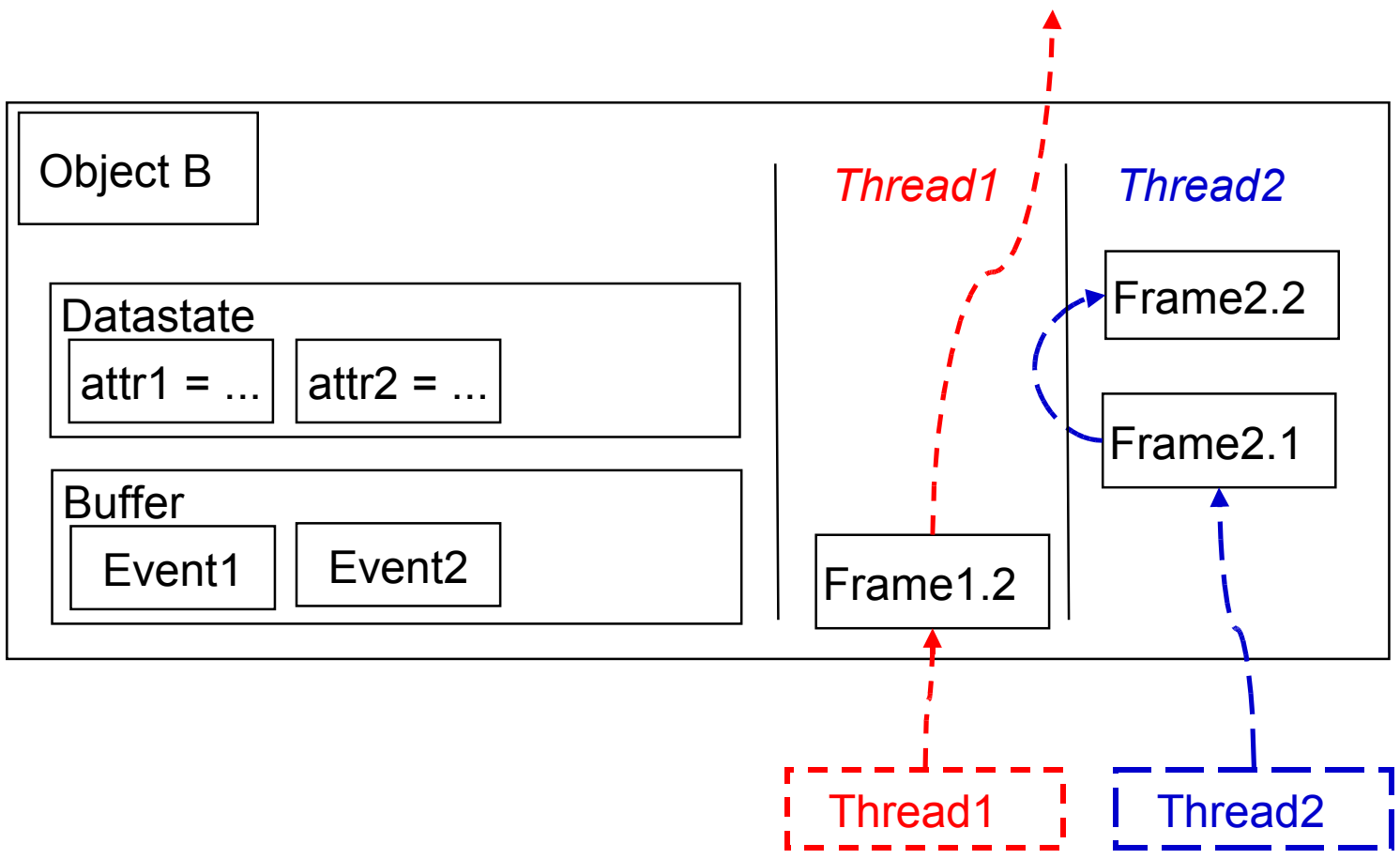

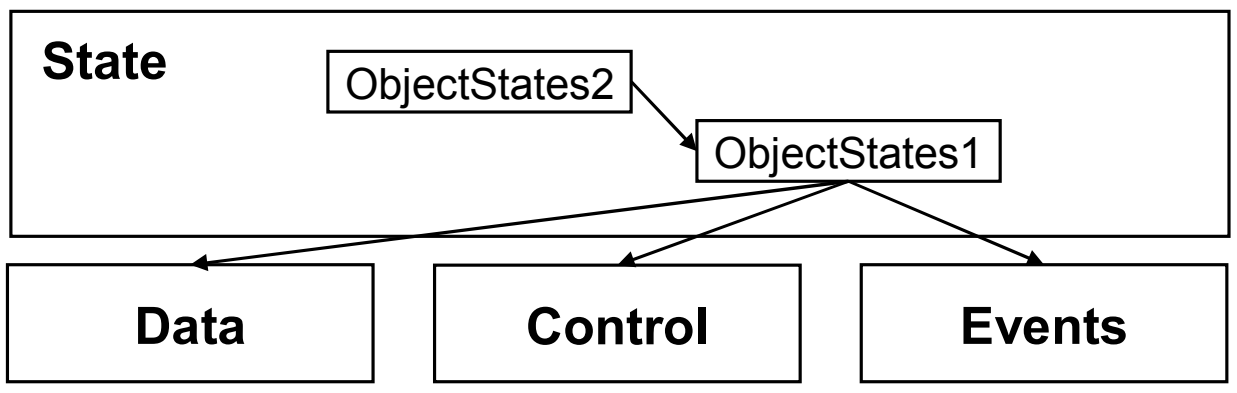

Figure 5.1: Theory *State* and its dependencies.

# 6 Event-based Object Behavior

Based on the notions of state for each object and the corresponding incoming and outgoing events, we can now specify the behavior of an object in form of a state transition system. For this purpose we use the theory of state transition systems (STS) as defined in Appendix B.3.

We start with a STS-based representation of basic actions. For that we use an ordinary programming language such as Java as basis instead of special actions that are defined in UML (c.f., [OMG07b, Chap. 11]) because of the better expressiveness of Java. However, this STS-based representation of method bodies is defined as an instantiation of a variation point and can be omitted or modified at will.

## 6.1 Variation Point: Method Definitions

Methods usually do not bother with the handling of the buffer, priority of events, etc., but provide a sequential realization within their body. Only the scheduler interrupts execution and prioritizes concurrently running methods.

If the method is described as ordinary implementation, we can use the following control flow STS (CF-STS) that is suited to describe the control flow within a method. This is a use of the STS defined in Appendix B.3.

When using normal methods bodies, each method has exactly one starting point, but possibly several exit points. Variation Point 6.1.1 introduces starting points, exits and waiting points for method bodies. A waiting point is a point where the method execution is suspended and waiting for another method to finish and return.

**Variation Point 6.1.1 (Classification of program counters)**

[*ControlFlowPCs*]
extend *State*

---

$StartPC, FinishPC, WaitReturnPC \subseteq UPC$

---

$\forall\, m \in UMETH :$
$\quad \exists^1\, start \in StartPC : (*, nameOf(m), *, *, start, *) \in FRAME \wedge$
$\quad \exists\, end \in FinishPC : (*, nameOf(m), *, *, end, *) \in FRAME$

---

Program counters are classified according to their meaning in the control flow of method bodies.

### 6.1.1 Control Flow State Transition Systems

As objects react to incoming events, an STS describing object behavior is basically event-based and does not necessarily describe timing aspects. To trigger the next execution step for thread *th* within an object, we use $\dagger(th)$ as a pseudo-event as given in Definition 6.1.2.

With this trigger as explicit input of an STS, we can define the scheduling in a separate entity.

Transitions within the CFSTS are regarded as atomic actions. A CFSTS is defined in such a way that an object has no direct access to an attribute of any other object, but may call methods and send events

**Definition 6.1.2 (The stepper for an STS)**

*STSStepper*

extend *State*

$\dagger : UTHREAD \rightarrow \dagger(UTHREAD)$

$injective(\dagger)$

$\dagger(th)$ is used as trigger for the next execution step in thread *th*.

as desired. The state of a CFSTS is defined by the objects own attributes and the currently active frame. Variation Point 6.1.3 introduces CFSTS, and uses STS as introduced in Definition B.3.1.

**Variation Point 6.1.3 (Control flow STS for methods)**

[*CFSTS*]

extend *ControlFlowPCs* use *STSStepper*

$cfsts : UMETH \times UOID \times UTHREAD \rightharpoonup STS(S, I, O)$

$$
\begin{aligned}
&\forall\, m \in UMETH, oid \in UOID, th \in UTHREAD :\\
&classOf(oid)\ sub\ classOf(m) \wedge cfsts(m, oid, th) = (S, I, O, \delta, s0) \Rightarrow\\
&\quad S = \{(o, fr) \in objects(oid) \times framesOf(m) \mid\\
&\qquad fr = (oid, *, *, *, *)\} \wedge\\
&\quad s0 = \{(o, fr) \in S \mid \exists\, start \in StartPC : fr = (*, *, *, start, *)\} \wedge\\
&\quad I = \{MsgEvent\ call \mid call \in callsOf(oid, m, *, th)\} \cup \{\dagger(th)\} \wedge\\
&\quad O = eventsOut(oid)
\end{aligned}
$$

*cfsts* assigns a possibly underspecified CFSTS to each method. This describes the implemented behavior of that method in form of a state machine.

Note that there are alternatives to describe the result of method execution, e.g., using actions as defined in [OMG07b, Chap. 11]. An action language may encompass an ordinary programming language but allow additional actions that, e.g., deal with manipulating associations, timing and scheduling, etc.

Indeed, we believe it is useful to define such high-level, "model-aware" actions and allow to specify them directly, as otherwise such concepts need to be emulated through lower-level concepts or cannot be handled at all. This would mean associations are encoded as attributes, scheduling is managed through an API of an ordinary object that serves as scheduler (in Java this would be a Thread object).

The above variation point is again not precisely constraining CFSTS. Indeed many states of the CFSTS will never be reached, many outputs that are included in *O* will not be made. However, it is relatively accurate on the input, as it describes all information that we know about the context.

Note that we have decided to attach one CFSTS for each method implementation to each object individually. This gives a lot of freedom, even allowing different behaviors for the objects of the same class. However, in practice objects of one class will be assigned the same CFSTS. Furthermore, objects of subclasses whose methods are not overridden shall be assigned the same CFSTS as their parent class objects. This resembles method inheritance on the level of behavior through CFSTS.

### Example: CFSTS for a Method

Figure 6.1 shows an example how a method might be implemented. With this example, we study how a CFSTS is composed. We define a rather fine grained CFSTS to describe the implementation of the two methods, using a program counter (here of form *P00-P99*) at least between each two statements. This is done under the assumption that statements are the units of atomicity. If the methods are synchronized, larger atomic actions can be found and the STS would be less fine grained.

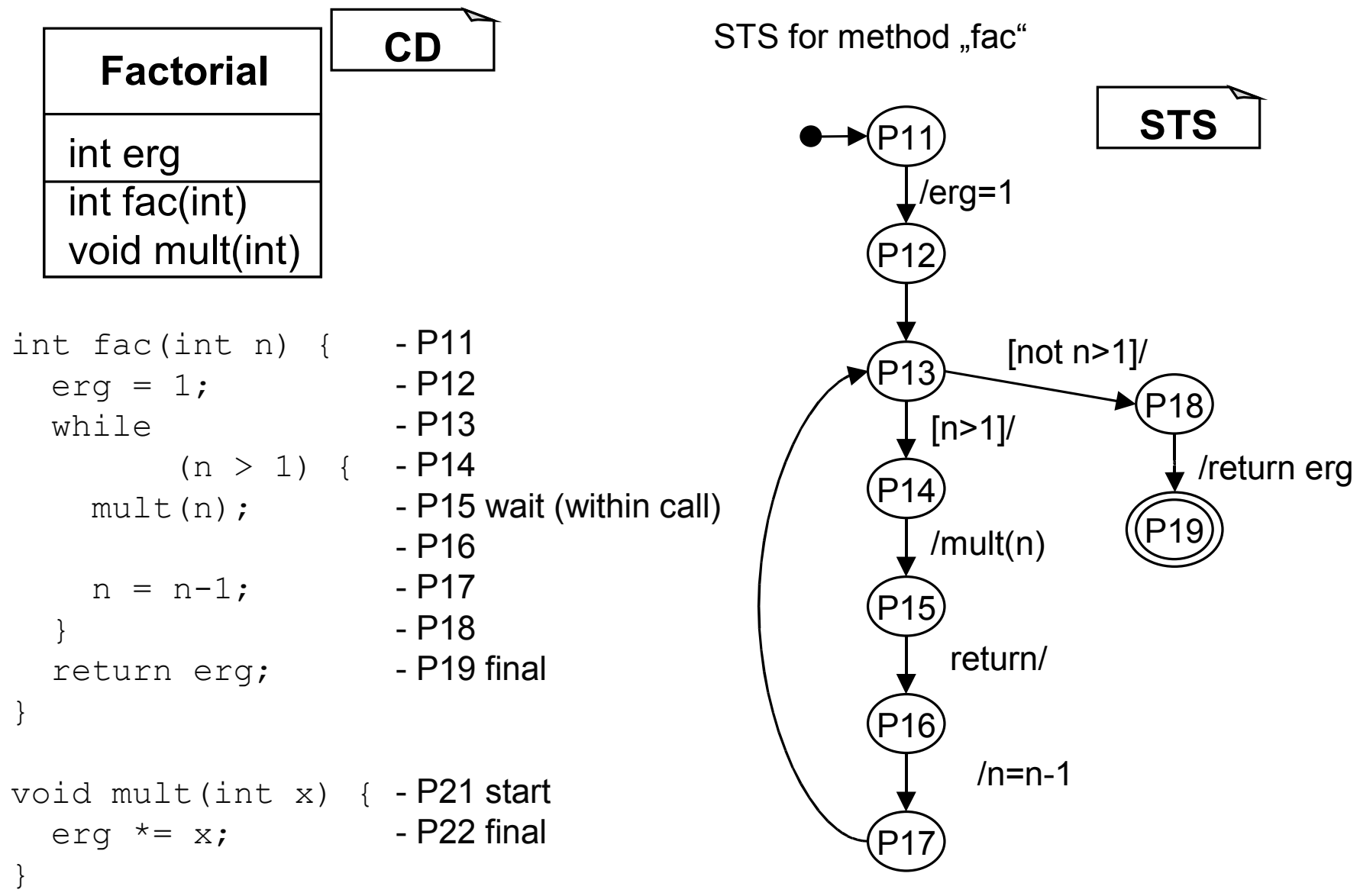

Figure 6.1: Factorial Example

This example is partially translated into mathematical terms in Example 6.1.4 and 6.1.5 containing the structural parts and the CFSTS as mathematical formalization of the Example in Figure 6.1.

Note that although its perfectly precise, a formalization of a compact model, such as a state transition diagram in a mathematical model as given in Example 6.1.5 is always awkward to read. We do not expect a system developer to handle such a formalization. But it is a possibility for a language developer to study the meaning of a construct using mathematical terms: If this meaning pleases us, then have this meaning in mind when developing an efficient code generator etc. Example 6.1.5 basically is nothing more than a mathematical form of the control flow diagram. As it is easier to write such parts (copy-and-paste) than to read, we have underlined interesting parts for convenience of reading.

## 6.2 Event-Based State Transition Systems

As mentioned above, objects react to incoming events and therefore can be described by an STS. This behavior does not describe timing aspects. Definition 6.2.1 describes a type of event-based STS (ESTS) that handles execution in a single object.

### 6.2.1 ESTS Definition

Definition 6.2.1 specifies the general structure and signature of ESTS. The ESTS operates on the full object state and is triggered either by real events or by steps indicated by a †. Again, those steps denote only scheduling of steps, not timing.

The nondeterministic transition function $\delta$ of an ESTS supports underspecification and thus multiple possible behaviors within the STS. This underspecification may be totally or partially resolved during design time by the developer or during runtime by the system itself choosing transitions according to some circumstances, sensor input, etc.

Compared to the above defined CFSTS this notion of ESTS is rather general. It embodies all data, control, and event states on a very general level and thus can describe interference of parallel executions as well as

**Example 6.1.4 (Factorial example, Part 1)**

extend *CFSTS*

$erg, n, x \in UVAR,\ \ vtype(erg) = vtype(n) = vtype(x) = int,$
$fac, mult \in UOMNAME$
$Factorial \in UCLASS,\ \ erg \in attr(Factorial),$
$Ofac, Omult \in UOPN,$
$nameOf(Ofac) = fac,\ \ nameOf(Omult) = mult,$
$classOf(Ofac) = Factorial = classOf(Omult),$
$parTypes(Ofac) = [int] = parTypes(Ofac),$
$resType(Ofac) = int,\ \ resType(Omult) = Void,$

$Mfac, Mmult \in UMETH$
$definedIn(Mfac) = Factorial = definedIn(Mmult),$
$parNames(Mfac) = [n],\ \ parNames(Mmult) = [x],$
$localsOf(Mfac) = [] = localsOf(Mmult),$
$impl(Ofac) = Mfac,\ \ impl(Omult) = Mmult,$
$\{P11, \ldots, P22\} \subseteq UPC,$
$pcOf(Mfac) = \{P11, \ldots, P19\},\ \ pcOf(MMult) = \{P21, P22\},$
$(pcOf(Mfac) \cup pcOf(MMult)) \cap StartPC = \{P11, P21\},$
$(pcOf(Mfac) \cup pcOf(MMult)) \cap FinishPC = \{P19, P22\},$
$(pcOf(Mfac) \cup pcOf(MMult)) \cap WaitReturnPC = \{P15\}$

This is a system model based definition of the class diagram (part 1).

handling of incoming events in the buffer. In contrast to an CFSTS, an ESTS embodies the complete object state including control state and event buffer.

### 6.2.2 Variation Point: Deterministic ESTS

An ESTS makes it possible to further constrain behavior, allowing underspecification at desired places. Firstly, the system model itself is highly underspecified, including behavioral underspecification. Secondly, ESTS are themselves nondeterministic and underspecified.

When we know that objects behave in a deterministic way, we may constrain ESTS to be deterministic as shown in Variation Point 6.2.2; Definition B.3.2 introduces DSTS. Underspecification then only arises from the fact that we do not know exactly which ESTS is the correct implementation.

A deterministic object behavior also means that the scheduling of threads and the handling of events are deterministic. This is not necessarily the case in all realizations. For example, a time sliced scheduling of several threads can better be handled nondeterministically.

### 6.2.3 Variation Point: Composing CFSTS to ESTS

If the methods of an object are described using CFSTS, we can compose those together to form the overall ESTS. Such an ESTS also contains the scheduling, the start and end of method executions, and the reception of events of any form. As an ESTS is relatively complex, we define the transition function of the ESTS in several smaller steps:

- $ESTS_{in}$: reception of an event (see Variation Point 6.2.3),

**Example 6.1.5 (Factorial example, Part 2)**

extend *CFSTS*

$\forall\, op \in UOPN, oid \in classOf(op)^{\&} : impl(op) = Mfac \wedge$
$cfsts(Mfac, oid, th) = (S, I, O, \delta, s0) \Rightarrow$
$\quad S = \{(o, fr) \in objects(oid) \times framesOf(m) \mid fr = (oid, *, *, *, *, *)\} \wedge$
$\quad s0 = \{(o, (*, *, *, start, *)) \in S \mid start \in StartPC\} \wedge$
$\quad I \supseteq \{\dagger(th), (oid, fac, *, *, *), (oid, mult, *, *, *), (oid, void, *, *)\}$
$\quad O \supseteq \{(oid, mult, *, oid, *), (*, void, oid, *), (*, r, oid, *) \mid r \in \mathbb{N}\}$
$\quad \delta((o, (oid, fac, pars, lv, \underline{P11}, c)), \dagger(th)) \ni$
$\qquad ((o, (oid, fac, pars, \underline{lv \oplus [erg = 1]}, \underline{P12}, c)), \epsilon)$
$\quad \delta((o, (oid, fac, pars, lv, \underline{P12}, c)), \dagger(th)) \ni$
$\qquad ((o, (oid, fac, pars, lv, \underline{P13}, c)), \epsilon)$
$\quad \delta((o, (oid, fac, pars, lv, \underline{P13}, c)), \dagger(th)) \ni$
$\qquad ((o, (oid, fac, pars, lv, \underline{P14}, c)), \epsilon) \;\text{ when }\; pars.n > 1$
$\quad \delta((o, (oid, fac, pars, lv, \underline{P13}, c)), \dagger(th)) \ni$
$\qquad ((o, (oid, fac, pars, lv, \underline{P18}, c)), \epsilon) \;\text{ when }\; pars.n \leq 1$
$\quad \delta((o, (oid, fac, pars, lv, \underline{P14}, c)), \dagger(th)) \ni$
$\qquad ((o, (oid, fac, pars, lv, \underline{P15}, c)), \underline{(oid, mult, (pars.n), oid, th)})$
$\quad \delta((o, (oid, fac, pars, lv, \underline{P15}, c)), \underline{(oid, void, oid, th)}) \ni$
$\qquad ((o, (oid, fac, pars, lv, \underline{P16}, c)), \epsilon)$
$\quad \delta((o, (oid, fac, pars, lv, \underline{P16}, c)), \dagger(th)) \ni$
$\qquad ((o, (oid, fac, \underline{pars \oplus [n = pars.n - 1]}, lv, \underline{P17}, c)), \epsilon)$
$\quad \delta((o, (oid, fac, pars, lv, \underline{P17}, c)), \dagger(th)) \ni$
$\qquad ((o, (oid, fac, pars, lv, \underline{P13}, c)), \epsilon)$
$\quad \delta((o, (oid, fac, pars, lv, \underline{P18}, c)), \dagger(th)) \ni$
$\qquad ((o, (oid, fac, pars, lv, \underline{P19}, c)), \underline{(c, o.erg, oid, th)})$

This is a system model based definition of the CFSTS for methods *mult* and *fac* (part 2). An error completion is not shown.

**Definition 6.2.1 (Event-based STS for objects)**

*EventSTS*

extend *STSStepper*

$ests : UOID \to STS(S, I, O)$

$\forall\, oid \in UOID :$
$\quad ests(oid) \in STS(states(oid), eventsIn(oid) \cup \dagger(UTHREAD), eventsOut(oid))$

*ests* assigns a possibly underspecified STS to each *oid*, thus allowing to describe externally visible behavior for an object as a state machine.

- $ESTS_{step}$: execution of a step within a method specified through an CFSTS (see Variation Point 6.2.4),
- $ESTS_{call}$: start of a method execution based on method call (see Variation Point 6.2.5), and
- $ESTS_{end}$: end of a method execution (after return was sent; see Variation Point 6.2.6).

**Variation Point 6.2.2 (Deterministic ESTS for deterministic objects)**

[*DeterministicESTS*]
extend *EventSTS*

$\forall\, oid \in UOID : ests(oid) \in DSTS(*, *, *)$

Deterministic object behavior described by deterministic ESTS.

In this list we have omitted the handling of events other than method calls like e.g. asynchronous signals. For these there are a number of realization possibilities, and we refrained from stating formal definitions since we want to concentrate on conventional method calls in this variation point. The following Variation Points 6.2.3 to 6.2.6 belong together and describe the above mentioned aspects of the behavior of one object.
Variation Point 6.2.3 specifies the signatures of the $\delta$ transition functions as well as the processing of an

**Variation Point 6.2.3 (ESTSin)**

[$ESTS_{in}$]
extend *CFSTS*, *ESTS*
context $oid \in UOID$

$$\delta_{in}, \delta_{step}, \delta_{call}, \delta_{end}, \delta_{idle}, \in$$
$$states(oid) \times (eventsIn(oid) \cup \dagger(UTHREAD)) \rightarrow$$
$$\mathbb{P}(states(oid) \times eventsOut(oid)^*)$$

$$\delta_{in}((ds, cs, es), m) = \{(ds, cs, addlast(es, m))\}$$
$$\delta_{in}((ds, cs, es), \dagger(th)) = \varnothing$$

$$\delta_{idle}((ds, cs, es), m) = \varnothing$$
$$cs(th) = Stack[] \wedge (\lambda\, m.thread(m) = th) \copyright es = Buffer[] \Rightarrow$$
$$\delta_{idle}((ds, cs, es), \dagger(th)) = \varnothing$$

$\delta_{in}$ takes an incoming event $m$ and puts it into the event buffer $es$ of object $oid$. $\delta_{idle}$ is used if no other step can happen.

incoming event (stored in the event buffer). Variation Point 6.2.4 runs one step within a currently active method execution; it adapts the object state, changes the currently active frame, e.g., pointing to another program counter, etc. This step heavily relies on the definition of the CFSTS of the currently running method.
Variation Point 6.2.5 takes a call event from the buffer and starts the execution of the call by adding an appropriate frame on the stack. The scheduling decides through sending repeated steps of form $\dagger(th)$ which thread shall proceed and therefore also which events to process from the buffer.

Variation Point 6.2.6 checks whether the program counter *pc* denotes the end of the computation in a method. In this case, the frame is removed from the stack. Note that in this step no return is sent to the caller: It is assumed that such a return has been sent already. Therefore, the same mechanism can be used to finish the asynchronous invocation of a method execution that does not return an event at all.

In Variation Point 6.2.7, the parts of the transition function for the ESTS are composed together in a complete definition of what an ESTS may do. Note that this is a union of state transition functions and not a composition of state transition systems.

The *ests* defined so far is not fully specified but open to additional clarifications, especially to define scheduling, start states of objects and similar things. In order to further constrain the actual ESTS, we can

**Variation Point 6.2.4 (ESTSstep)**

$[ESTS_{step}]$

extend $ESTS_{in}$
context $oid \in UOID$

---

$\forall (ds, cs, es) \in states(oid), ms \in eventsIn(oid)$ :
$\delta_{step}((ds, cs, es), ms) = \varnothing$
$\delta_{step}((ds, cs, es), \dagger(th)) =$
$\{((ds \oplus [oid = obj'], setframe(cs, th, fr'), es), out') \mid$
$cfsts(methodOf(fr), oid, th) = (S, I, O, \delta, s0) \wedge$
$((obj', fr'), out') \in \delta((ds(oid), top(cs(th))), \dagger(th))\}$

---

Helper functions:
$methodOf((callee, mn, *, *, *)) = m$ `where`
$nameOf(m) = mn \wedge classOf(m) = classOf(callee)$
$setframe(cs, th, fr') = cs \oplus [th = push(pop(cs(th)), fr')]$

---

$\delta_{step}$ makes an execution step on object *oid* in a thread *th* (which is not further specified).
$obj'$ is the new object state, $fr'$ the new frame state on the stack, $out'$ the output, *m* the currently running method and $cfsts(m, oid, th)$ the CFSTS describing the step.
*methodOf* gives the method of a frame and *setFrame* updates a frame on the stack of the appropriate thread.

**Variation Point 6.2.5 (ESTScall)**

$[ESTS_{call}]$

extend $ESTS_{step}$
context $oid \in UOID$

---

$\forall (ds, cs, es) \in states(oid), ms \in eventsIn(oid), th \in UTHREAD$ :
$\delta_{call}((ds, cs, es), ms) = \varnothing$
$\delta_{call}((ds, cs, es), \dagger(th)) =$
$\{((ds, addframe(cs, th, fr'), es'_1 ++ es'_2), \epsilon) \mid$
$es = es'_1 ++ [ms] ++ es'_2 \wedge$
$ms = MsgEvent(oid, mn, par, caller, th) \wedge$
$fr' = (oid, mn, * \oplus rec_{parNames(m)}(par), *, caller) \wedge$
$m = methodOf(fr') \wedge nameOf(m) = mn \wedge$
$cfsts(m, oid, th) = (S, I, O, \delta, s0) \wedge$
$(ds(oid), fr') \in s0\}$

---

Helper functions:
$addframe(cs, th, fr') = cs \oplus [th = push(cs(th), fr')]$

---

$\delta_{call}$ invokes a method call with new frame $fr'$ on object *oid* by removing a call event *ms* with appropriate thread *th* from the buffer, setting up frame $fr'$ such that it starts in a start state of $cfsts(m, oid, th)$.

enforce the transition function $\delta$ to be a subset of the behavior defined above. A further constraint is for instance the definition of a scheduling strategy.

By construction, each arriving step forces the ESTS to do something useful on an existing thread. Only if all threads and the event buffer are empty, the state transition diagram is allowed to idle.

**Variation Point 6.2.6 (ESTSend)**

$[ESTS_{end}]$

extend $ESTS_{call}$
context $oid \in UOID$

$\forall (ds, cs, es) \in states(oid), ms \in eventsIn(oid), th \in UTHREAD$ :
$\delta_{end}((ds, cs, es), ms) = \varnothing$
$\delta_{end}((ds, cs, es), \dagger(th)) =$
$\quad \{((ds, popframe(cs, th), es), \epsilon) \mid$
$\quad\quad fr = top(cs(th)) = (oid, *, *, pc, *) \wedge pc \in FinishPC\}$

Helper functions:
$popframe(cs, th) = cs \oplus [th = pop(cs(th))]$

$\delta_{end}$ removes a stack frame *fr* after all computation has actually finished, because the program counter *pc* is a finishing one.

**Variation Point 6.2.7 (Composing an ESTS)**

$[ESTSComp]$

extend *EventSTS*
extend $ESTS_{end}$

$\forall\, oid \in UOID$ :
$ests(oid) = (S, I, O, \delta, s0) \Rightarrow$
$\quad \forall\, s \in S, i \in I : \delta(s, i) \subseteq$
$\quad\quad \delta_{in}(s, i) \cup \delta_{step}(s, i) \cup \delta_{call}(s, i) \cup \delta_{end}(s, i) \cup \delta_{idle}(s, i) \wedge$
$\quad s0 \subseteq \{(ds, (\lambda\, th.Stack[]), Buffer[]) \mid ds \in objects(oid)\}$

*ests* assigns a possibly underspecified ESTS to an object. This describes the behavioral implementation of that object as a state machine.

### 6.2.4 Variation Point: Basic Scheduling in the ESTS

We have so far not clarified who is deciding which object and which thread is going to make the next execution step. This can be decided individually for each object, or in a regional or centralized manner. In the following Variation Point 6.2.8, we show the signature of a Scheduling-STS that can be used to determine the order of execution.

Internally, such a Scheduling-STS may use round robin, priorities, and a variety of other techniques to determine which thread to activate next. It may also handle synchronization and blocking between objects and threads.

The pseudo-event $\dagger(th)$ which can be consumed by all ESTS resembles a trigger for an execution step of the ESTS. The state the scheduler has access to or is storing information is left unspecified; it should at least comprise the buffer and the control state of the object. This allows, e.g., to identify the set of threads that can execute a next step by examining the pc of the active stack frame. The pc should not be an element of *FinishPC* or a *WaitReturnPC*.

As context switching is relatively expensive in today's computer architectures, one might include a counter into a state and while counting down on each tick, the scheduler just keeps the last thread activated until zero is reached and another thread is selected.

**Variation Point 6.2.8 (Scheduling in the ESTS)**

[*ESTSScheduling*]
extend *ESTSComp*

$scheduler : UOID \to STS(S, \{\dagger\}, \dagger(UTHREAD))$

$scheduler(oid) = (S, \{\dagger\}, \dagger(UTHREAD), \delta, s0) \Rightarrow$
$\forall\, \delta : s \xrightarrow{i/o} t \Rightarrow \#o = 1$

## 6.3 Summary for Event STS

The theory built so far is summarized in Definition C.1.7. Figure 6.2 illustrates the theory dependencies and variation points.

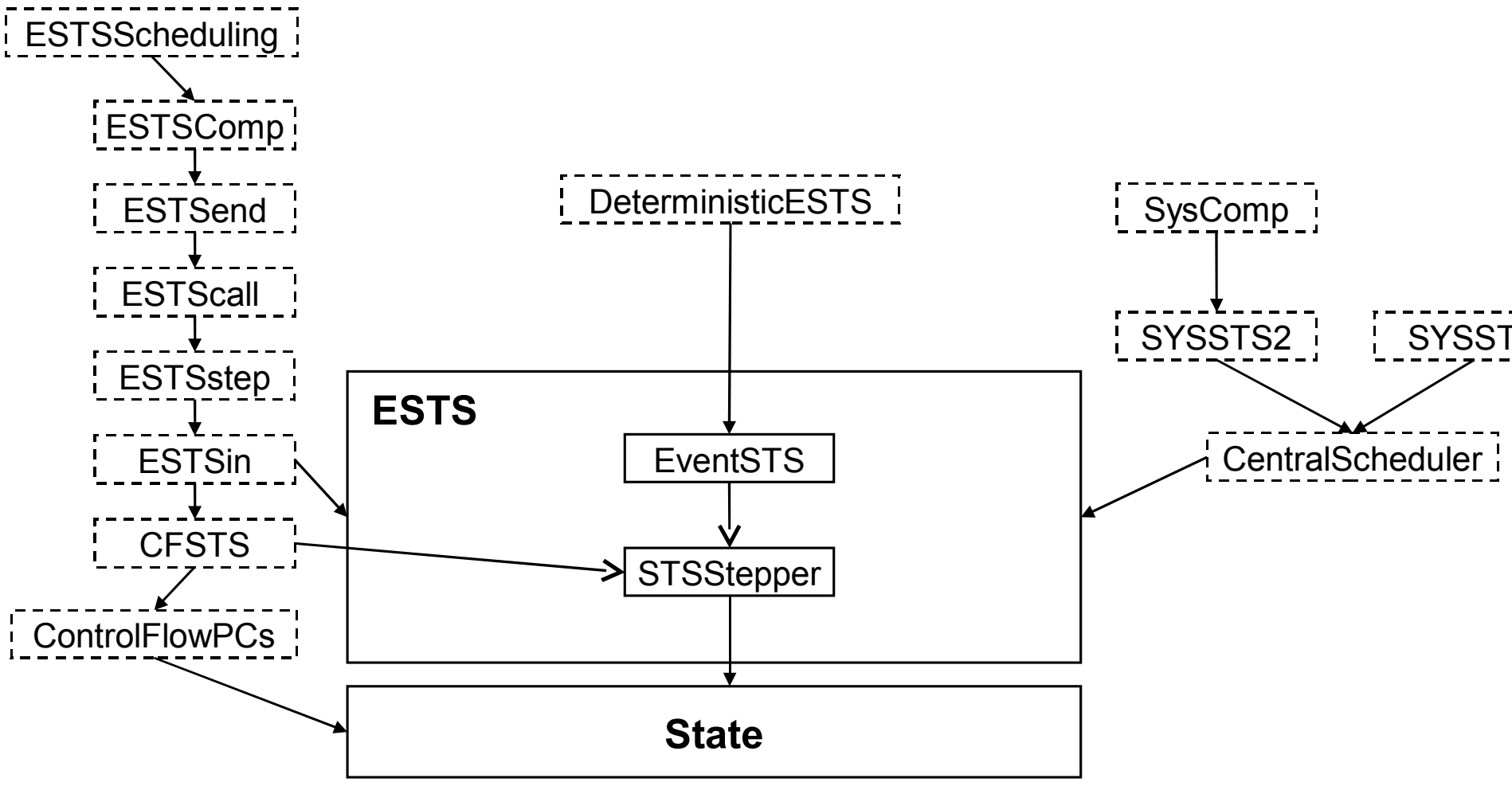

Figure 6.2: Theory *ESTS* and its dependencies.

Although this chapter contains rather detailed descriptions of how to derive an ESTS from a method implementation given as a CFSTS, the actual definitions are rather few. This is largely credited to the fact that CFSTS are defined as variation points and need not be used. In particular they need not be incorporated when another type of action language should be used and mapped to an ESTS through a different kind of mechanism.

## 6.4 Variation Point: Composing ESTS to System STS

According to Definition 6.2.1, each object is equipped with an ESTS. All these ESTS can be composed to form a transition system for the whole system that, among other things, explicitly describes scheduling. In order do this, we assume a scheduler that selects runnable threads in objects. The scheduler, in principle, may examine the whole system state in order to schedule the next object and thread combinations. This central scheduler is defined in Variation Point 6.4.1.

In the scheduler, a variety of scheduling strategies can be encoded. The following list sketches some (and not necessarily all) possibilities.

- In UML terms, run-to-completion execution can be achieved if the scheduler schedules the same thread that is handling a call in some object as long as the call is not completely processed (partially

**Variation Point 6.4.1 (Central system scheduling)**

$[CentralScheduler]$
extend $ESTS$

$scheduler : STATE \rightarrow \mathbb{P}(UOID \times UTHREAD)$

also depends on $\delta_{call}, \delta_{step}$ in case of object recursion). If state-relevant and state-irrelevant calls can be distinguished, more concurrency may be allowed in one object.

- If we assume that objects are mapped to resources (e.g., processors), the number of concurrently schedulable threads may additionally be restricted.
- A simple variant of round-robin scheduling would correspond to alternately scheduling all runnable threads (possibly selecting one thread for each resource).
- (Dynamic) priorities may be assigned to threads, so priority-based scheduling becomes possible.
- More sophisticated strategies can be encoded if we assume a system clock and that each step in the system run consumes time. Each step of the system corresponds to a "call" or "interrupt" to the scheduler that, given the current time, may allow the last thread to continue or to schedule a new thread (because, e.g., the time in a time slice is up, a higher- priority thread was scheduled periodically, preempting a low-priority thread). Current time may also be specific for individual resources in case of imprecise clocks for which a maximum deviation could be specified. This current time could then also be made available to objects using an API.
- If each object had a separate thread, this could be interpreted as task or process. If furthermore the number of tasks in the system were fixed, classical scheduling strategies from real-time systems (rate monotonic, earliest deadline) together with their schedulability analysis could be realized.

In Variation Point 6.4.2 a definition for a state machine for the whole system is given. Basically, the system transitions from one state to another while storing not transmitted events in a buffer. The scheduler selects runnable threads in objects. If there is an event in the buffer whose receiver belongs to the scheduled objects, the object's state machine is executed, contributing to the resulting state $s'$. If no such method is found, the scheduled object's state machine is also executed but its input is $\dagger(th)$, resulting in a execution step of the state machine. All objects not scheduled this time, do not change their state. All events that have be produced are stored in the modified buffer $buf'$ in which all consumed events have been removed.

In Variation Point 6.4.3 an alternative definition for the system state machine is given. In this definition it is assumed that the individual state machines for objects may process a *list* of inputs. The system instantly delivers all produced events and does not need to carry them in an extra buffer as in Variation Point 6.4.2. If an object was scheduled by the scheduler, it is allowed to make a transition as it received $\dagger(th)$ as an input. Additionally, all objects receive the produced events in the same step. This does not lead to an inconsistent feedback because events are stored in the object's buffer anyway. The last condition ensures that objects have only stored the events and did not make further state changes.

The Variation Point 6.4.4 introduces a notion of component in the context of a system. Any subset of objects (e.g., a subset of objects with the same scheduling strategy, or a group of objects realizing a certain system behavior) can be regarded as a component.

**Variation Point 6.4.2 (System STS)**

$$
\begin{array}{l}
[\mathit{SYSSTS}] \\
\text{extend } \mathit{CentralScheduler} \\
\hline
\mathit{SYSSTS} = (\delta_{sys}, \mathit{INITS}) \\
\hline
\mathit{INITS} \subseteq \mathit{STATE} \\
\mathit{INITS} \neq \varnothing \\
\delta_{sys} \in \mathit{STATE} \times \mathit{Buffer}(\mathit{UEVENT}) \rightarrow \mathbb{P}(\mathit{STATE} \times \mathit{Buffer}(\mathit{UEVENT})) \\
\delta_{sys}(s, \mathit{buf}) = \{(s', \mathit{buf}') \mid \\
\quad \mathit{oat} = \mathit{scheduler}\ s \wedge \\
\quad \exists\, \mathit{out} : \\
\qquad \exists (\mathit{oid}, \mathit{th}) \in \mathit{oat} \ : \ \exists\, m \in \mathit{buf} \ : m \notin \mathit{buf}' \wedge \mathit{receiver}(m) = \mathit{oid} \Rightarrow \\
\qquad\quad (s' \mid_{oid}, *) \in \mathit{ests}(\mathit{oid}).\delta(s \mid_{oid}, m) \\
\qquad \wedge \exists (\mathit{oid}, \mathit{th}) \in \mathit{oat} \ : \ \nexists m \in \mathit{buf} \ : \mathit{receiver}(m) = \mathit{oid} \Rightarrow \\
\qquad\quad (s' \mid_{oid}, \mathit{out} \mid_{oid}) \in \mathit{ests}(\mathit{oid}).\delta(s \mid_{oid}, \dagger(\mathit{th})) \\
\qquad \wedge \forall\, \mathit{oid} \in \mathit{oids}(\pi_1(s)) \ : \ (\mathit{oid}, *) \notin \mathit{oat} \Rightarrow \\
\qquad\quad s' \mid_{oid} \ = s \mid_{oid} \\
\qquad \wedge \forall\, m \in \mathit{buf}' \ : \ m \in \mathit{buf} \vee m \in \mathit{out}\}
\end{array}
$$

**Variation Point 6.4.3 (System STS alternative)**

$$
\begin{array}{l}
[\mathit{SYSSTSalt}] \\
\text{extend } \mathit{CentralScheduler} \\
\hline
\mathit{SYSSTS} = (\delta_{sys}, \mathit{INITS}) \\
\hline
\mathit{INITS} \subseteq \mathit{STATE} \\
\mathit{INITS} \neq \varnothing \\
\delta_{sys} \in \mathit{STATE} \rightarrow \mathbb{P}(\mathit{STATE}) \\
\delta_{sys}(s) = \{s'' \mid \\
\quad \mathit{oat} = \mathit{scheduler}\ s \wedge \\
\quad \forall\, \mathit{oid} \in \mathit{oids}(\pi_1(s)) \ : \ \exists\, \mathit{out}, s' \ : \\
\qquad \exists\, \mathit{th} \ : \ (\mathit{oid}, \mathit{th}) \in \mathit{oat} \Rightarrow (s' \mid_{oid}, \mathit{out} \mid_{oid}) \in \mathit{ests}(\mathit{oid}).\delta(s \mid_{oid}, [\dagger(\mathit{th})]) \\
\qquad \wedge (s'' \mid_{oid}, *) \in \mathit{ests}(\mathit{oid}).\delta(s' \mid_{oid}, \mathit{out} \mid_{oid}) \\
\qquad \wedge \pi_i(s'') = \pi_i(s'), i = 1, 2 \\
\quad \}
\end{array}
$$

**Variation Point 6.4.4 (Components in a system context)**

$$
\begin{array}{l}
[\mathit{SysComp}] \\
\text{extend } \mathit{SYSSTSalt} \\
\hline
\mathit{COMPONENT} = (\mathit{comp}, \delta_{comp}) \\
\hline
\forall\, s \in \mathit{STATE}, \mathit{comp} \subseteq \mathit{oids}(\pi_1(s)) \ : \\
\quad \delta_{comp}(s \mid_{comp}) = \{(s' \mid_{comp}) \mid s' \in \delta_{sys}(s)\}
\end{array}
$$

# 7 Timed Object Behavior

One of the main features of the system model is its compositionality. This means that we can describe object behavior on an individual basis as well as in any (meaningful) group. So we can define behavior for compositions of groups of objects into larger components. For this purpose we use the time-aware version of STS, called timed STS (or TSTS) as defined in Appendix B.4.

We assume a discrete global time available. Each step of transition of the TSTS corresponds to a progress of one time unit. A system executes in steps, each consuming a fixed amount of time. Timed state transition systems (TSTS) are transition systems that deal with this kind of paradigm. Roughly speaking, in each step a finite set of input events is provided to a TSTS, and a finite set of output events is produced by the TSTS.

As a further mechanism, we introduce communication channels which allow us to model the interaction (communication flow) between parts of the objects and thus describe the behavior of objects on a very fine grained level.

As a general result of this report, we have a complete description of how systems are decomposed into objects, what states objects may have, and how objects interact.

## 7.1 Object Behavior in the System Model

One crucial question is the choice of the appropriate communication or interaction mechanism. Two basic flavors are asynchronous and synchronous communication. While there is still debate ongoing, we argue in Appendix B.1 that the asynchronous approach seems to be the more abstract. As both approaches can model each other, we have already encoded synchronous method calls into an asynchronous message passing[1] mechanism. In particular, our time-based approach allows us to use a simple abstraction on the time scale to look at communication as being synchronous.

In our system model the object and component instances cooperate by asynchronous message passing. Method invocation is already modeled by the exchange of two events, the method invocation event and the method return event.

As Appendix B describes, communication between objects is dealt with by channels. Channels, on the one hand, help to compose groups of objects into larger units and hide their internal communication. On the other hand, UML provides linguistic constructs like "pins" in some of its diagrams; these pins resemble communication lines between objects and can be mapped to these channels.

A communication channel is a unidirectional communication connection between two objects. We model the channels in the system model as a universe and leave open how many channels are used between objects.

Each channel has a name, e.g., $c \in UCN$, and the type of events that may flow through $c$ is given by $csort(c)$. Each object has a number of incoming and outgoing channels and each event knows through which channel it flows; see Definition 7.1.1.

The existence of the sender and the receiver function has an important consequence. Each event knows on which channel it flows and from which object it originates. From that we can conclude in Lemma 7.1.2 that each channel can be in the output signature of only one object. This lemma ensures the applicability of composition techniques for TSTS that only work if the output channels of composed objects are disjoint (see the Definition of composition in B.2.13).

Based on the definition of channels and their type, the behavior of a single object is defined in Definition 7.1.3.

[1] Message passing is the general term; in our case events (which include message events) are passed.

**Definition 7.1.1 (Channel signatures of objects)**

*Channels*

extend *State*
use *Focus*

---

$UCN$
$sender, receiver : UCN \to UOID$
$channel : UEVENT \to UCN$
$inC, outC : UOID \to \mathbb{P}(UCN)$
$csort : UCN \to \mathbb{P}(UEVENT)$

---

$\forall\, m \in UEVENT, oid \in UOID :$
$\quad sender(m) = oid \Rightarrow sender(channel(m)) = oid$
$\quad receiver(m) = oid \Rightarrow receiver(channel(m)) = oid$
$\forall\, c \in UCN :$
$\quad inC(oid) = \{c \mid receiver(c) = oid\}$
$\quad outC(oid) = \{c \mid sender(c) = oid\}$
$\quad csort(c) = \{m \in UEVENT \mid channel(m) = c\}$

---

*UCN* denotes the universe of channel names.
*sender* and *receiver* assign a sending and a receiving object to each channel.
*channel* assigns a channel to each event.
*inC*, *outC* denote the channel signature of each object.
The type of each channel $csort(c)$ describes the possible events flowing over that channel.

**Lemma 7.1.2 (Channels disjoint)**

use *Channels*

---

$\forall\, a, b \in UOID : a \neq b \Rightarrow outC(a) \cap outC(b) = \varnothing$

---

follows from the definition of *outC* and *channel*.

**Definition 7.1.3 (Behavior of individual objects)**

*ObjBehavior*

extend *Channels*

---

$beh : UOID \to \mathcal{B}^{csort}(I, O)$

---

$\forall\, oid \in UOID :$
$beh(oid) \in \mathcal{B}^{csort}(inC(oid), outC(oid))$

---

$beh(oid)$ denotes the behavior of one single object.

Definition 7.1.3 is based on the assumption of a fine enough time granularity, i.e., so fine that the output in a step does not depend on the input received in that step. This way, strong causality between input and output is preserved. The composition of state machines is moreover simplified since feedback within one time unit is ruled out and thus causal inconsistencies are avoided. Even so, we are able to abstract away from the actual (real-)time occurrence of events and only consider the untimed behavior of objects.

**Definition 7.1.4 (Behavior of object compositions)**

*CompBehavior*

extend *ObjBehavior*

$beh : \mathbb{P}(UOID) \to \mathcal{B}^{csort}(I, O)$
$inC, outC : \mathbb{P}(UOID) \to \mathbb{P}(UCN)$

$\forall\, os \subset UOID :$
$I = inC(os) = \{c \mid receiver(c) \in os \wedge sender(c) \notin os\}$
$O = outC(os) = \{c \mid sender(c) \in os \wedge receiver(c) \notin os\}$
$beh(os) = \bigoplus_{oid \in os} beh(oid)$

$beh(os)$ denotes the behavior of a group of objects where internal communication is not visible anymore. *inC* describe the incoming channels for a group of objects, *outC* the outgoing channels.

Definition 7.1.4 provides a flexible concept of components including, e.g., classical sequential systems (in this case, there is only one input and one output channel). For instance, we may restrict the input and output events in such a way that in each step at most one input event is received or one output event is dispatched. At the other extreme, we can model highly concurrent systems with a large number of input and output events in one state transition step.

## 7.2 State-based Object Behavior

While the behavior of an object *oid* is precisely defined as $beh(oid)$, its relationship to a state-based view on object behavior still needs to be clarified. For this purpose, we attach a timed state transition system to each object in Definition 7.2.1.

**Definition 7.2.1 (Behavior as TimedSTS)**

*TimedSTS*

extend *CompBehavior*

$tsts : UOID \to TSTS^{csort}(S_1, I_1, O_1)$
$tsts : \mathbb{P}(UOID) \to TSTS^{csort}(S, I, O)$

$\forall\, oid \in UOID :$
$\quad tsts(oid) \in TSTS^{csort}(states(oid), inC(oid), outC(oid))$
$\quad \mathbb{S}(tsts(oid)) = beh(oid)$
$\forall\, os \subset UOID :$
$\quad tsts(os) = \bigoplus_{oid \in os} tsts(oid)$

$tsts(oid)$ denotes the TSTS based description of behavior of one single object.
The definition is then generalized to a set of objects.

According to Definition 7.2.1, each object $oid \in UOID$ can be described by a nondeterministic TSTS as introduced in Appendix B.4. $\mathbb{S}(tsts(oid)) = beh(oid)$ states that the behavior of each object is defined by an appropriate $tsts(oid)$. Appendix B.4.3 shows that the composition of TSTS and of I/O-behaviors is compatible, which means that we can switch between a state-based and a purely I/O-based view of object behavior and specify individual objects or meaningful groups (components) at will.

Note that each object *oid* has exactly one single timed state transition system $tsts(oid)$. However, as $tsts(oid)$ is a nondeterministic state machine, it allows various forms of underspecification. Therefore, there is no need to add a further concept of underspecification by, e.g., assigning a set of possible TSTS to each object. Any UML model, however, may have an impact on the elements of a TSTS. For instance, the sets of reachable states can be constrained, the initial states restricted to be a singleton, or the nondeterminism reduced by enforcing a behavior that is deterministic in reaction and time.

With this last part of the system model, we now have a TSTS for the whole system that includes all snapshots and all system states and thus is capable of describing any behavioral and structural restrictions by $tsts(UOID)$.

Note that we have a closed world assumption now: The overall system transition system $tsts(UOID)$ does not have external channels anymore, but incorporates all "objects". This also includes objects that have direct connections to interfaces to other systems, mechanical devices or users and thus can act as surrogates for the context of the system. In [Rum96] we have discussed how to deal with this to model open, reactive systems in a closed world, and also what the advantages are.

## 7.3 Mapping Event STS to Timed STS

As a final step, we investigate variants of mapping the event-based STS as given in Definition 6.2.1 to timed STS from Definition 7.2.1.

### 7.3.1 General Mapping of Event STS to Timed STS

Definition 7.3.1 gives a general description of the relationship between the ESTS and the TSTS of any object. We can concentrate on the relation between the transition function of each TSTS, as the input and output signatures as well as the state space are already defined through $ests(oid)$ and $tsts(oid)$. In particular both have the same state basis and the I/O-signatures are given by the channel sorting function *csort*.

The idea of the mapping between $\delta_e$ and $\delta_t$ is that one step in $\delta_t$ equals a sequence of steps in $\delta_e$. This sequence, of length say $k$, produces intermediate states $s_j$, a sequence of outputs $o_j$ and is triggered by a sequence of inputs $i_j$ $(1 \leq j \leq k)$.

These inputs also contain †-pseudo events that resembles stepwise progress of the ESTS. In the used definition for *tsts*, these steps are not coming from outside but are freely chosen within the definition of $\delta_t$. The scheduling in this version is embedded in the transition function of the TSTS.

An important constraint of the mapping is that $\delta_t$ is input enabled. This means, $\delta_t$ describes a reaction on any sequence of input events. However, Definition 7.3.1 only states that some of the paths of $\delta_e$ need to be taken, but not necessarily all of them. This gives us freedom to choose appropriate scheduling strategies by disregarding certain sequences of steps (which are modelled through $\dagger(th)$).

Whatever scheduling strategy will be used, however and by construction, the resulting TSTS must be time guarded. This means, e.g., that the return event for a method call can occur at the earliest one time step after the method call was received. That is why we store the method call in the buffer before we actually start processing it.

### 7.3.2 Variation Point: Constraining the Timed STS

The above definitions permit the definition of various scheduling strategies. We can constrain $\delta_t$ to, e.g., allow only one step per thread per time unit, or handle a series of steps in one thread, but none in any other, etc.

**Definition 7.3.1 (General mapping between ESTS and TSTS)**

*ESTStoTSTS*

use *ESTS*
extend *TimedSTS*

$\forall\, oid \in UOID,$
$ests(oid) = (S, I^e, O^e, \delta^e, s0^e),$
$tsts(oid) = (S, \mathcal{T}(I^t), \mathcal{T}(O^t), \delta^t, s0^t)$ :
$S_e = S_t \wedge s0^e = s0^t \wedge$
$\delta_t : s \xrightarrow{i/o} t \Rightarrow$
$\quad \exists\, k \in \mathbb{N}_0, m_j \in I^e, s_j \in S^e, o_j \in (O^e)^*$ :
$\quad i = cdist(UEVENT \copyright m_j) \wedge$
$\quad o = cdist(\frown_j\, o_j) \wedge$
$\quad s_0 = s \wedge s_{k+1} = t \wedge$
$\quad \forall\, 0 \leq j < k : \delta_e : s_j \xrightarrow{m_j/o_j} s_{j+1}$

$cdist : (I \rightarrow \mathbb{P}(M)) \rightarrow M^* \rightarrow \mathcal{T}(I)$

$\forall\, s \in M^*, c \in I$ :
$\quad cdist(s).c = csort(c) \copyright s$

Clarifies the relationship between $ests(oid)$ and $tsts(oid)$ by describing the set of possible TSTS that may belong to an ESTS.
*cdist* is used to distribute a stream of events over their channels.

**Variation Point 7.3.2 (Scheduling in the TSTS)**

[*TSTSScheduling*1]

extend *ESTStoTSTS*

$\forall\, tsts(oid) = (S, \mathcal{T}(I^t), \mathcal{T}(O^t), \delta^t, s0^t)$ :
$\delta_t : s \xrightarrow{i/o} t \Rightarrow$
$\quad \#(\dagger(UTHREAD) \copyright i) = 1$

Variation Point 7.3.2 constrains the scheduling to exactly one step per time frame and thus aligns the ESTS stepping with the TSTS time frames. However, it does not yet specify any fairness strategies for the threads waiting to execute.

We omit further specifications of these strategies here as they are known from operating systems.

## 7.4 The System Model Definition

The last definition in this report finally introduces the universe of system models in Definition 7.4.1.

## 7.5 Summary for Object Behavior with Timed STS

The final part of the theory is summarized in theory *TSTS* in Definition C.1.8. Figure 7.1 illustrates the theory dependencies and variation points. The complete picture, including the Definition of *SYSMOD*, was already presented in Figure 1.2.

**Definition 7.4.1 (Definition of the system model as a universe)**

*SYSMOD*

extend *ESTS*, *TSTS*

---

*SYSMOD*

---

$sm \in SYSMOD \Rightarrow$
$sm =$
$(UTYPE, UVAL, CAR,$
$UVAR, vtype, vsort,$
$UCLASS, UOID, attr, oids, classof,$
$sub, \&,$
$UASSOC, classes, extraVals, relOf,$
$UOPN, UOMNAME, nameOf, classof, parTypes, params, resType,$
$UMETH, UPC, nameof, definedin, parNames,$
$localNames, resType, pcOf, impl,$
$UTHREAD,$
$UVENT, eventIn, eventsOut,$
$UMESSAGE, MsgEvent, USIGNAL,$
$ests,$
$UCN,$
$tsts)$
such that all constraints defined in this document are fulfilled.

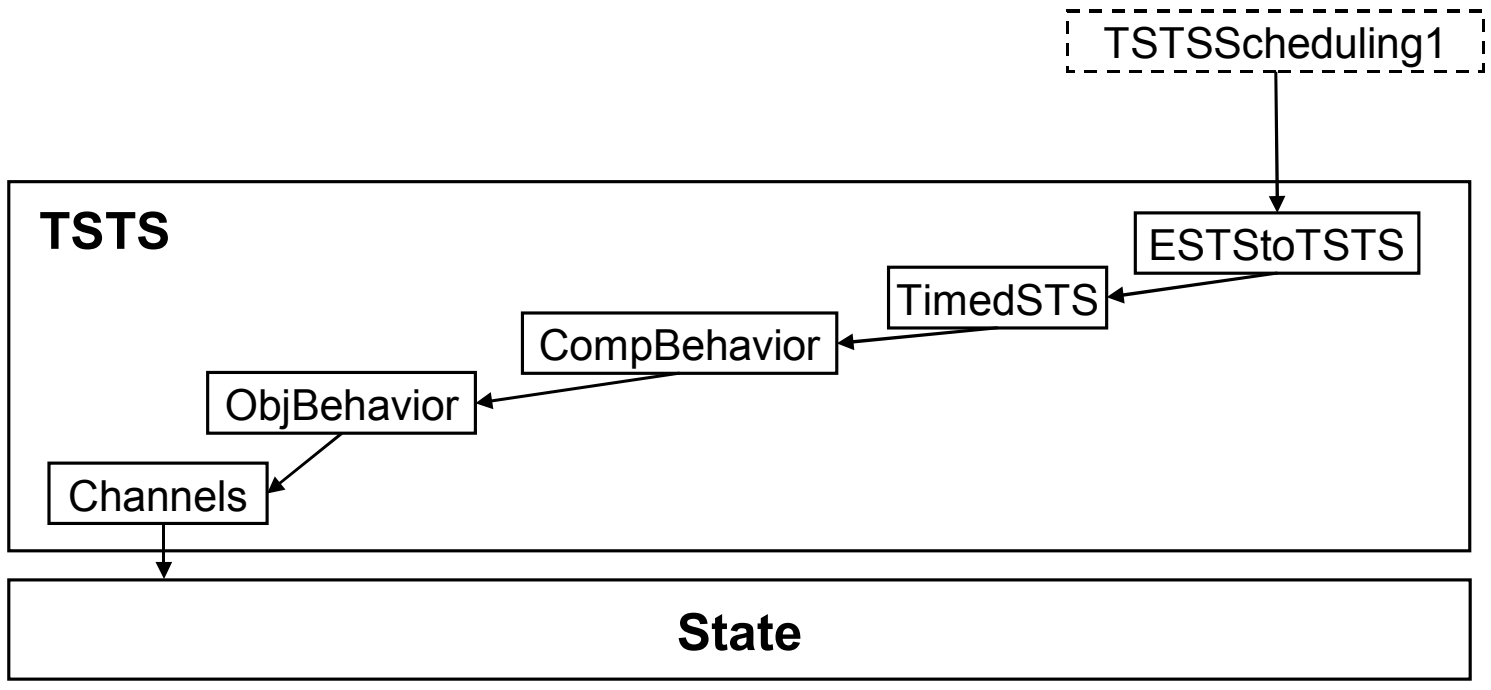

Figure 7.1: Theory *TSTS* and its dependencies.

# 8 Concluding Remarks

In this report, we have introduced the system model as a mathematically defined semantic domain for the UML.

It describes structure and behavior of object systems on a very detailed and fine-grained basis. It uses the general notion of timed state transition systems and I/O-behaviors which both are integrated with the data, control and event stores. As a general result of the theory of system models introduced in this document, we have a complete description of how systems are decomposed into objects, what states objects may have and how objects interact. As motivated in the introduction, we have developed the mathematical theory in layers, each building up an algebra that introduces some universe of elements, functions and laws for these functions.

We have chosen this approach because we want a semantics that is not biased by the choice of a concrete formal language or tool. Even the use of mathematical theories probably will bias the semantics a little but we hope as little as possible. Such bias easily creeps in and we carefully tried to avoid it. In particular, we did not address executability, because this includes one of the biggest biases a modeling language can have: A model shall have the ability for underspecification. It shall be open for a specification of many different implementations. An executable semantics for an underspecified UML model must therefore necessarily contain implicit choices added by the semantic mapping.

To prevent the executability bias, we have chosen a specific style of description. The form of description used throughout this document allows us to leave quite a number of definitions open. We have usually introduced a universe and then characterized the properties of its elements without fully determining how many elements it has or how these elements look like. Sometimes, we only described a subset of the elements and allow other kinds of elements to be in the universe as well (e.g., the universes of events, messages and values are defined in such a way).

This gives us the chance to specialize variation points according to specific situations. To put it in UML jargon, we could for example define a "system model profile" that specializes the general definitions to sequential, single threaded systems, to static systems without creation of new objects, or to systems without subclassing, etc.

While the system model is an underlying basis for these kinds of systems, it does not provide such specialization directly; this is matter of further work. Indeed, as one of the results of this work, we have been able to make a number of variation points explicit. Although there are a lot more variation points to explore and their bandwidth to clarify, we regard this approach as a first important step to the formalization and clarification of variation points.

On the other hand, the complexity of larger parts of the system model has shown that the integration of objects, threads, state-based behavior and concurrency is complex, has many variations and is therefore somewhat arbitrary. It is particularly complex to model the possible interactions between these, leading us to the assumption that it is particularly difficult to master these not so well integrated concepts. It might be worthwhile rethinking a better integration.

The system model defined in [BCR06, BCR07a, BCR07b] has actively been used to define the semantics of UML sublanguages like class diagrams [CGR08a] and Statecharts [CGR08b]. In [CDGR07] a simulator for UML models has been developed based on the system model definitions. This work has been carried out in the context of the DFG rUML project. In [CD08] UML action are formalized using the system model as a semantic domain. The system model also forms the basis for characterizing the semantics of model composition [HKR$^+$07] as part of the MODELPLEX project. Using the system model definitions revealed strengths and potentials for improvement. The experiences made led to the system model, version 2, defined in this document.

## 8.1 Further Extensions

Of course this system model that can be seen as a hierarchy of algebras may and probably should be extended by adding further functional machinery to ease description of the mapping of UML constructs to the system model. However, we wanted to keep the system model rather simple and therefore did not concentrate on this additional machinery very much. “Users” of the system model are really invited to add whatever they feel appropriate.

There are also a number of loopholes and particular variation points that can be further investigated by providing additional machinery to clarify a mapping of UML concepts to the system model.

A number of higher-level concepts could be added to the system model more or less directly. As we have demonstrated with associations, which are manifested as retrieval functions on the object store, we might add a basic set of actions and activities or components thereof like the “pins” of the activity diagrams, features like in [KPR97], or workflow elements as in [RT98].

We wish to thank a number of colleagues, and especially Bran Selic, Michelle Crane, Jürgen Dingel, Gregor von Bochmann, Gregor Engels, Alain Faivre, Christophe Gaston, Sébastien Gérard and Martin Schindler for their valuable help.

# A Basic mathematics used

In this report we have used various mathematical theories. In the following some sub-theories used are shortly described.

## A.1 Functions, Logic, Sets

Mathematical theory is quite compact, however some shorthand notations make mathematical formulas still more compact. In the definitions A.1.1, A.1.2 and A.1.3 we have the most basic theories at hand.

**Definition A.1.1 (Functions)**

*Functions*

| | |
|---|---|
| $f : X \to Y$ | total functions |
| $f : X \rightharpoonup Y$ | partial functions |
| $f(a), f.a$ | application in normal and selector-style |
| $dom(f) \subseteq X$ | domain of $f$ (useful if f is partial) |
| $[a = b]$ | singleton $f$ mapping $a$ to $f(a) = b$ |
| $[a = b, c = d, \ldots]$ | |
| $(\lambda x.f(x))$; | lambda abstraction |
| $(\lambda x \in S.f(x))$ | |
| $f \oplus g$ | overriding union of functions: $(f \oplus g)(a) = g(a)$ if $a \in dom(g)$, else $= f(a)$ |
| $f\mid_S$ | restricts $f$: $dom(f\mid_S) = dom(f) \cap S$ |

Please note that $g$ has precedence in $f \oplus g$.

**Definition A.1.2 (Logic)**

*Logic*

| | |
|---|---|
| $P \wedge Q, P \vee Q, \neg P,$ | logic operators |
| $P \Rightarrow Q, P \Leftrightarrow Q$ | |
| $\mathbb{B} = \{tt, ff\}$ | truth values |
| $\exists x : P(x)$; $\exists^{=1}, \exists^{\geq 1}, \exists^{\leq 1}$ | variants of $\exists$ operators |
| $P(a, *, *)$ | wildcard $*$ is shorthand for $\exists x, y : P(a, x, y)$ |
| $\forall x : P(x)$ | |

Wildcard $*$ is a convenient form to specify the existence of a value that isn't otherwise of interest. It can, e.g., be used in set comprehensions. The existence operator applies at the innermost possibility and uses fresh variables: $\forall a : P(a, *)$ is equivalent to $\forall a : \exists x : P(a, x)$

**Definition A.1.3 (Sets)**

*Sets*

| | |
|---|---|
| $\varnothing, \{1, 2, 3\}$ | sets |
| $A \cup B, A \cap B, A \setminus B$ | set operators |
| $x \in A, \mathbb{P}(A), \mathbb{P}_f(A)$ | |
| $\{x \mid P(x)\}, \{f(x) \mid P(x)\}$ | set comprehension |
| $\{(x, *) \mid P(x)\}$ | wildcard means $\{(x, y) \mid P(x), \exists\, y\}$ |

## A.2 Collections (or Containers)

We will in the forthcoming also use a variety of collections, such as lists, tuples and records which provide some intuitive, but also some sophisticated functionality to handle. Please note that we intend these theories to be basic theories and do not (necessarily) intend these data structures to become part of a typing system.

Definition A.2.1 provides a number of functions that apply to collections in general. We use $\mathbb{C}(X)$ to denote a collection such as sets, lists etc. over set $X$.

**Definition A.2.1 (Collections)**

*Collections*

| | |
|---|---|
| $. \in . : X \times \mathbb{C}(X) \to \mathbb{B}$ | element in? |
| $\#. : \mathbb{C}(X) \to \mathbb{N} \cup \{\infty\}$ | size |
| $map : (X \to Y) \times \mathbb{C}(X) \to \mathbb{C}(Y)$ | pointwise application |
| $filter : \mathbb{P}(X) \times \mathbb{C}(X) \to \mathbb{C}(X)$ | set-based filter |
| $filter : (X \to \mathbb{B}) \times \mathbb{C}(X) \to \mathbb{C}(X)$ | predicate-based filter |
| $set : \mathbb{C}(X) \to \mathbb{P}(X)$ | as set |

Notation:
$P$ⓒ$c$ is shorthand for $filter(P, c)$
$f^*(c)$ is shorthand for $map(f, c)$

Definitions A.2.2, A.2.3 and A.2.4 describe lists and stacks useful, e.g., for parameter lists and the implementation stack when describing behavior. While structurally equivalent, they serve different intentions and therefore have different access and modifier operations.

**Definition A.2.2 (List)**

*List*

| | |
|---|---|
| $List[], List[1, 2, 3]$ | lists |
| $List(S)$ | Lists over set S |
| $l_1 + +l_2$ | concatenation |
| $nub(l)$ | remove doubles from list |
| $List[f(x) \mid x \in ls, P(x)]$ | list comprehension for list $ls$ |

**Definition A.2.3 (Stack)**

*Stack*

| | |
|---|---|
| $Stack[], Stack[1,2,3]$ | stacks |
| $Stack(S)$ | stacks over set S |
| $push(st, el)$ | push an element onto the stack |
| $pop(st)$ | remove top element from the stack |
| $top(st)$ | top element of the stack |
| $update(st, el)$ | replace top element of the stack |
| $Stack[f(x) \mid x \in ls, P(x)]$ | stack comprehension for stack *ls* |

**Definition A.2.4 (Buffer)**

*Buffer*

| | |
|---|---|
| $Buffer[], Buffer[1,2,3]$ | buffers (like lists) |
| $Buffer(S)$ | all buffers over set S |
| $b_1 + +b_2$ | concatenation |
| $fst(b)$ | first element |
| $rst(b)$ | rest of the nonempty buffer b |
| $addfirst(b, el), addlast(b, el)$ | add |
| $Buffer[f(x) \mid x \in b, P(x)]$ | buffer comprehension |

## A.3 Records

Record types are a classical concept to represent the state space of classes and their objects as well as local variables and parameters. In particular, in name-tagged records, these functions do have appropriate names, resembling attribute names. In addition to constructing new values from given ones, record values (i.e., values in the carrier set of a record type name) also provide selection functions for each part of a record value. In name-tagged records, these tag names provide proper names for the selection functions. We start with Definition A.3.1 describing the nature of a record as a function.

For access of record variables we define auxiliary functions in A.3.2.

The variables $a_i$ are called the attributes of the record. Notice that *Rec* is defined on functions and thus, the definition of *Rec* does not rely on the ordering of its attributes. Therefore $Rec\{a : T, b : S\}$ and $Rec\{b : S, a : T\}$ describe the same records.

## A.4 Tuples: Cartesian Products

Cartesian products (also called "cross products" or "tuples") over values and the carrier sets of the record type names introduced above do share some common structure. However, they also differ significantly, so that we do not identify them. Records have indexed value entries, whereas Cartesian products have an ordered list of values. Although in programming Cartesian products can be mimicked by records, in our system model we need a Cartesian product for example to model parameters of messages and methods,because these parameters are submitted by position and not by name. Therefore, we unfortunately need both, records and cartesian products.

Due to their similarity, records and tuples can be translated into each other. There are two mappings between tuples and corresponding records given in Definition A.4.2.

**Definition A.3.1 (Basic structure of records)**

*Record*1

$SRec : D \times (D \to \mathbb{P}(V)) \to \mathbb{P}(D \to V)$

Notation:
$Rec\{a_1 : V_1, \ldots, a_n : V_n\}$ is shorthand for
    $SRec(\{a_1, \ldots, a_n\}, S)$ with $S(a_i) = V_i$
$[a_1 = v_1, \ldots, a_n = v_n]$ is a shorthand for the value (function)
    $r : \{a_1, \ldots, a_n\} \to V$ with $r(a_i) = v_i \in V_i$

$SRec(D, S)$ is the set of records with domain $D$ (its attributes) and its attribute values constrained by sort function $S$. The notations used for record values $[\ldots]$ and for record types $Rec\{\ldots\}$ help accommodate an easier reading.

Derived:

$r_i \in SRec(D_i, S_i) \Rightarrow r_1 \oplus r_2 \in SRec(D_1 \cup D_2, S_1 \oplus S_2)$

**Definition A.3.2 (Attribute selection for records)**

*Record*2

$RECORD(D, S) = \bigcup_{E \subseteq D} SRec(E, S)$
$attr : RECORD(D, S) \to D$

Notation:
$r.a_k$ is the record denotation for $r(a_k) \, \forall \, a_k \in D$

$attr(Rec\{a_1 : S_1, \ldots, a_n : S_n\}) = \{a_1, \ldots, a_n\}$
$attr([a_1 = x_1, \ldots, a_n = x_n]) = \{a_1, \ldots, a_n\}$

*attr* is the list of attribute names.
$RECORD(D, S)$ contains all records.

The mappings between tuples and records are inverse (Lemma A.4.3). Note that both, tuples and records, need the list of attribute names: *rec* needs the ordered list $[a_1, .., a_n]$ to map the values to the appropriate attributes. *tuple* needs the list to restore the order given in the tuple (which is not present in the record).

**Definition A.4.1 (Cartesian products (tuples))**

*CartesianProduct*

$Tuple : List(V) \to \times_{i \leq n} V$
$STuple : List(\mathbb{P}(V)) \to \mathbb{P}(\times_{i \leq n} V)$
$\#. : Tuple(V) \to \mathbb{N}$

Notation:
$(a_1, \ldots, a_n)$ for $Tuple[a_1, \ldots, a_n]$

$STuple[S_1, \ldots, S_n] = \times_{(1 \leq k \leq n)} S_k$
$STuple[] = \{()\}$
$\#(a_1, \ldots, a_n) = n$

$TUPLE(V) = \{Tuple(lt) \mid lt \subseteq V, \#lt \in \mathbb{N}\}$
$(a_1, \ldots, a_n) \in STuple[S_1, \ldots, S_n] \Rightarrow a_i \in S_i$

$\pi_k : Tuple(V) \to V$
$\pi_k(a_1, \ldots, a_n) = a_k$

$Tuple[]$ with the empty list is special unit type whose carrier has one value $()$.
*STuple* describes the set of all tuples of given sorts $V_i$. $TUPLE(V)$ contains all tuples over set $V$ and $\pi_k$ is used as selector.

**Definition A.4.2 (Cartesian products and records)**

*CartRec*

$rec_{d=[a_1,\ldots,a_n] \in List(D)} : TUPLE[V_1, \ldots, V_n] \to RECORD(d, \oplus_i [a_i = V_i])$
$tuple_{d=[a_1,\ldots,a_n] \in List(D)} : RECORD(d, \oplus_i [a_i = V_i]) \to TUPLE[V_1, \ldots, V_n]$

$rec_{[a_1,\ldots,a_n]}(v_1, \ldots, v_n) = [a_1 = v_1, \ldots, a_n = v_n]$
$tuple_{[a_1,\ldots,a_n]}([a_1 = v_1, \ldots, a_n = v_n]) = (v_1, \ldots, v_n)$

*rec* produces a record from a tuple. The list of attributes determines where the values are assigned to.
*tuple* produces a Cartesian product from a record. The list of variables determines the order of the Cartesian product, as records are unordered.

**Lemma A.4.3 (Mappings between tuples and records are inverse)**

use *CartRec*

$rec_{[a_1,\ldots,a_n]}(tuple_{[a_1,\ldots,a_n]}(r)) = r$
$tuple_{[a_1,\ldots,a_n]}(rec_{[a_1,\ldots,a_n]}(t)) = t$

# B Central Model of Interaction: Streams, Components, STS

This chapter describes a closed, well-defined theory on stream processing components and state transition systems that define two different views on distributed, interacting systems. Both, streams processing components and the state transition systems are well related through appropriate mapping functions and their refinement and composition techniques are compatible.

## B.1 Types of Models for Interactive Systems

There are a number of different theories and fundamental models of interactive systems. Most significant for them are their paradigms of interaction and composition. We identify the following basic concepts of communication in distributed systems that interact by message exchange:

**Asynchronous communication** (message asynchrony): a message is sent as soon as the sender is ready, independent of the fact whether a receiver is ready to receive it or not. Sent messages are buffered (by the communication mechanism) and can be accepted by the receiver at any later time. If a receiver wants to receive a message but no message was sent it has to wait. However, senders never have to wait (see [Kah74, EHS97]) until receivers are ready since messages may be buffered.

**Synchronous communication** (message synchrony, rendezvous, handshake communication): a message can be sent only if both the sender and the receiver are simultaneously ready to communicate; if only one of them (receiver or sender) is ready for communication, it has to wait until a communication partner gets ready (see [Hoa83, Mil82]).

**Time synchronous communication** (perfect synchrony): several interaction steps (signals or atomic events) are conceptually gathered into one time slot; in this way, systems are modeled with the help of sequences of sets of events (see [BG92] as a well-known example).

**Traditional method call:** It combines some characteristics of all three previously described approaches. For sequential method calls, progress of time is not such a big issue, which allows programmers to think of synchronous message passing and even of perfect synchrony. The receiver, however, cannot prevent the sender (caller) to start the method call like with asynchronous message passing. The receiver must accept the call and react somehow.

Any of the first three models can be used to encode any other, and method calls can be simulated in all of them. Furthermore, a general purpose modeling language, like the UML, attempts to provide mechanisms for all of these communication paradigms. Moreover, these communication paradigms shall be used within one system and work together. It is therefore necessary to integrate all communication paradigms, e.g., by encoding one within the other. It is a matter of taste to choose one of these paradigms as underlying mechanism. In the following, we work with asynchronous message passing since this model has, according to our experience, the finest properties for our purpose. We follow the system model given in [BS01] basing our approach on a concept of a component that communicates messages asynchronously with its environment via named channels within a synchronous time frame.

## B.2 Streams

For a convenient specification of object behavior, it is of interest to look not only at the currently incoming message, but at the overall sequence of messages that has arrived on a channel so far. We thus use channel histories to model traces of behavior.

### B.2.1 Basic Streams

A stream is a finite or infinite sequence of elements of a given set to describe object behavior. In interactive systems streams are built over sets of messages or actions. A stream describes an observation that an observer can make when sitting on a directed communication channel. The behavior of a component can then be modeled through a relation between its observed input and output streams. Streams are therefore used to represent interaction patterns by communication histories for channels or histories of activities.

Technically, a container structure and thus the functions from Definition A.2.1 apply here as well. Let $M$ be a set (of messages). By $M^*$ we denote the set of finite sequences of elements of $M$, and by $M^\infty$ the set of infinite sequences of elements of $M$. The set $M^\omega$ of streams over $M$ are finite or infinite sequences of elements of the set $M$. Thus $M^\omega = M^* \cup M^\infty$. If desired, streams over $M$ can be understood as partial functions of form $x : [1..n] \to M$ with "length" $\#x = n \in \mathbb{N} \cup \{\infty\}$, where infinite streams are exactly the total functions $(\mathbb{N} \to M)$. We write $x.t$ instead of $x(t)$ as shorthand for selection of the element in $x$ at position $t$. A finite stream $x$ of elements $x.1, \ldots, x.n$ (in this ordering) is also written $\langle x.1, \ldots, x.n \rangle$. A special case is the empty stream, denoted by $\epsilon$ (sometimes $\langle\,\rangle$ is used to denote the empty stream). The set of streams has an adequate set of mathematical operations, forming a rich algebraic and topological structure as given in Definition B.2.1.

Based on the Definition B.2.1 we can conclude a number of properties in Lemma B.2.2.

Further operators can be defined on streams allowing easier specification; see Definition B.2.3, additional properties are presented in B.2.4.

$(M^\omega, \sqsubseteq)$ is a partial order, i.e., the relation $\sqsubseteq$ is reflexive, transitive and antisymmetric. This partial order is well-founded, i.e., it contains no countable infinite descending chains and complete (it has a least element, namely $\epsilon$, and each of its chains has a least upper bound). This property is very useful, as it allows the description of finite prefixes of streams and the use of inductive (or recursive) techniques for full stream characterization. The theory can be found in [BS01].

### B.2.2 Timed Streams

A stream represents the sequence of messages sent over a channel during the lifetime of a system. Of course, in concrete systems this communication takes place in a time frame. Hence, it is often convenient to be able to refer to this time. Moreover, as we will see, the theory of feedback gets much simpler. Therefore we work with timed streams as given in Definition B.2.5.

Streams are used to represent histories of communications of data messages transmitted within a time frame. Given a message set $M$, we define a (infinite) timed stream as the elements in the set $(M^*)^\infty$, or, equivalently, as the functions of form $s : \mathbb{N} \to M^*$.

As timed streams are just a special form of streams, we can use those operators again. $s.t$ denotes the sequence of messages observed in the stream $s$ at time slot $t$. That is, a timed stream $s \in (M^*)^\infty$ expresses which messages are transmitted at which time(s). To map between timed and untimed streams, we use $(\diamond s)$ that extracts all messages. Note that $(\diamond s)$ is finite iff $s$ carries only a finite number of nonempty sequences.

### B.2.3 Channels and Histories

Timed streams are used to model the communication histories of sequential, unidirectional communication media (e.g., between two objects) that we call channels. Since a system usually has a larger number of communication streams, we work with channels to refer to individual communication streams. Accordingly,

**Definition B.2.1 (Definition of streams)**

*Streams*1

$M^\omega = M^* \cup M^\infty$

| | |
|---|---|
| $\epsilon, \langle 1,2,3\rangle, \langle 1,2,3,\ldots\rangle$ | streams |
| $x : s$ | prefix an element |
| $s.i$ | selection for $i \in \mathbb{N}$ |
| $\# : M^\omega \to \mathbb{N} \cup \{\infty\}$ | length |
| $\frown : M^\omega \times M^\omega \to M^\omega$ | stream concatenation |
| $fst : M^\omega \to M$ | first element |
| $rst : M^\omega \to M^\omega$ | rest of stream (if nonempty) |
| $\langle f(i) \mid 1 \le i \le n\rangle,$ | stream comprehension |
| $\langle f(i) \mid i \in \mathbb{N}\rangle,$ | infinite stream comprehension |
| $\langle f(x) \mid P(x), x \in st\rangle,$ | stream comprehension for stream *st* |

$m : s = \langle m\rangle \frown s$

$\forall \langle s.1,\ldots,s.n\rangle, \langle r.1,\ldots,r.m\rangle \in M^*, z \in M^\infty :$

$\quad \langle s.1,\ldots,s.n\rangle \frown \langle r.1,\ldots,r.m\rangle = \langle s.1,\ldots,s.n,r.1,\ldots,r.m\rangle$

$\quad i \le n \Rightarrow (\langle s.1,\ldots,s.n\rangle \frown z).i = s.i$

$\quad i > n \Rightarrow (\langle s.1,\ldots,s.n\rangle \frown z).i = z.(i-n)$

$\forall s \in M^\infty, r \in M^\omega :$

$\quad s \frown r = s$

$\#\epsilon = 0$

$\#\langle s.1,\ldots,s.n\rangle = n$

$\#(s \frown r) = \#s + \#r$

$s \neq \epsilon \Rightarrow fst(s) \frown rst(s) = s$

$M^\omega$ is the set of all finite and infinite streams.

**Lemma B.2.2 (Properties for streams 1)**

use *Streams*1

$\#(x : s) = 1 + \#s$

$fst(x : s) = x \wedge rst(x : s) = s$

$\#s = 0 \Rightarrow s = \epsilon$

$s \neq \epsilon \Rightarrow fst(s \frown r) = fst(s)$

$s \neq \epsilon \Rightarrow rst(s \frown r) = rst(s) \frown r$

$\#s = \infty \Rightarrow s \frown r = s$

$s \neq \epsilon \Rightarrow \#rst(s) = \#s - 1$

$s = r \Leftrightarrow s = \epsilon = r \vee (fst(s) = fst(r) \wedge rst(s) = rst(r))$

$s = r \Leftrightarrow \#s = \#r \wedge \forall 1 \le i \le \#n : s.i = r.i$

a channel is simply an identifier (channel name) which is associated with a stream observation in every execution of the system.

As the messages that can be observed on a channel are constrained, each channel is given a “type” of

**Definition B.2.3 (More operations on streams)**

*Streams*2

extend *Streams*1

| | |
|---|---|
| $s \sqsubseteq r$ | prefix relation |
| $s \downarrow t$ | prefix of length $t \in \mathbb{N}$ |
| $P©s, S©s$ | filter (predicate or set) |
| $P\#s, S\#s$ | number of messages in S / P |
| $f^*(s)$ | map function element-wise |

$\forall\, s, r \in M^\omega$ :

$s \sqsubseteq r \Leftrightarrow \exists\, z \in M^\omega : s \frown z = r$

$\#(s \downarrow t) = min(\#s, t)$

$(s \downarrow t) \sqsubseteq s$

$f^*(x : s) = f(x) : f^*(s)$

$P(x) \Rightarrow P©(x : s) = x : (P©s)$

$\neg P(x) \Rightarrow P©(x : s) = P©s$

$P\#s = \#(P©s) \wedge S\#s = \#(S©s)$

$s \sqsubseteq r$ denotes the prefix relation between streams, $x \downarrow t$ selects the prefix of length $t$, $P©s$, $S©s$ and $f^*(s)$ denote the usual filters and mapping functions.

**Lemma B.2.4 (Properties for streams 2)**

use *Streams*2

$\#s = \infty \wedge s \sqsubseteq r \Rightarrow s = r$

$\#t < \infty \Rightarrow s \sqsubseteq r \Leftrightarrow (t \frown s) \sqsubseteq (t \frown r)$

$f^*(s \frown r) = f^*(s) \frown f^*(r)$

$\#r < \infty \Rightarrow P©(r \frown s) = (P©s) \frown (P©r)$

$(X \cup Y) \downarrow t = (X \downarrow t) \cup (Y \downarrow t)$

messages that flow on the channel. The concept of a stream is then used to define the concept of a channel history. A channel history is given by the messages communicated over a channel. Such a history describes an observation on a channel, when recording the flow of messages over time. See Definition B.2.6 for the basic structure and Definition B.2.7 for operations.

All operations and notational conventions introduced for streams generalize in a straightforward way to histories applying them element-wise. As we deal with piecewise composed behavior, we extend our notion of channels histories to partial histories of the form $s : D \to (M^*)^*$ and to time slices of the form $u : D \to M^*$. The latter is explicitly given by Definitions B.2.8 (where $\mathcal{T}$ is used in place of $\mathcal{H}$):

### B.2.4 Interfaces, I/O-Behaviors and Time

In this section we introduce a theory of component behaviors and interface abstraction. Then we discuss issues of time and causality.

Definition B.2.9 contains a signature view on components in terms of syntactic interfaces and continue with a behavioral view.

**Definition B.2.5 (Timed streams)**

*TStreams*1

extend *Streams*2

| | |
|---|---|
| $(M^*)^\infty$ | set of timed streams |
| $P©^t s, S©^t s$ | filter (predicate or set) |
| $P\#^t s, S\#^t s$ | number of messages in *S* / *P* |
| $f^{*t}(s)$ | map function on messages |
| $\diamond s \in M^\omega$ | untimed message stream of *s* |

$\forall\, t \in \mathbb{N}, s \in (M^*)^\infty$ :
$(P©^t s).t = (P©s.t)$
$(S©^t s).t = (S©s.t)$
$P\#^t s = \#(P©^t s)$
$S\#^t s = \#(S©^t s)$
$f^{*t}(s).t = f^*(s.t)$
$(\diamond s) = \frown_{t\in\mathbb{N}} s.t$

$(M^*)^\infty$ is the set of all infinite, timed streams. Filters $P©s$, $P\#s$ and mapping $f^*(s)$ are adapted to work on individual messages rather than time slices. If its clear from the context, suffix$^t$ is omitted. $\diamond s$ is the list of untimed messages of *s*, where all timing information was removed.

**Definition B.2.6 (Channels)**

*Channel*1

extend *TStreams*1

$SHist : D \times (D \to \mathbb{P}(M)) \to \mathbb{P}(D \to (M^*)^\infty)$

Notation:
$\mathcal{H}^S(C)$ is shorthand for $SHist(C, S)$
with $S(c_i) = (M_i^*)^\infty$
$\mathcal{H}\{c_1 : M_1, \ldots, c_n : M_n\}$ is shorthand for
$SHist(\{c_1, \ldots, c_n\}, S)$ with $S(c_i) = (M_i^*)^\infty$
$[c_1 = s_1, \ldots, c_n = s_n]$ is a shorthand for the channel history
$r : \{c_1, \ldots, c_n\} \to (M^*)^\infty$ with $r(c_i) = s_i \in (M_i^*)^\infty$

$SHist(D, S)$ describes the set of channel histories for channels *D* with messages of sort $S(c)$ for each channel $c \in D$. The notations used for channel assignments $[\ldots]$ and for channel types $\mathcal{H}$ help accommodate an easier reading. The types of channels are omitted, if clear from the context.

The syntactic interface $(I \rhd O)$ does not say much about the behavior of a component. Basically it only fixes the basic steps of information exchange possible for the component and its environment.

For input/output information processing devices there is a crucial dependency of output on input. Certain output messages depend on certain input messages. A crucial notion for interactive systems is therefore causality. Causality indicates dependencies between the messages exchanged within a system and can be expressed through the order of occurrence in the timing. It describes, which output message is a reaction on which input. I/O-behaviors generate their output and consume their input in a time frame. This time frame is useful to characterize causality between input and output. Output that depends causally on certain input cannot be generated before this input has been received.

**Definition B.2.7 (Channel operations)**

*Channel*2

extend *Channel*1

| | |
|---|---|
| $HIST(D,S) = \bigcup_{E \subseteq D} SHist(E,S)$ | all histories |
| $channels(s)$ | set of channels |
| $s.c \in (M^*)^\infty$ | stream on channel $c \in E$ |
| $(s \oplus r)$ | sum of the histories $s$ and $r$ |
| $s \mid_C$ | restriction on $C$ |

$\forall\, s \in \mathcal{H}^S(D_1), r \in \mathcal{H}^S(D_2), c \in D_1 \cup D_2$ :
$channels(s) = dom(s) = D_1$
$c \in channels(s) \Rightarrow s.c = s(c)$
$(s \oplus r) \in \mathcal{H}^S(D_1 \cup D_2)$
$channels(s \oplus r) = channels(s) \cup channels(r) = D_1 \cup D_2$
$c \in D_1 \wedge c \notin D_2 \Rightarrow (s \oplus r).c = s.c$
$c \in D_2 \Rightarrow (s \oplus r).c = r.c$
$channels(s \mid_C) = channels(s) \cap C$
$(s \oplus r) \mid_{channels(r)} = r$
$D_1 \cap D_2 = \varnothing \Rightarrow (s \oplus r) \mid_{channels(s)} = s$

$HIST(D,S)$ is the set of all channel histories for subsets of $D$. $s \oplus r$ and $s \mid_C$ are defined as known from functions. Please note that $r$ has precedence in $s \oplus r$.

**Definition B.2.8 (Definition of channels time slices)**

*Channel*3

extend *Channel*2

| | |
|---|---|
| $s.t \in (E \to M^*)$ | time slice on time $t \in \mathbb{N}$ |
| $\mathcal{T}^S(D) = \{s.t \mid s \in \mathcal{H}^S(D), t \in \mathbb{N}\}$ | set of slices |

$\forall\, s, r \in HIST(E,S), t \in \mathbb{N}, c \in E, C \subseteq E$ :
$channels(s.t) = channels(s)$
$(s.c).t = (s.t).c$
$(s \oplus r).t = (s.t) \oplus (r.t)$
$(s \mid_C).t = (s.t) \mid_C$

$\mathcal{T}^S(D)$ is in analogy to $\mathcal{H}^S(E)$ the set of channel time slices. All operations on histories carry forward to time slices, as dealing with channels and time slicing are orthogonal: $(s.c).t = (s.t).c$.

Both predicates *properlytimed* and *timeguarded* describe these constraints between input and output. A function $F$ is properly timed if the output up to the $t$-th time interval does not depend on input that is received after time $t$. This ensures that there is a proper time flow for the component modeled by $F$. $F$ cannot predict the future input and react on it.

If F were not properly timed, there would exist a time $t$ and input histories $x$ and $x'$ such that $x \downarrow t = x' \downarrow t$ and $(F.x) \downarrow t \neq (F.x') \downarrow t$. A difference between $x$ and $x'$ occurs only after time $t$, but at time $t$ the reactions of $F$ in terms of output messages are already different. Thus $F$ then could predict the future.

Nevertheless, the defined notion of causality permits instantaneous reaction [BG92]: the output at time $t$ may depend on the input at time $t$. This may lead to problems with causality between input and output,

**Definition B.2.9 (Component interfaces and behavior)**

*Behavior*

extend *Channel*3

| | |
|---|---|
| $I \rhd^S O$ | syntactic interface of a component |
| $\mathcal{B}^S(I, O)$ | behaviors of a component |

$timeguarded, properlytimed : \mathcal{H}^S(I) \to \mathbb{P}(\mathcal{H}^S(O)) \to Bool$

$\mathcal{B}^S(I, O) = \{F : \mathcal{H}^S(I) \to \mathbb{P}(\mathcal{H}^S(O)) \mid timeguarded(F)\}$
$properlytimed(F) =$
$\quad \forall t \in \mathbb{N}, s, r \in \mathcal{H}^S(I) : s \downarrow t = r \downarrow t \Rightarrow (F.s) \downarrow t = (F.r) \downarrow t$
$timeguarded(F) =$
$\quad \forall t \in \mathbb{N}, s, r \in \mathcal{H}^S(I) : s \downarrow t = r \downarrow t \Rightarrow (F.s) \downarrow (t+1) = (F.r) \downarrow (t+1)$

$I \rhd^S O$ describes the syntactic interface of a component, namely the channels $I$ it reads and the channels $O$ it writes on. $\mathcal{B}^S(I, O)$ is the set of time guarded behaviors for that component. $F.i$ describes the output histories that may be returned for any input history $i \in \mathcal{H}^S(I)$. The set $F.i$ can be empty. Furthermore $I$ and $O$ can overlap allowing feedback loops.
$S$ is again typing information for the channels (and will be omitted for brevity in the following).

if we consider in addition delay free feedback loops known as causal loops. To avoid these problems we strengthen the concept of proper time flow to the notion of strong causality, or so called "time guardedness". As Lemma B.2.10 states, *timeguarded* is stronger than *properlytimed*.

**Lemma B.2.10 (Properties of behavior functions)**

use *Behavior*

$timeguarded(F) \Rightarrow properlytimed(F)$

Strong causality simply enforces components to introduce a delay of one time unit before it can react. If the granularity of time units is fine enough, we can always detect such a delay. If time units are defined at a proper granularity, we can always observe some delay for the output to happen. So time guardedness correctly reflects what we see in reality.

In general, an I/O-behavior $F : \mathcal{H}(I) \to \mathbb{P}(\mathcal{H}(O))$ allows many implementations, as it allows many reactions to one input. Each one of the possible implementations can be described as a deterministic descendant of this behavior. Such an implementation is given through a deterministic function $f : \mathcal{H}(I) \to \mathcal{H}(O)$ as given in Definition B.2.11.

Even though the definition of $F$ is complete ($\forall s : F.s \neq \emptyset$), there need not necessarily be a deterministic descendant. However, such deterministic descendants describe proper implementations. *realizable* ensures the existence of such implementations. Lemma B.2.12 describes properties of behaviors.

## B.2.5 Composition of Interface Behavior

In this section, we introduce an operator for the composition of components. We prefer to introduce only one very general, powerful composition operator in Definition B.2.13. This operator generalizes sequential and parallel composition as well as introduction of feedback loops.

**Definition B.2.11 (Deterministic component implementation)**

*Implementation*

extend *Behavior*

$\subseteq: \mathcal{B}(I,O) \rightarrow \mathcal{B}(I,O)$
$complete, deterministic, realizable : \mathcal{B}(I,O) \rightarrow Bool$
$det : \mathcal{B}(I,O) \rightarrow (\mathcal{H}(I) \rightarrow \mathcal{H}(O))$

$G \subseteq F \Leftrightarrow \forall\, s \in \mathcal{H}(I) : F.s \subseteq G.s$
$complete(F) =$
$\quad \forall\, s \in \mathcal{H}(I) : F.s \neq \varnothing$
$deterministic(F) =$
$\quad \forall\, s \in \mathcal{H}(I) : \#(F.s) = 1$
$deterministic(F) \Rightarrow \exists^1 f : det(F) = f$ where
$\quad \forall\, s \in \mathcal{H}(I) : F.s = \{f.s\}$
$realizable(F) =$
$\quad \exists\, G \in \mathcal{B}(I,O) : G \subseteq F \wedge deterministic(G)$

$G \subseteq F$ describes refinement of $F$ into a more deterministic version $G$. *deterministic* determines whether there is exactly one function. *det* translates to that function and *realizable* determines whether there are such deterministic descendants.

**Lemma B.2.12 (Properties of realizability)**

use *Implementation*

$deterministic(F) \wedge complete(F)$
$deterministic(F) \wedge realizable(F)$
$deterministic(F) \wedge complete(G) \wedge G \subseteq F \Rightarrow F = G$
$deterministic(F) \wedge realizable(G) \wedge G \subseteq F \Rightarrow F = G$
$deterministic(G) \wedge G \subseteq F \Rightarrow realizable(F)$
$deterministic(G) \wedge G \subseteq F \Rightarrow complete(F)$
$realizable(F) \Rightarrow complete(F)$

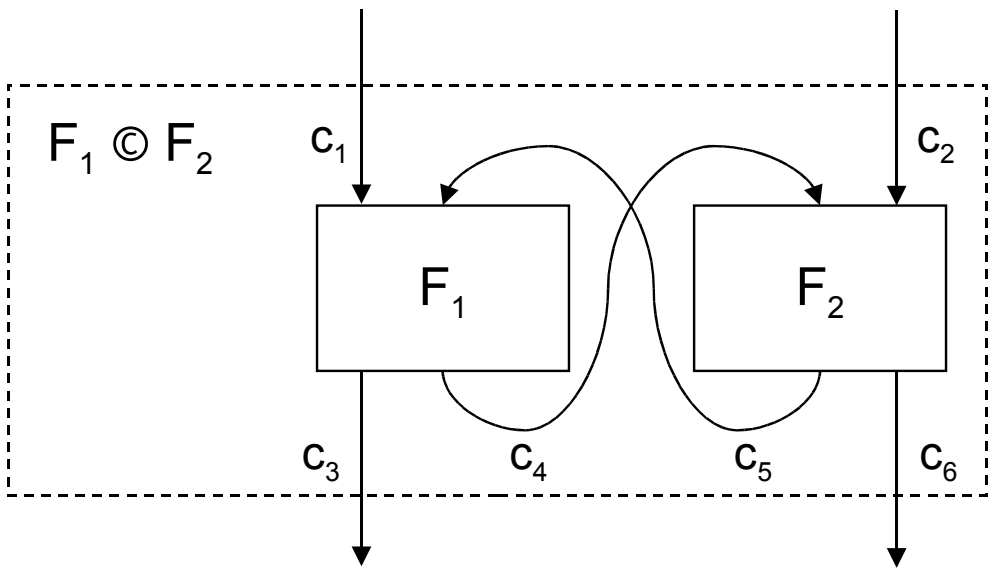

Figure B.1: Parallel composition with feedback

Here, $y$ denotes the history for all the internal, input and output channels in the composition. The composition formula essentially says that all the streams on output channels of the components $F_1$ and $F_2$ are feasible output streams of these components.

**Definition B.2.13 (Component composition)**

*Focus*

extend *Implementation*

$$\otimes : \mathcal{B}(I_1, O_1) \times \mathcal{B}(I_2, O_2) \to \mathcal{B}(I, O)$$

$$\begin{array}{l}
\forall F_1 \in \mathcal{B}(I_1, O_1), F_2 \in \mathcal{B}(I_2, O_2), \\
\quad I = (I_1 \cup I_2) \setminus (O_1 \cup O_2), \\
\quad O = (O_1 \cup O_2) \setminus (I_1 \cup I_2) : \\
O_1 \cap O_2 = \varnothing \Rightarrow \\
\quad F_1 \otimes F_2 \in \mathcal{B}(I, O) \wedge \\
\quad \forall x \in \mathcal{H}(I) : (F_1 \otimes F_2).x = \\
\qquad \{(y|_O) \mid y \in \mathcal{H}(I \cup O_1 \cup O_2) \wedge y|_I = x|_I \wedge \\
\qquad\qquad y|_{O_1} \in F_1(y|_{I_1}) \wedge y|_{O_2} \in F_2(y|_{I_2})\}
\end{array}$$

$F_1 \otimes F_2$ is the central composition of behaviors. In the composition $y$ contains all channel histories involved.

Figure B.1 illustrates parallel composition with feedback where the feedback channels are not externally visible in the composed system.

Please note that the only restriction for composition is that output channels are disjoint, since each channel is output of one component only.

Please note that there are slightly generalized versions of composition operators possible. E.g., the actual feedback resp. the channels that are not hidden can be added as a parameter, such that some internally used channels still are available outside and $I \cap O = \varnothing$ not necessarily holds. Second, the typing information $S$ is assumed to be globally the same, but could be defined as joinable. Third, it would be possible to define composition in such a way that not every component needs to be time guarded, but in each feedback loop, at least one time guarded component needs to be involved (a technique known from electrical circuit design).

It is straightforward to prove that the composition is strongly causal, if one component is strongly causal. If both components are deterministic, then so is the composition. If both components are realizable, then so is the composition. See Lemma B.2.14.

**Lemma B.2.14 (Properties of behavior function composition)**

use *Focus*

$$\begin{array}{l}
\forall F_i, G_i \in \mathcal{B}(I_k, O_k) : \\
complete(F_1, F_2) \Rightarrow complete(F_1 \otimes F_2) \\
deterministic(F_1, F_2) \Rightarrow deterministic(F_1 \otimes F_2) \\
realizable(F_1, F_2) \Rightarrow realizable(F_1 \otimes F_2) \\
G_1 \subseteq F_1 \Rightarrow (G_1 \otimes F_2) \subseteq (F_1 \otimes F_2) \\
G_2 \subseteq F_2 \Rightarrow (F_1 \otimes G_2) \subseteq (F_1 \otimes F_2) \\
F_1 \otimes F_2 = F_2 \otimes F_1 \\
O_1 \cap O_3 = \varnothing \Rightarrow \\
\quad (F_1 \otimes F_2) \otimes F_3 = F_1 \otimes (F_2 \otimes F_3)
\end{array}$$

But as the most important result, the composition is designed in such a way, that it is compatible with independent development of its parts. This means, given a composition, we can chose a deterministic implementation for each part individually, compose these and get a deterministic implementation of the composition. Again see Lemma B.2.14.

And finally, the composition is associative and commutative, which allows generalizing the composition operator to any (signature compatible) set of components - including infinite sets.

# B.3 State Transition Systems

As objects react on incoming messages, state transition systems are an appropriate way of describing object behavior. Several forms of state transition systems and their compositions are used in the report. Therefore, we introduce the basics of STS as a general technique here.

## B.3.1 STS-Definition

The theory used here is based on the theory of automata, but was partly enhanced in [Rum96] to describe a form of automata, called $I/O^*$-automata, where transitions are triggered by one incoming message and the effect of this message, namely a sequence of possible outputs is the output of the same transition. In contrast to $I/O$-automata [LT89], this form allows to abstract away from many internal states of the automaton, which are necessary, if each output is triggered by an individual transition. The application of $I/O^*$-automata to our description of objects is given in Definition B.3.1.

**Definition B.3.1 ($I/O^*$-STS)**

*STS*

$$
\begin{aligned}
& \mathit{STS}(S, I, O) = \\
& \quad \{(S, I, O, \delta, s0) \mid s0 \subseteq S \wedge s0 \neq \varnothing \\
& \quad\quad \wedge\, \delta \in S \times I \to \mathbb{P}(S \times O^*) \\
& \quad\quad \wedge\, \forall\, s \in S, i \in I : \delta(s, i) \neq \varnothing\}
\end{aligned}
$$

Notation:

$\delta : s \xrightarrow{i/o} t$ is shorthand for $(o, t) \in \delta(s, i)$

$\mathit{STS}(S, I, O)$ is the set of all, possibly underspecified STS with given state, input and output sets. An STS has a complete transition relation as $\delta(s, i) \neq \varnothing$ for all $s, i$.

As can be seen from the definition the transition function is nondeterministic. This allows to model underspecification and thus multiple behaviors in the STS. As discussed in [Rum96], this underspecification may be resolved during design time by the developer or during runtime by the system itself taking the choice according to some random circumstances, sensor input, etc.

The semantics of such an STS is defined in [Rum96] using stream processing functions in the form of [BDD$^+$93]. These stream processing functions allow composition, behavioral refinement etc.

However, STS themselves are not fully compositional regarding compositionality of the state space. But there are quite a number of techniques to combine smaller STS to a larger STS.

## B.3.2 Deterministic STS

STS allow us to constrain behavior, but keep underspecification at desired places. If a behavior can be fully defined, underspecification is unnecessary and we can use a deterministic STS as given in Definition B.3.2.

**Definition B.3.2 (Deterministic STS)**

*DSTS*

$$
\begin{array}{l}
DSTS(S, I, O) = \\
\quad \{(S, I, O, \delta, s0) \in STS \mid \\
\quad\quad \#s0 = 1 \land \forall s \in S, i \in I : \#\delta(s, i) = 1\}
\end{array}
$$

## B.4 Timed State Transition Systems

Timed state transition systems do not directly use events to make their steps, but time progress. A timed state machine equidistantly performs its steps as time progresses and consumes all messages arriving at that time. As a big advantage, we cannot only integrate time into the specification technique, but also have composition operators at hand that are compatible with the composition on streams.

### B.4.1 Definition of Timed State Transition Systems

A timed state transition system (TSTS) is a STS where each transition resembles a time step. Such a time step can handle several input messages and produce several outputs. TSTS are therefore defined in B.4.1. Here $I$ and $O$ play the roles of channels, which are typed by the channel typing function $c : (I \cup O) \to \mathbb{P}(M)$.

**Definition B.4.1 (Timed STS)**

*TSTS1*

$$
\begin{array}{l}
TSTS^c(S, I, O) = \{(S, \mathcal{T}^c(I), \mathcal{T}^c(O), \delta, s0) \in STS(S, \mathcal{T}^c(I), \mathcal{T}^c(O)) \mid \\
\quad \forall \delta : s \xrightarrow{i/o} t \Rightarrow \#o = 1 \land \\
\quad \forall \delta : s \xrightarrow{i/o} t, i' : \exists t' : \delta : s \xrightarrow{i'/o} t' \\
\}
\end{array}
$$

$TSTS^c(S, I, O)$ is the set of all, possibly underspecified STS that resemble timed object behavior. A TSTS has a complete transition relation.

The restriction $\#o = 1$ in TSTS is not a real one, as by definition $o \in (\mathcal{T}(O))^*$ which can be regarded as equivalent to $o \in \mathcal{T}(O)$. Instead, we could also use a flatting operator on $o$. The simplified representation of the timed transition function $\delta$, which will now be used is thus

$$\delta : (S \times \mathcal{T}^c(I)) \to \mathbb{P}(S \times \mathcal{T}^c(O))$$

where $\mathcal{T}^c(I)$ denotes the set of channel time slices for the channels in $I$ like described in Definition B.2.8.

The second restriction models the fact that the state transition function describes the behavior of a Moore machine [Kat93]. The output $o$ therefore only depends on the start state $s$, but not on the input $x$ as for all other inputs $x'$ the same output $y$ is happening, too.

One way to interpret this rule is that the granularity of time is fine enough to trace state changes in such a detailed way that the reaction to an input is always delayed by at least one time unit (one state transition step). As an immediate consequence, feedback cycles include a time step and thus preserve causality. Another consequence is that the output of a transition is independent of the input of this transition and, therefore, intermediate storage for either the input before being processed or the resulting output in the state space is inevitable.

### B.4.2 Composition of TSTS

Timed state transition systems are compositional. The following Definition B.4.2 describes an appropriate composition operator.

**Definition B.4.2 (Component composition)**

*TSTSComposition*

extend *TSTS*1

$\otimes : TSTS(S_1, I_1, O_1) \times TSTS(S_2, I_2, O_2) \to TSTS(S, I, O)$

$$
\begin{aligned}
&\forall A_k = (S_k, \mathcal{T}(I_k), \mathcal{T}(O_k), \delta_k, s0_k) \in TSTS(S_k, I_k, O_k),\\
&\quad I = (I_1 \cup I_2) \setminus (O_1 \cup O_2),\\
&\quad O = (O_1 \cup O_2) \setminus (I_1 \cup I_2),\\
&\quad S = S_1 \times S_2 \wedge s0 = s0_1 \times s0_2 :\\
&O_1 \cap O_2 = \varnothing \Rightarrow\\
&\quad A_1 \otimes A_2 = (S, \mathcal{T}(I), \mathcal{T}(O), \delta, s0) \in TSTS(S, I, O) \wedge\\
&\quad \forall x \in \mathcal{T}(I), (s_1, s_2) \in S : \delta((s_1, s_2), x) =\\
&\qquad \{((t_1, t_2), y|_O) \mid y \in \mathcal{T}(I \cup O_1 \cup O_2) \wedge y|_I = x|_I \wedge\\
&\qquad\qquad (t_1, y|_{O_1}) \in \delta_1(s_1, y|_{I_1}) \wedge (t_2, y|_{O_2}) \in \delta_2(s_2, y|_{I_2})\}
\end{aligned}
$$

$A_1 \otimes A_2$ composes TSTS. In the composition $y$ contains the time slices of all channels involved.

Given that the output only depends on the state, the composition formula is always uniquely fulfilled and thus yields a precisely defined TSTS. The formula expresses that the input to the composed machine is split into input to the first machine and input to the second machine. With this input, and possibly additional input from feedback, both machines carry out their transition and produce output. The new states of the small machines define the new state of the composed machine; the output of the composed machine is built using the output of the small machines.

When comparing the composition for behaviors as given in B.2.13 and this composition, is becomes apparent that both operations are structurally very similar. The Figure B.1 for behaviors is the same for state machines is one replaces $F_k$ by the state machines $A_k$. This becomes even more apparent, when comparing the properties of TSTS-composition with behavior composition. Lemma B.4.3 exhibits just commutativity and associativity. Thus composition can be generalized to any finite and (by induction) also infinite set of TSTS.

A proof of well-foundedness of this composition can, e.g., be found in [GR95].

**Lemma B.4.3 (Properties of TSTS composition)**

use *TSTSComposition*

$$
\begin{aligned}
&\forall A_i \in TSTS(S_k, I_k, O_k) :\\
&A_1 \otimes A_2 = A_2 \otimes A_1\\
&O_1 \cap O_3 = \varnothing \Rightarrow\\
&\quad (A_1 \otimes A_2) \otimes A_3 = A_1 \otimes (A_2 \otimes A_3)
\end{aligned}
$$

Preservation of determinism etc. are also given, but not further investigated here, as they come for free when we connect TSTS with interface behaviors.

### B.4.3 Interface Behavior and Interface Abstraction

State machines can provide detailed models for systems, because the structure of the state is shown explicitly. However, if a state is encapsulated, a representation of the system behavior without considering the structure of states seems most appropriate. This is essentially what we call the interface of a system. The interface abstraction of a state machine is made explicit using the translator in Definition B.4.4.

**Definition B.4.4 (Interface abstraction for a TSTS)**

*TSTSAbstraction*

extend *TSTSComposition*

$\mathbb{S}[.] : TSTS(S,I,O) \to \mathcal{B}(I,O)$
$B : TSTS(S,I,O) \to S \to \mathcal{B}(I,O)$

$\forall A = (S, \mathcal{T}(I), \mathcal{T}(O), \delta, s0) \in TSTS(S,I,O):$
$\forall s \in S, i \in \mathcal{T}(I), x \in \mathcal{H}(I):$
$B^A(s)(i \frown x) = \{o \frown y \mid \exists\, o,t : \delta : s \xrightarrow{i/o} t \wedge y \in B^A(t)(x)\}$
$\mathbb{S}[A] \in \mathcal{B}(I,O)$
$\forall x \in \mathcal{H}(I) : \mathbb{S}[A](x) = \bigcup_{s \in s0} B^A(s)(x)$

$B^A$ is a recursively defined functions which equals the behavior of the TSTS. Instantiated with a start state, it results in the I/O-behavior $\mathbb{S}[A]$ that describes the behavior of TSTS $A$.

Definition B.4.4 basically consists of a recursively defined I/O-behavior function $B^A$ that exactly equals the behavior induced by the transitions of $\delta$. $B^A$ therefore provides the interface abstraction of the state transition function $\delta$. This interface abstraction produces a behavioral description that excludes internal states and transition steps, such that the overall behavior becomes easier to grasp.

Please note that the recursive definition of $B^A$ has several solutions. We use the inclusion maximal solution, which is uniquely defined, because the right hand side of the equation is inclusion monotonic in $B^A$. Thus $B^A$ is recursively defined by an inclusion monotonic function which even is guarded. Hence there exists a unique inclusion maximal solution. $B^A(s)$ defines an I/O-behavior for any initial state $s \in s0$, which represents the behavior of the component described by the state machine $\delta$ if initialized by the state $s$. The element-wise union of the resulting I/O-behaviors for all $s \in s0$ yields again an I/O-behavior.

The fact that the TSTS is input enabled and provides the behavior of a Moore machine (output reactions have at least one time unit delay) guarantees that $\mathbb{S}(A)$ is a time guarded I/O-behavior. Proofs for this can be found in [GR95].

Lemma B.4.5 states that composing TSTS and deriving their abstraction to I/O-behaviors gives exactly the same result as deriving the abstraction to I/O-behaviors and composing them.

**Lemma B.4.5 (TSTS composition is fully abstract)**

use *TSTSAbstraction*

$\forall A_i \in TSTS(S_k, I_k, O_k):$
$\mathbb{S}(A_1 \otimes A_2) = \mathbb{S}(A_1) \otimes \mathbb{S}(A_2)$

The proof is done by induction over the time intervals and can also be found for example in [GR95] on a variation of this approach.

Now we have two models for systems available: state machines and interface behaviors that are fully compatible. This is an important property of the construction of this system model, as it demonstrates that mapping UML constructs to either interface behaviors or state machines is both possible.

# C Glossary: Summary of all Signatures

This chapter contains a repetition of all signatures defined in previous chapters and thus serves as a glossary.

**Definition C.1.1 (Type infrastructure summary)**

*Type*

extend $Type1, Variable, BoolInt, Void$

---

$UTYPE$
$UVAL$
$CAR : UTYPE \rightarrow \mathbb{P}(UVAL)$

---

$Bool, Int \in UTYPE$
$true, false \in UVAL$

---

$Void \in UTYPE$
$void \in UVAL$

---

$UVAR$
$vtype : UVAR \rightarrow UTYPE$
$vsort : UVAR \rightarrow \mathbb{P}(UVAL)$
$VarAssign = RECORD(UVAR, vsort)$

---

This is a collection of all basic elements for types, variables and values.

**Definition C.1.2 (Object infrastructure summary)**

*Object*

extend *Type*;
extend *Class*, *Attribute*, *Nil*, *Subclassing*

---

$UCLASS, \quad UOID, \quad INSTANCE$
$attr : UCLASS \rightarrow \mathbb{P}_f(UVAR)$
$oids : UCLASS \rightarrow \mathbb{P}(UOID)$
$objects : UCLASS \rightarrow \mathbb{P}(INSTANCE)$
$objects : UOID \rightarrow \mathbb{P}(INSTANCE)$
$classOf : INSTANCE \rightarrow UCLASS$
$classOf : UOID \rightarrow UCLASS$

---

$this : INSTANCE \rightarrow UOID$
$getAttr : INSTANCE \times UVAR \rightharpoonup UVAL$
$attr : INSTANCE \rightarrow \mathbb{P}_f(UVAR)$
$attr : UOID \rightarrow \mathbb{P}_f(UVAR)$

---

$Nil \in UOID$

---

$sub \subseteq UCLASS \times UCLASS$
$.^{\&} : UCLASS \rightarrow UTYPE$

---

This theory contains all elements for objects, identifiers and classes.

**Definition C.1.3 (The DataStore with all its elements)**

*Data*

extend *Association*;

---

$DataStore \subseteq (UOID \rightarrow INSTANCE)$
$oids : DataStore \rightarrow \mathbb{P}(UOID)$

---

$val : DataStore \times UOID \times UVAR \rightharpoonup UVAL$
$setval : DataStore \times UOID \times UVAR \times UVAL \rightharpoonup DataStore$
$addobj : DataStore \times INSTANCE \rightarrow DataStore$

---

$UASSOC$
$classes : UASSOC \rightarrow List(UCLASS)$
$extraVals : UASSOC \rightarrow \mathbb{P}(UVAL)$
$relOf : UASSOC \times DataStore \rightarrow \mathbb{P}(UVAL \times UVAL)$

---

This theory defines the data state of the system.

**Definition C.1.4 (Object infrastructure summary)**

*Control*

extend $\mathit{ControlStore}, \mathit{TypeSafeOps}$

---

$\mathit{UOPN}, \mathit{UOMNAME}$
$\mathit{nameOf} : \mathit{UOPN} \to \mathit{UOMNAME}$
$\mathit{classOf} : \mathit{UOPN} \to \mathit{UCLASS}$
$\mathit{parTypes} : \mathit{UOPN} \to \mathit{List}(\mathit{UTYPE})$
$\mathit{params} : \mathit{UOPN} \to \mathit{TUPLE}(\mathit{UVAL})$
$\mathit{resType} : \mathit{UOPN} \to \mathit{UTYPE}$

---

$\mathit{UMETH}, \mathit{UPC}$
$\mathit{nameOf} : \mathit{UMETH} \to \mathit{UOMNAME}$
$\mathit{definedIn} : \mathit{UMETH} \to \mathit{UCLASS}$
$\mathit{parNames}, \mathit{localNames} : \mathit{UMETH} \to \mathit{List}(\mathit{UVAR})$
$\mathit{parOf}, \mathit{localsOf} : \mathit{UMETH} \to \mathit{VarAssign}$
$\mathit{resType} : \mathit{UMETH} \to \mathit{UTYPE}$
$\mathit{pcOf} : \mathit{UMETH} \to \mathbb{P}_f(\mathit{UPC})$

---

$\mathit{impl} : \mathit{UOPN} \rightharpoonup \mathit{UMETH}$

---

$\mathit{FRAME} = \mathit{UOID} \times \mathit{UOMNAME} \times \mathit{VarAssign} \times \mathit{UPC} \times \mathit{UOID}$
$\mathit{framesOf} : \mathit{UMETH} \to \mathbb{P}(\mathit{FRAME})$

---

$\mathit{UTHREAD}$
$\mathit{CentralControlStore} \subseteq (\mathit{UTHREAD} \to \mathit{Stack}(\mathit{FRAME}))$

---

$\mathit{ControlStore} \subseteq (\mathit{UOID} \to \mathit{UTHREAD} \to \mathit{Stack}(\mathit{FRAME}))$
$. \sim . \subseteq \mathit{CentralControlStore} \times \mathit{ControlStore}$

---

This theory describes the stacks and threads within the control store.

**Definition C.1.5 (Event and message summary)**

*Events*

extend *Control*;
extend *EventStore*, *Message*, *Signal*;
extend *MethodCall*, *MethodReturn*

---

$UEVENT$
$eventsIn : UOID \to \mathbb{P}(UEVENT)$
$eventsOut : UOID \to \mathbb{P}(UEVENT)$
$EventStore \subseteq (UOID \to Buffer(UEVENT))$

---

$UMESSAGE$
$MsgEvent : UMESSAGE \to UEVENT$
$sender, receiver : UMESSAGE \to UOID$
$msgIn, msgOut : UOID \to \mathbb{P}(UMESSAGE)$

---

$callsOf : UOID \times UOPN \times UOID \times UTHREAD \to \mathbb{P}(UMESSAGE)$
$callsOf : UOID \to \mathbb{P}(UMESSAGE)$

---

$returnsOf : UOID \times UOPN \times UOID \times UTHREAD \to \mathbb{P}(UMESSAGE)$
$returnsOf : UOID \to \mathbb{P}(UMESSAGE)$

---

$USIGNAL \subseteq UMESSAGE$

---

This theory describes events, messages, and the event store.

**Definition C.1.6 (Object state summary)**

*State*

extend *ObjectStates*2

---

$STATE \subseteq DataStore \times ControlStore \times EventStore$
$oids : STATE \to \mathbb{P}(UOID)$
$OSTATE = INSTANCE \times (UTHREAD \to Stack(FRAME))$
$\quad \times Buffer(UEVENT)$
$state : STATE \times UOID \to OSTATE$
$states : UOID \to \mathbb{P}(OSTATE)$

---

$state : STATE \times \mathbb{P}(UOID) \to (UOID \to OSTATE)$
$states : \mathbb{P}(UOID) \to \mathbb{P}(UOID \to OSTATE)$

---

This theory describes the object states.

**Definition C.1.7 (ESTS summary)**

*ESTS*

extend *EventSTS*;

---

$\dagger : UTHREAD \rightarrow \dagger(UTHREAD)$

---

$ests : UOID \rightarrow STS(S, I, O)$

---

ESTS are event driven STS and describe object behavior. They stepwise handle incoming events and allow to describe interleaving and concurrency on a rather fine grained level.

**Definition C.1.8 (TSTS summary)**

*TSTS*

extend *ESTStoTSTS*;

---

$UCN$
$sender, receiver : UCN \rightarrow UOID$
$channel : UEVENT \rightarrow UCN$
$inC, outC : UOID \rightarrow \mathbb{P}(UCN)$
$csort : UCN \rightarrow \mathbb{P}(UEVENT)$
$beh : UOID \rightarrow \mathcal{B}^{csort}(I, O)$
$beh : \mathbb{P}(UOID) \rightarrow \mathcal{B}^{csort}(I, O)$
$inC, outC : \mathbb{P}(UOID) \rightarrow \mathbb{P}(UCN)$

---

$tsts : UOID \rightarrow TSTS^{csort}(S_1, I_1, O_1)$
$tsts : \mathbb{P}(UOID) \rightarrow TSTS^{csort}(S, I, O)$

---

TSTS are time driven STS. They describe object behavior in a timed form. A step handles all incoming events of that time frame.

# D List of Figures

# E List of Definitions

# F List of Lemmata and Variation Points

# G List of Example Definitions

Technische Universität Braunschweig
Informatik-Berichte ab Nr. 2004-01

| | | |
|---|---|---|
| 2004-01 | T.-P. Fries, H. G. Matthies | A Review of Petrov-Galerkin Stabilization Approaches and an Extension to Meshfree Methods |
| 2004-02 | B. Mathiak, S. Eckstein | Automatische Lernverfahren zur Analyse von biomedizinischer Literatur |
| 2005-01 | T. Klein, B. Rumpe, B. Schätz (Herausgeber) | Tagungsband des Dagstuhl-Workshop MBEES 2005: Modellbasierte Entwicklung eingebetteter Systeme |
| 2005-02 | T.-P. Fries, H. G. Matthies | A Stabilized and Coupled Meshfree/Meshbased Method for the Incompressible Navier-Stokes Equations — Part I: Stabilization |
| 2005-03 | T.-P. Fries, H. G. Matthies | A Stabilized and Coupled Meshfree/Meshbased Method for the Incompressible Navier-Stokes Equations — Part II: Coupling |
| 2005-04 | H. Krahn, B. Rumpe | Evolution von Software-Architekturen |
| 2005-05 | O. Kayser-Herold, H. G. Matthies | Least-Squares FEM, Literature Review |
| 2005-06 | T. Mücke, U. Goltz | Single Run Coverage Criteria subsume EX-Weak Mutation Coverage |
| 2005-07 | T. Mücke, M. Huhn | Minimizing Test Execution Time During Test Generation |
| 2005-08 | B. Florentz, M. Huhn | A Metamodel for Architecture Evaluation |
| 2006-01 | T. Klein, B. Rumpe, B. Schätz (Herausgeber) | Tagungsband des Dagstuhl-Workshop MBEES 2006: Modellbasierte Entwicklung eingebetteter Systeme |
| 2006-02 | T. Mücke, B. Florentz, C. Diefer | Generating Interpreters from Elementary Syntax and Semantics Descriptions |
| 2006-03 | B. Gajanovic, B. Rumpe | Isabelle/HOL-Umsetzung strombasierter Definitionen zur Verifikation von verteilten, asynchron kommunizierenden Systemen |
| 2006-04 | H. Grönniger, H. Krahn, B. Rumpe, M. Schindler, S. Völkel | Handbuch zu MontiCore 1.0 - Ein Framework zur Erstellung und Verarbeitung domänenspezifischer Sprachen |
| 2007-01 | M. Conrad, H. Giese, B. Rumpe, B. Schätz (Hrsg.) | Tagungsband Dagstuhl-Workshop MBEES: Modellbasierte Entwicklung eingebetteter Systeme III |
| 2007-02 | J. Rang | Design of DIRK schemes for solving the Navier-Stokes-equations |
| 2007-03 | B. Bügling, M. Krosche | Coupling the CTL and MATLAB |
| 2007-04 | C. Knieke, M. Huhn | Executable Requirements Specification: An Extension for UML 2 Activity Diagrams |
| 2008-01 | T. Klein, B. Rumpe (Hrsg.) | Workshop Modellbasierte Entwicklung von eingebetteten Fahrzeugfunktionen, Tagungsband |
| 2008-02 | H. Giese, M. Huhn, U. Nickel, B. Schätz (Hrsg.) | Tagungsband des Dagstuhl-Workshop MBEES: Modellbasierte Entwicklung eingebetteter Systeme IV |
| 2008-03 | R. van Glabbeek, U. Goltz, J.-W. Schicke | Symmetric and Asymmetric Asynchronous Interaction |
| 2008-04 | M. V. Cengarle, H. Grönniger, B. Rumpe | System Model Semantics of Statecharts |
| 2008-05 | M. V. Cengarle, H. Grönniger, B. Rumpe | System Model Semantics of Class Diagrams |
| 2008-06 | M. Broy, M. V. Cengarle, H. Grönniger, B. Rumpe | Modular Description of a Comprehensive Semantics Model for the UML (Version 2.0) |